\documentclass[useAMS,usenatbib,usegraphicx]{mn2e}
\usepackage[usenames,dvipsnames]{color}

\title[Jet-regulated cooling catastrophe]
  {Jet-regulated cooling catastrophe}

\author[Y. Dubois et al.]
  {Yohan~Dubois,$^1$\thanks{E-mail: yohan.dubois@physics.ox.ac.uk} Julien~Devriendt,$^{1,2}$ Adrianne~Slyz,$^1$ Romain Teyssier$^{3,4}$ \\
  $^1$Astrophysics, University of Oxford, Denys Wilkinson Building, Keble Road, Oxford, OX13RH, United Kingdom\\
    $^2$Centre de Recherche Astronomique de Lyon, 9 Avenue Charles Andr\'e, 69561, St-Genis-Laval Cedex, France\\
  $^3$Universit\"at Z\"urich, Institute f\"ur Theoretische Physik, Winterthurerstrasse 190, CH-8057 Z\"urich, Switzerland\\
  $^4$CEA Saclay, DSM/IRFU/SAP, B\^atiment 709, F-91191 Gif-sur-Yvette, Cedex, France
}
\date{Accepted 2010 July 09. Received 2010 June 04; in original form 2010 April 11}

\pagerange{\pageref{firstpage}--\pageref{lastpage}} \pubyear{2010}

\def\LaTeX{L\kern-.36em\raise.3ex\hbox{a}\kern-.15em
    T\kern-.1667em\lower.7ex\hbox{E}\kern-.125emX}

\begin{document}

\label{firstpage}

\maketitle

\begin{abstract}
We present the first implementation of Active Galactic Nuclei (AGN) feedback in the form of momentum driven jets 
in an Adaptive Mesh Refinement (AMR) cosmological resimulation of a galaxy cluster.
The jets are powered by gas accretion onto Super Massive Black Holes (SMBHs) which also grow by mergers. Throughout its formation, the cluster experiences different 
dynamical states: both a morphologically perturbed epoch at early times and a relaxed state at late times allowing us to study the different modes
of BH growth and associated AGN jet feedback. BHs accrete gas efficiently at high redshift ($z > 2$),  significantly pre-heating proto-cluster halos.
Gas-rich mergers at high redshift also fuel strong, episodic jet activity, which transports gas from the proto-cluster core to its outer regions. 
At later times, while the cluster relaxes, the supply of cold gas onto the BHs is reduced leading to lower jet activity. Although the cluster is still heated
by this activity as sound waves propagate from the core to the virial radius, the jets inefficiently redistribute gas outwards and a small cooling 
flow develops, along with low-pressure cavities similar to those detected in X-ray observations. Overall, our jet implementation
of AGN feedback quenches star formation quite efficiently, reducing the stellar content of the central cluster galaxy by a factor 3 compared
to the no AGN case. It also dramatically alters the shape of the gas density profile, bringing it in close agreement with the $\beta$ model favoured
by observations, producing quite an isothermal galaxy cluster for gigayears in the process. However, it still falls short in matching the lower than 
Universal baryon fractions which seem to be commonplace in observed galaxy clusters.   
\end{abstract}


\section{Introduction}

The well-known over--cooling problem in galaxy formation is encountered in numerical simulations for a range of galaxy sizes, spanning dwarves to red ellipticals.
Stellar feedback mechanisms, such as winds from young stars and supernovae explosions are potentially good candidates for expelling large amounts of cold gas from galaxies.
Unfortunately, these mechanisms are inadequate for the most massive galaxies, where the energy liberated by star formation activity is too low to unbind material from their
gravitational potential wells.

Different theories find ways to partially or totally prevent the cooling catastrophe in the most massive structures: galaxy groups and clusters.
Propagating heat with a Spitzer conductivity from the outskirts to the central parts of a cluster, thermal conduction could solve the cooling problem for the more
massive clusters \citep{voigt&fabian04}.
However, due to the propagation of electrons along magnetic field lines, the conduction is essentially anisotropic leading to instabilities (such as the Heat flux Buoyancy Instability,
 HBI, see \citealp{quataert08, parrish&quataert08})
in the cluster core.
These can reorient the magnetic field lines and stop the net inflow of heat towards the centre \citep{parrish09, bogdanovic09}. Pre-heating of the gas destined to fall into
a cluster potential well during proto-galactic stages has also been proposed as a means to empty gas on galactic scales in these massive halos \citep{babuletal02}.

However, in most models of massive galaxy formation, AGN play a crucial role in regulating their gas content.
Red and dead galaxies commonly exhibit the signatures of SMBHs which are thought to power high velocity jets
into the hot surroundings of the galaxies. Observational evidence for strong AGN activity in groups and clusters is plenty
\citep{arnaudetal84, carillietal94, mcnamaraetal01, mcnamaraetal05, fabianetal02, birzanetal04, formanetal07}.
When spatially resolved, this activity often takes the form of radio lobes or cavities \citep{boehringeretal93, owenetal00, birzanetal04, 
mcnamaraetal05, fabianetal06, tayloretal06, dongetal10, dunnetal10}, or thin extended jets \citep{bridleetal94}.
The radio lobes or cavities are associated with low AGN activity, i.e. the radio mode, and jets with the quasar modes.

AGN feedback is invoked to efficiently suppress star formation
in the most massive galaxies either by ejecting gas from their Interstellar Medium (ISM) into the Intra-Cluster Medium (ICM), or by preventing the ICM gas from
collapsing into galactic discs \citep{binney&tabor95, rephaeli&silk95}. The powerful ejection of gas by AGN is also supposed to suppress the formation of cool cores in some
fraction of galaxy clusters, and in turn to make the ICM turbulent \citep{duboisetal09}.

Many of the previous numerical studies that attempt to simulate the growth of BHs and their associated AGN feedback have been performed using the Smoothed Particle
Hydrodynamics (SPH) technique \citep{sijackietal07, dimatteoetal08, booth&schaye09}. As shown by \cite{mitchelletal09}, SPH codes suffer from underestimating
the true entropy\footnote{Here and later, the entropy is defined as $K=T \rho^{-2/3}$, where $T$ is the gas temperature and $\rho$ is the gas mass density} profile in cluster cores 
because of their trouble resolving Kelvin-Helmoltz instabilities in regions of strong density contrast \citep{agertzetal07}.
As it is generally assumed that BHs accrete gas at a Bondi rate which is related to the local entropy
$\dot M_{\rm BH}\propto K^{-3/2}$, a poor estimate of this entropy leads to an incorrect calculation of the accretion rate onto a BH and, thus, its energy release.
To circumvent this issue, we use a sink particle approach to follow the growth and AGN feedback of BHs within an Adaptive Mesh Refinement (AMR) code.

Moreover, in previous cosmological simulations, AGN feedback is modeled with a thermal input of energy. Meanwhile, theoretical work and observations suggest that AGN feedback  
is mostly mechanical, not thermal. Numerous numerical simulations have implemented and tested the formation and propagation of AGN jets on cluster scales and their impact on the ICM using either 
idealized simulations \citep{churazovetal01, quilisetal01, reynoldsetal01, basson&alexander03, ommaetal04, ruszkowskietal04, vernaleo&reynolds06, cattaneo&teyssier07, simionescuetal09, gaibleretal09, 
oneill&jones10}, or cosmological simulations \citep{heinzetal06, morsonyetal10}, but none so far have followed the BH growth self-consistently (i.e. resolving both the BH growth and jet-AGN feedback
 in cosmological simulation over a Hubble time). As the AGN feedback is tightly linked to its BH growth history, it is of crucial importance to model both the BH evolution through time and its 
jet-energy release.
In this paper, we propose to bridge this gap by performing the first cosmological simulation including a self-consistent treatment of BH evolution and its associated AGN jet energy release.

The paper is organized as follows. In section~\ref{physics}, we describe the physical ingredients of our simulation. We start by presenting our cooling and star formation prescriptions and 
then we introduce the scheme for following the formation and mergers of BHs, gas accretion onto them, and their energy release in the form of jets. In section~\ref{IC}, we describe our initial 
condition set-up and our simulation run. In section~\ref{BHgrowth}, we show how the BH growth is linked to the galaxy cluster formation history, and what drives the different modes of AGN feedback 
(radio versus quasar). In section~\ref{Cooling_catastrophe}, we demonstrate that this type of anisotropic mechanical AGN feedback is able to suppress the cooling catastrophe occuring in the
 galaxy cluster. Finally, in section~\ref{discussion}, we discuss our results.

\section{Modeling the physics of galaxy formation}
\label{physics}

\subsection{Modeling star formation}
\label{sec:sf}
Gas in our simulation is allowed to radiate energy by atomic collisions in a H/He primordial gas \citep{sutherland&dopita93} so that it can collapse into
dark matter potential wells to form galaxies \citep{silk77}. To model reionization from $z=8.5$, heating from a UV background is followed with the prescriptions from \cite{haardt&madau96}.
Star formation occurs in high density regions $\rho> \rho_0$ ($\rho_0=0.1\, \rm H.cm^{-3}$). When the density threshold is surpassed, a random Poisson process spawns
star cluster particles according to a Schmidt law $\dot \rho_*= \epsilon \rho/t_{\rm ff}$, where $t_{\rm ff}$ is the gas free-fall
time and $\epsilon$ is the star formation efficiency, taken to be $\epsilon=0.02$ in order to reproduce the observational surface density
laws \citep{kennicutt98}. The reader can consult \cite{rasera&teyssier06} and \cite{dubois&teyssier08winds} for more information on the star formation implementation.

In this work, we do not model supernova feedback. Several authors have argued that supernovae can only have a dynamical impact on 
low mass galaxies \citep{springel&hernquist03, dubois&teyssier08winds}. Thus, as a first order approximation, we assume that supernova feedback has
very little effect on the growth of BHs in massive galaxies as they alone appear incapable of removing a substantial fraction of gas from the ISM. 
Whilst this simplification allows us to properly isolate the effect of
AGN feedback on the surrounding gas from any other galactic feedback mechanism, it does not allow us to study the role of metal enrichment
on cooling in the centre of massive halos. However, we will see that the main features of a cooling catastrophe are already captured with zero metallicity cooling.

Finally it is possible to view the modification of the temperature at high density $\rho> \rho_0$ by a
polytropic equation of state (EoS) that we introduce for numerical reasons in section~\ref{sec:accretion} as a way  
to take into account the thermal effect of the heating of the ISM by supernovae.
As a matter of fact, a similar EoS approach is used by other authors (e.g. \citealp{springel&hernquist03}) as a simple model for the unresolved multiphase 
structure of the ISM in cosmological simulations. More specifically, the minimum temperature in dense regions becomes $T_{\rm min}=T_0 (\rho/\rho_0)^{n-1}$, 
with $T_0=10^4K$, and $n=4/3$ which leads to a constant Jeans mass $M_{\rm J}=1.3\, 10^9 \, \rm M_{\odot}$. Such a value of the polytropic index $n$ roughly compares with 
the complex functional form of the EoS obtained by analytical considerations on the multiphase structure of the ISM in \cite{springel&hernquist03}.

\subsection{SMBHs as sink particles}
\label{sink_algorithm}
Sink particles were first introduced by \cite{bateetal95} in a SPH code. Sinks are massive particles that capture gas particles in their surroundings.
They  mimic the formation of unresolved compact objects, e.g. proto-stellar cores in the ISM, black holes in the ISM,
central super-massive black holes in galaxies, etc. Due to the very Lagrangian nature of the sink particle technique, it had been extensively and
exclusively used in SPH codes until \cite{krumholzetal04} extended its use to grid codes. The version in {\sc ramses} \citep{teyssier02} is strongly inspired by
the \cite{krumholzetal04} numerical implementation.

Sink particles are created in regions where the Jeans criterion is violated, i.e. in regions where the maximum level of refinement is reached and
where the gas density is large enough to potentially produce a numerical instability, in other words where:
\begin{equation}
{\Delta x\over 4} >  \lambda_{\rm J}=\sqrt{\pi c_s^2\over G \rho}\, .
\label{L_Jeans}
\end{equation}
Here $\Delta x$ is the size of the smallest cell, $\lambda_{\rm J}$ the Jeans length, $c_s$ the sound speed and $\rho$ the gas density. According to
\cite{trueloveetal97}, the numerical stability of a gravitationally bound object is ensured if it is resolved with at least 4 cells. With a mixed composition
of matter (dark matter, gas, stars), Jeans stability is not trivial anymore, but we can reasonably assume that gas is the dominant source of gravitational
potential inside dense collapsed objects, like galaxies,  in our case.

For numerical stability, each time that the Jeans criterion is violated we should spawn a sink particle with a mass corresponding to the depleted mass.
However, in cosmological simulations this leads to excessively large sink masses. The reason is that the gas is concentrated in structures (galaxies) that are poorly
resolved (kpc scale). As a result an entire galactic disk can be defined by only a few Jeans-violating cells leading to massive sinks.
To form sufficiently small seed BHs in the centres of the galaxies, we prefer to choose their initial mass, $M_{\rm seed}$, thereby introducing a free parameter. 
We set $M_{\rm seed}=10^5 \, \rm M_{\odot}$ in agreement with previous cosmological simulations (e.g. \citealp{booth&schaye09}).
However, BHs are still spawned only in cells belonging to the maximum level of refinement and that verify equation~(\ref{L_Jeans}) . One consequence
of this self-controlled formation of the sink BHs is that they are not allowed to accrete gas when the Jeans criterion is violated. The only way for them to
accrete gas is to do so by a reasonable physical process such as Bondi accretion. With this prescription for initializing the mass of the seed black hole, it is conceivable
that gas could be numerically violently Jeans unstable, but this issue is partially solved by the consumption of gas in the star forming process that temporarily
restores gravitational stability.

To get only one BH per massive galaxy, a halo finder is usually run on-the-fly during the simulation to check if candidate galaxies already host a BH \citep{dimatteoetal05, booth&schaye09}.
We prefer a simpler, more direct, and computationally cheaper approach. To avoid creating multiple sinks inside the same galaxy, we ensure that
each time a cell could potentially produce a sink (i.e. it verifies eq.~(\ref{L_Jeans})), it is farther than a minimum radius $r_{\rm min}$
from all other pre-existing sinks. This distance has to be larger than the typical size of galactic discs and smaller than the typical average
inter-galactic distance.
Test runs suggest that the choice $r_{\rm min}=100$ kpc produces very satisfactory results.

To avoid formation of sink particles in low density regions that are Jeans-unstable, we set a minimum threshold for the density $\rho>\rho_0$ of gas that
can create a new sink, where $\rho_0$ is the same density threshold that we use for star formation. In order to avoid the creation of sink black holes before 
the formation of the very first stars, we check that the star density $\rho_*$ verifies
\begin{equation}
f_*={\rho_*\over \rho_*+\rho} > 0.25 \, ,
\label{fraction_star}
\end{equation}
before a sink is spawned, where $\rho$ is the gas density.

When a sink particle is finally created it is split into several cloud particles with equal mass. Cloud particles are spread over a $4 \Delta x$ radius sphere
and positioned every $0.5 \Delta x$ in (x,y,z). The exact number of cloud particles in this configuration is therefore $n_{\rm cloud}=2109$ per sink. This splitting process is useful
in many ways. First, it keeps a heavy sink particle from becoming the dominant gravitational contribution in one single cell. The latter could catapult particles
far from their host galaxy by two body encounters. Second, it provides a simple canvas over which to compute averaged quantities around the sink, for example
we use it to determine the Bondi accretion rate.

Sinks are also allowed to merge together if they lie at a distance closer than $4\Delta x$ from each other. Mass is conserved in this process and momentum vectors of
the old sinks are simply added to compute the momentum of the new sink.

Finally we insist on the fact that sink positions and velocities are updated in the classical way used to update standard particles such as DM particles.
No correction on their positions and velocities is done to force them to stay near their host galaxy (as could be done with the Halo
finder approach). Thus, weakly bound BHs, such as BHs in satellite galaxies of large groups and clusters, may easily be
stripped from their host galaxy. These BHs behave like star particles that tidal forces compel to populate the stellar halo of massive galaxies.

\subsection{Accretion rate onto SMBHs}
\label{sec:accretion}
Since we fail to resolve the accretion disk around our SMBHs, whose size is sub-parsec even for the most massive ones ($\sim 10^{-3}$ pc according to \citealp{morganetal10} from micro-lensing estimates), we use the common prescription that these BHs accrete gas at a Bondi-Hoyle-Lyttleton rate \citep{bondi52}
\begin{equation}
\dot M_{\rm BH}={4\pi \alpha G^2 M_{\rm BH}^2 \bar \rho \over (\bar c_s^2+v^2) ^{3/2}}
\label{dMBH}
\end{equation}
where $\alpha$ is a dimensionless boost factor ($\alpha\ge 1$), $M_{\rm BH}$ is the black hole mass, $\bar \rho$ is the average gas density, $\bar c_s$ is the average sound speed,
and $v$ is the gas velocity relative to the BH velocity. One of the major difficulties encountered with the computation of the relative gas velocity is that in cosmological runs, the
ISM is poorly resolved and leads to a very thin scale height for galaxies (compared to the resolution). Moreover, due to poor sampling of the gravitational force in the
galactic disc, BHs can slightly oscillates in their host galaxy. For this reason a BH close to the centre of a galaxy can feel the infalling material coming from the halo
at a relative velocity much higher than the typical velocity inside the bulge. As $v$ is not a reliable quantity, we do not compute $v$ as the gas velocity relative
to the sink velocity but we prefer to set it to the average gas velocity dispersion in the ISM which is assumed constant and equal to $\sigma=10\, \rm km.s^{-1}$ \citep{dibetal06, agertzetal09}.

The average density $\bar \rho$ and sound speed $\bar c_s$ are computed around the BH using the cloud particles for this operation, as mentioned in section~\ref{sink_algorithm}.
To compute the averages, the cell in which each particle sits is assigned a weight given by a kernel function $w$, similar to the one used in \cite{krumholzetal04}:
\begin{equation}
w\propto\exp \left( -r^2/r_K^2\right )\, ,
\end{equation}
where $r$ is the distance from the cloud particle to the sink particle and $r_K$ is the radius defined as
\begin{equation}
r_K =
\left\{
\begin{array}{lr}
\Delta x/4 & \,\, r_{\rm BH}<\Delta x /4\, ,\\
r_{\rm BH}&\,\, \Delta x /4 \le r_{\rm BH} \le 2 \Delta x\, , \\
2\Delta x & \,\, r_{\rm BH}>2\Delta x\, .
\end{array}
\right.
\end{equation}
The Bondi-Hoyle radius $r_{\rm BH}$ is given by:
\begin{equation}
r_{\rm BH}={GM_{\rm BH} \over c_s^2}\, ,
\end{equation}
where $c_s$ is the exact sound speed in the cell where the sink lies.

The true accretion rate onto the sink is finally limited by its Eddington rate
\begin{equation}
\dot M_{\rm Edd}={4\pi G M_{\rm BH}m_{\rm p} \over \epsilon_{\rm r} \sigma_{\rm T} c}\, ,
\label{dMEdd}
\end{equation}
where $\sigma_{\rm T}$ is the Thompson cross-section, $c$ is the speed of light, $m_P$ is the proton mass, and $\epsilon_{\rm r}$ is the radiative efficiency, assumed to be equal to $0.1$ 
for the \cite{shakura&sunyaev73} accretion onto a Schwarzschild BH.

The accretion rate is computed at each time step and a fraction $\dot M_{\rm BH} \Delta t / n_{\rm cloud}$ of gas mass is depleted from the cell where the cloud particle lies and is added to
that cloud particle, and its sink mass is updated accordingly. At each coarse time step cloud particles are re-scattered with equal-mass $M_{\rm BH}/n_{\rm cloud}$.
As the timestep does not depend on the accretion speed onto BHs and as low density cells can be close to high density cells, a BH might remove more mass than is acceptable.
To avoid negative densities and numerical instabilities arising from this, we do not allow a cloud particle to deplete more than 25\% of the gas content in a cell.

In such large scale cluster simulations, it is impossible to resolve the scale and the clumpiness of the ISM. To prevent the collapse of the gas from numerical
instabilities and to take into account the mixing of the different phases in the ISM (cold and warm components), we use the polytropic EoS described in 
section \ref{sec:sf}. Applying this EoS means that it is impossible to know the ``true'' density and the ``true''
sound speed in the ISM, thus the accretion rate onto the BHs must be modified. Previous work modeling the accretion rate onto BHs with such a polytropic EoS
set the $\alpha$ parameter to a constant 100 \citep{springeletal05, sijackietal07, dimatteoetal08}. Here we follow the prescription from \cite{booth&schaye09} who show
that $\alpha = (\rho/\rho_0)^2$ is the best parametric choice to match observational laws.

We stress that this polytropic EoS has important consequences on the accretion rate onto BHs in gas rich systems: equation~(\ref{dMBH}) turns into
$\dot M_{\rm BH}\propto{M_{\rm BH}^2 \rho^{5/2}}$, and the temperature dependence is removed. On the other hand, as soon as the cold gas component
has been evaporated by star formation or feedback mechanisms from massive galaxies, the accretion rate of the black hole is, by definition,
the proper Bondi accretion rate. This $\alpha$ boost of the accretion rate is an artificial way of modeling the very fast accretion of gas within
cold and gas-rich galaxies at early epochs, where the clumpiness of the ISM is unresolved in large-scale cosmological simulations.

\begin{figure*}
  \centering{\resizebox*{!}{8.5cm}{\includegraphics{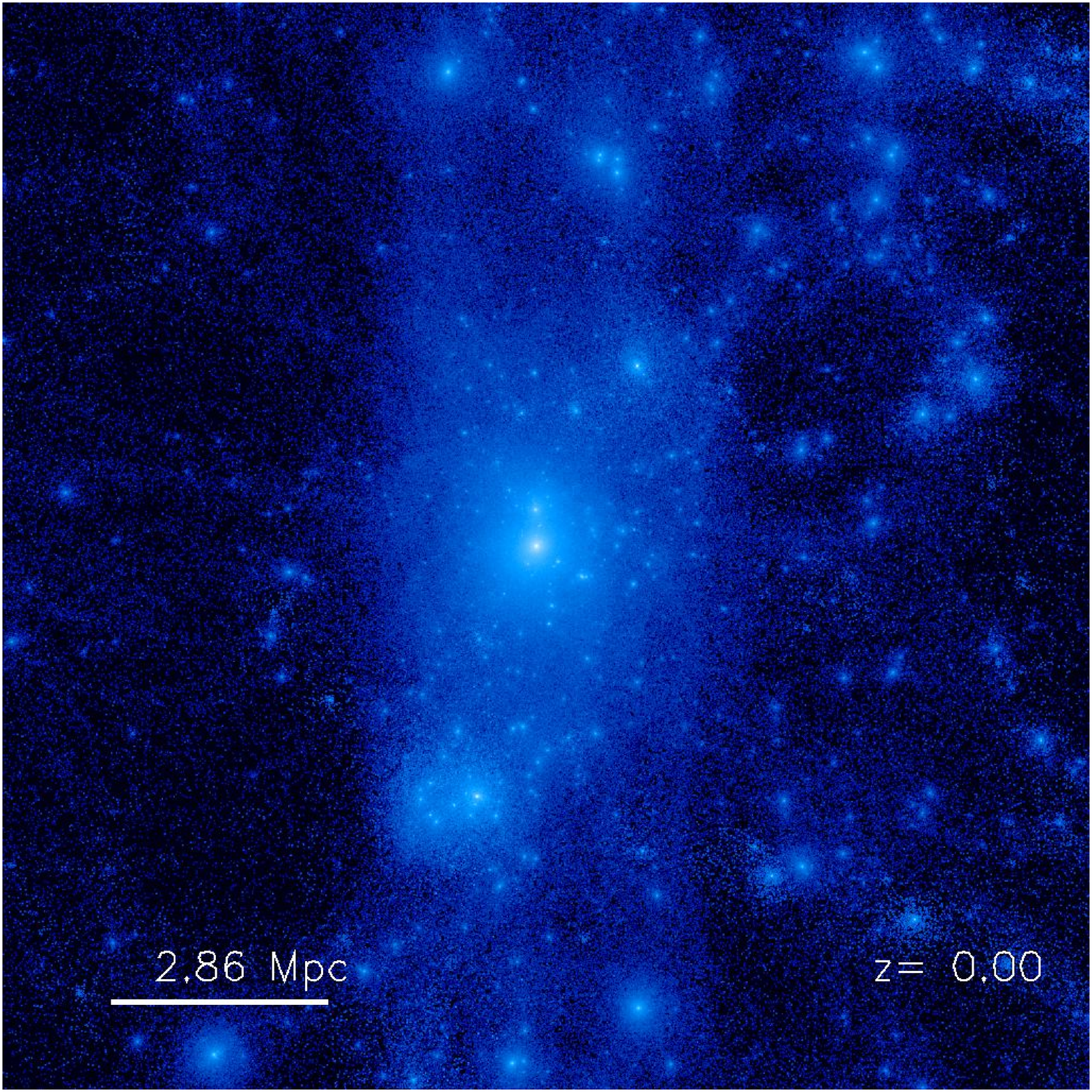}}}
  \centering{\resizebox*{!}{8.5cm}{\includegraphics{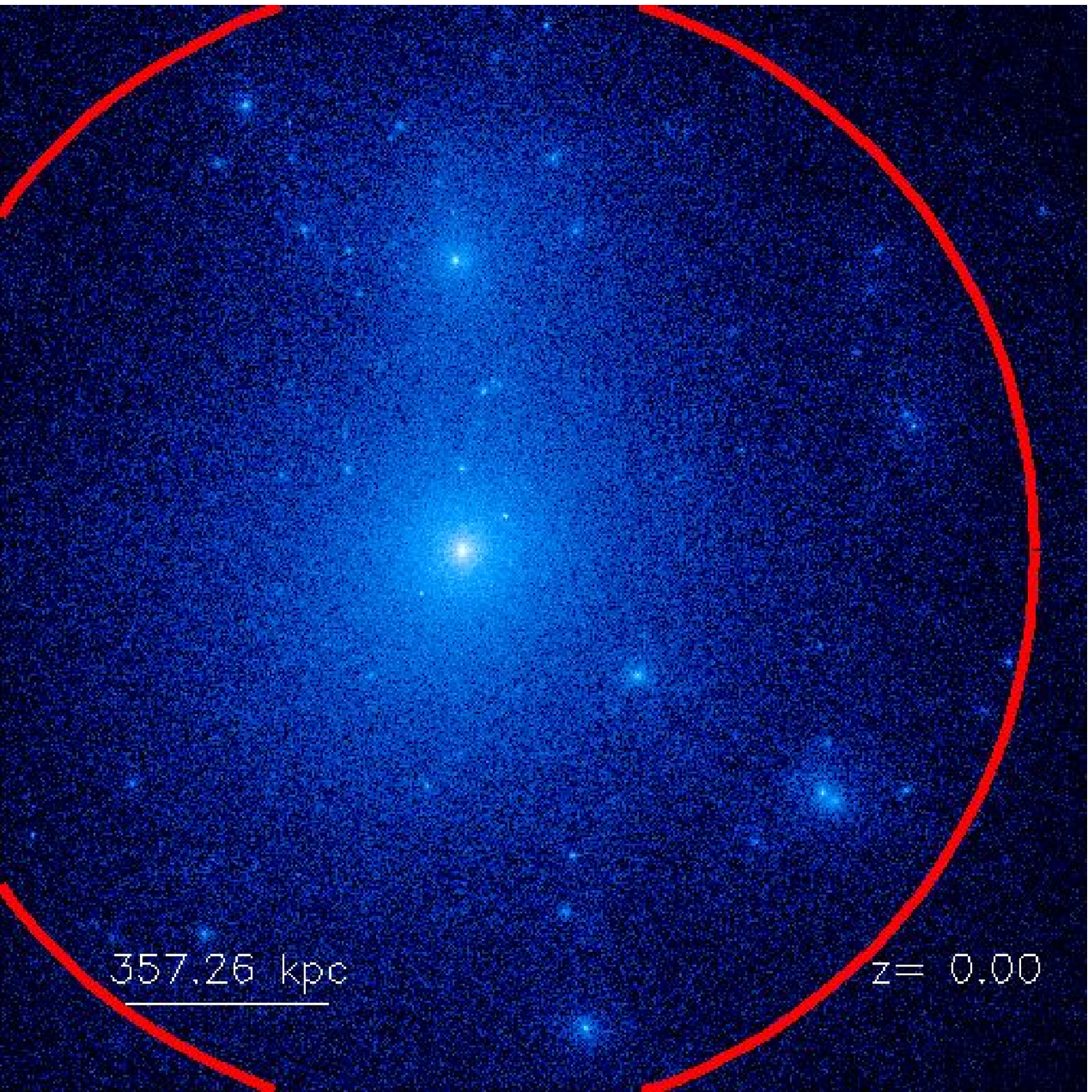}}}
  \centering{\resizebox*{!}{8.5cm}{\includegraphics{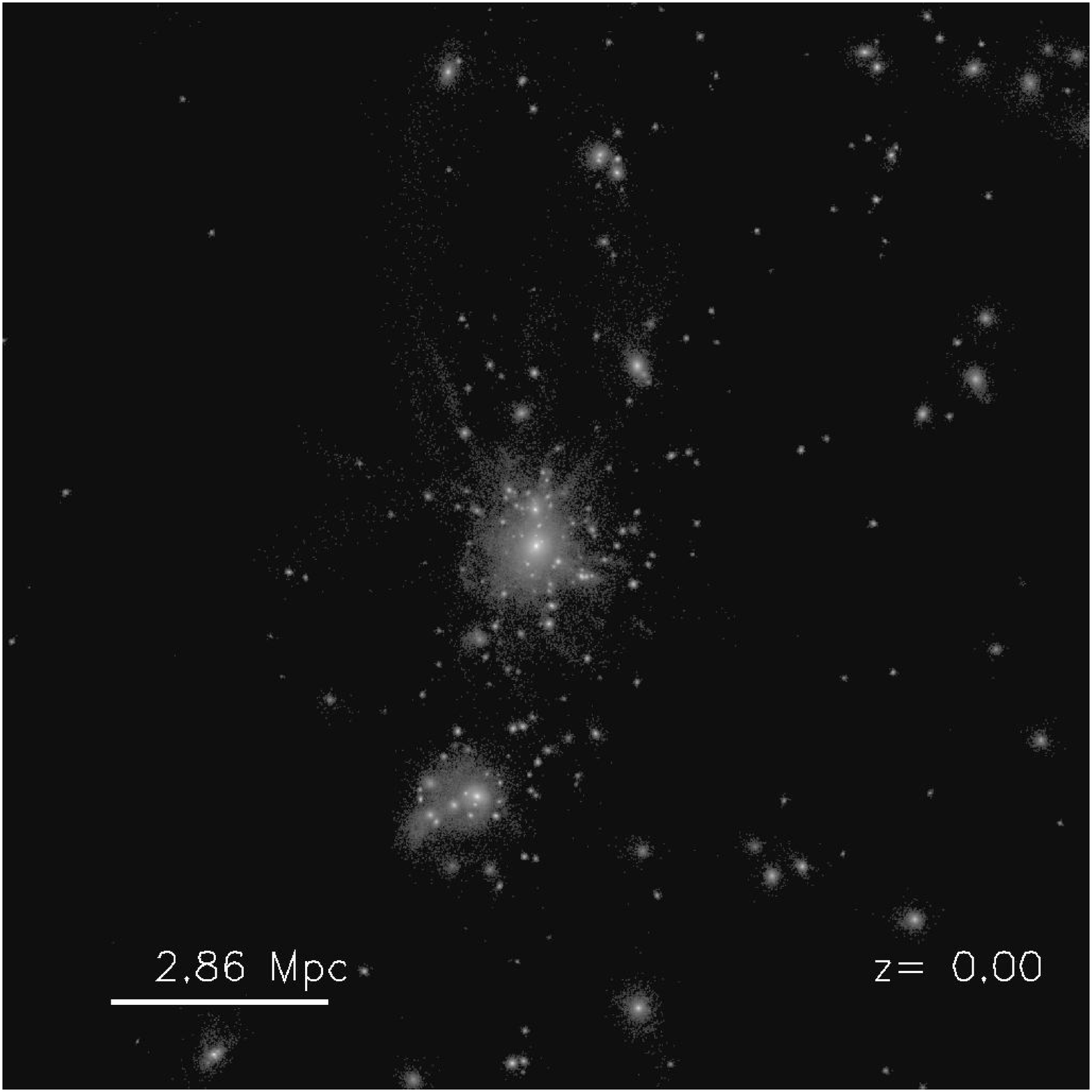}}}
  \centering{\resizebox*{!}{8.5cm}{\includegraphics{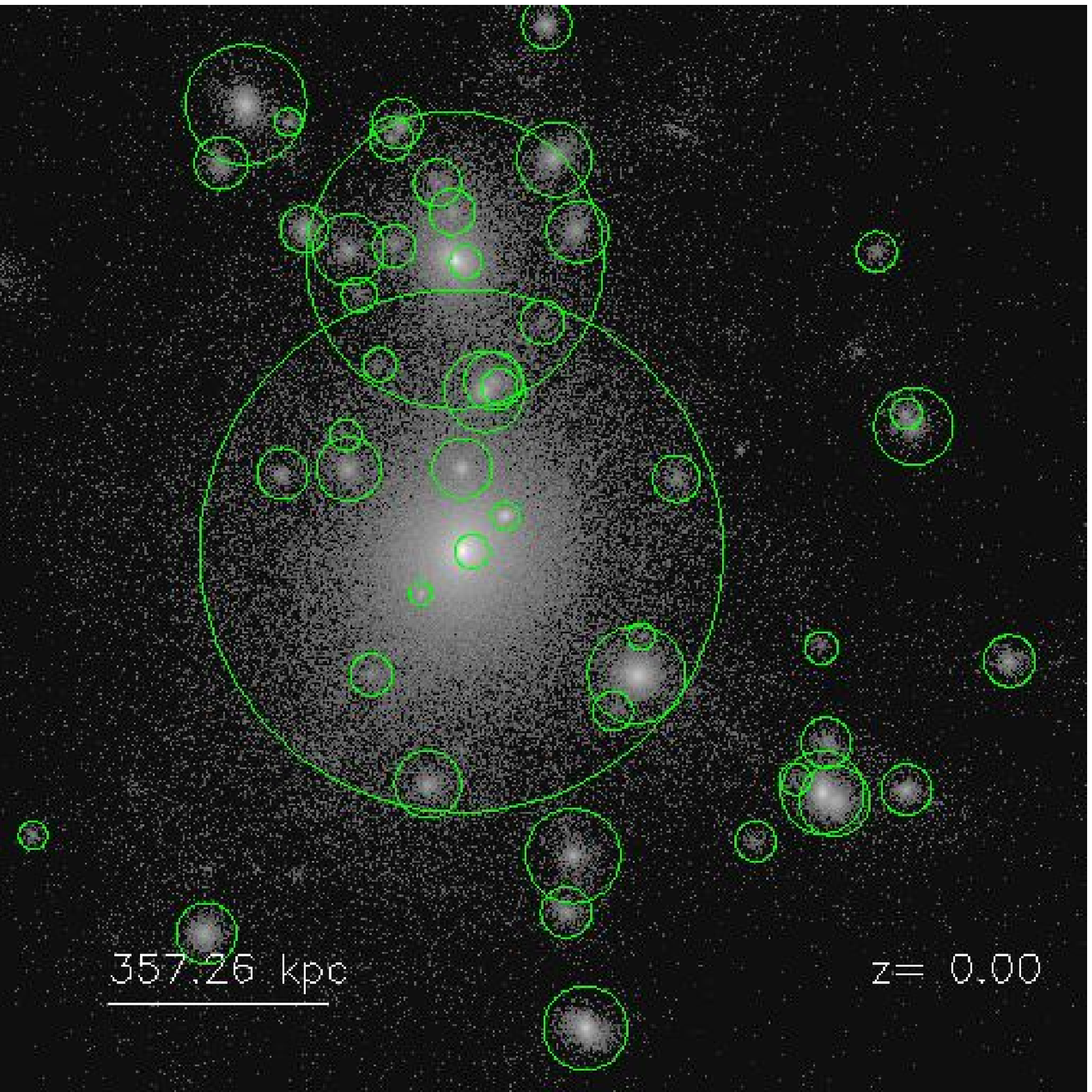}}}
  \caption{Projected DM density (upper panels) and stellar density (bottom panels) in the resimulated 14 Mpc on a side region (left-hand panels) and in the cluster, 1.8 Mpc on a side region (right-hand panels). 
The red circle in the upper right panel shows the $r_{500}=940$ kpc  radius of the DM halo at $z=0$, and green circles in the lower right panel show the virial radii of the stellar  
structures as detected by the halo finder.}
    \label{nice_dm_stars}
\end{figure*}

\subsection{AGN feedback modeling}

Following \cite{ommaetal04}, we assume that the primary source of AGN feedback is a sub-relativistic, momentum-imparting bipolar outflow.
As discussed in detail by these authors, there are many considerations which support this hypothesis, the most blatant one being that bipolar outflows 
are observed around virtually all accreting objects in the Universe: stars, black holes and galaxies. Along with these authors, we further 
assume that the advection dominated inflow-outflow solution (ADIOS) developed by \cite{blandford&begelman99} to explain the low luminosities
of AGNs compared to their estimated Bondi-Hoyle accretion rates is correct. 
The most important feature of the ADIOS model, as far as we are concerned, is that the bulk of the accretion energy (released as plasma falls into the SMBH) drives a (sub-relativistic) wind from the surface of the accretion disk.
We emphasize that, as pointed out in \cite{ommaetal04}, this bipolar wind is distinct from observed relativistic synchrotron jets which are probably powered by the spin of the SMBH itself, although both jets can simultaneously be present. However the synchrotron jet is very likely irrelevant in terms of AGN feedback 
since the mechanical luminosity of the sub-relativistic outflow is much higher than the synchrotron luminosity. Finally, we note that in very dense environments 
it is probable that much of the accretion energy is radiated away, driving outflows with velocities $\approx 0.1 c$ in objects with photon luminosities also on the order 
of the Eddington luminosity \citep{king&pounds03}. We argue that as more and more shock-heated rarefied gas fills the central parts of the growing host halo, 
more and more energy will come out as mechanical energy.        
More specifically, we assume that a fraction $\epsilon_{\rm f}$ of the radiated energy is imparted to the ambient gas
\begin{equation}
\dot E_{\rm AGN}=\epsilon_{\rm f} L_r=\epsilon_{\rm f} \epsilon_{\rm r} \dot M_{\rm BH}c^2\, ,
\label{E_BH}
\end{equation}
in the form of a kinetic jet.

Such an implementation has the advantage of continuously releasing energy without radiating it away by cooling on the scale of a hydro timestep.
Indeed this problem plagues supernovae feedback modelling where energy is generally injected in thermal form \citep{navarro&white93} and leads to the feedback having no 
dynamical impact on the surrounding gas. To bypass this issue, some authors release AGN thermal energy only after a sufficient amount of gas has been accreted onto the BH so as to 
more severely impact the ambient medium \citep{sijackietal07, booth&schaye09, teyssieretal10}. In contrast, we model the AGN feedback as a jet-like structure with the same
momentum profile defined in \cite{ommaetal04}. This model has already been used to follow the self-consistent BH growth and its energy release
in an isolated galaxy cluster \citep{cattaneo&teyssier07, duboisetal09}. Mass, momentum and energy are spread over a small cylinder of radius $r_{\rm J}$ and height 2$h_{\rm J}$ ($h_{\rm J}$ for one side of the jet) multiplied with a kernel window function
\begin{equation}
 \psi \left (r_{\rm cyl} \right)={1 \over 2\pi r_{\rm J}^2} \exp \left( -{r_{\rm cyl}^2\over 2 r_{\rm J}^2 } \right) \, ,
\end{equation}
where $r_{\rm cyl}$ is the distance to the axis of the cylinder. The mass deposition follows
\begin{equation}
\dot M_{\rm J} \left (r_{\rm cyl} \right)={\psi \left (r_{\rm cyl} \right) \over \| \psi \|} \eta \dot M_{\rm BH} \, ,
\end{equation}
where $\| \psi \|$ is the integrated value of $\psi$ over the whole cylinder, and $\eta=100$ is an arbitrary value that represents the mass loading factor of the jet
on unresolved scales. Mass is transferred from the central cell (where the sink lies) to all the enclosed cells within the jet. Momentum $\mathbf{q_{\rm J}}$ is deposited in opposite directions from
the centre along the jet axis, according to
\begin{equation}
\| \mathbf {\dot q_{\rm J} }\| \left (r_{\rm cyl} \right)
= \dot M_{\rm J} \left (r_{\rm cyl} \right)\| \mathbf{u_{\rm J}\|}
={\psi \left (r_{\rm cyl} \right) \over \| \psi \|} \dot M_{\rm BH} \sqrt{2 \epsilon_{\rm f}\epsilon_{\rm r}\eta} c {\mathbf{j}.{\rm d}\mathbf{r}\over \| {\rm d}\mathbf{r} \| } \, ,
\label{mom_input}
\end{equation}
where $\| \mathbf{u_{\rm J}} \| =(2\epsilon_{\rm f}\epsilon_{\rm r}/\eta)^{1/2} c$ is the velocity of the jet ($\| \mathbf{u_{\rm J}} \| \simeq 10^4 \, \rm km.s^{-1}$
for $\epsilon_{\rm f}=1$), $\mathbf{j}$ is the unit spin vector of the BH which defines the jet axis, and ${\rm d}\mathbf{r}$ is the distance vector from the centre of
the black hole. $\mathbf{j}$ is computed by adding the different contributions from the neighbouring cells (sampled with the cloud particles) to the total angular momentum
\begin{equation}
\mathbf{J}=\sum_{i=1}^{n_{\rm clouds}} m_{i} {\rm d}\mathbf{r_{i}}\times \mathbf{u_{i}} \, ,
\end{equation}
where $m_i$ and $\mathbf{u_i}$ are the mass and velocity of the gas in the cell harbouring the cloud particle $i$, 
so that $\mathbf{j}=\mathbf{J}/\| \mathbf{J}\|$. Finally the kinetic energy deposited within a cell is
\begin{equation}
\dot E_{\rm J}\left (r_{\rm cyl} \right)
={\mathbf {\dot q_{\rm J} }^2 \left (r_{\rm cyl} \right) \over 2 \dot M_{\rm J}\left (r_{\rm cyl} \right)}
={\psi \left (r_{\rm cyl} \right) \over \| \psi \|} \dot E_{\rm AGN} \, .
\end{equation}
Integrating this energy deposition over all the cells within the jet, we recover $\dot E_{\rm AGN}$.

We point out that our jet has no opening angle and should therefore propagate along a straight line as it is perfectly collimated. \cite{ommaetal04} have shown that this kind of jet stays collimated over quite long distances ($100$ kpc) compared to its initial broadness and length ($1$ kpc).  It broadens as it reaches equilibrium with the surrounding hot ambient medium. The same behavior is also pointed out by \cite{cattaneo&teyssier07} and \cite{duboisetal09}, but with the difference that when strong turbulent motions begin to develop in the cluster core due to the formation of a cooling flow, the jet is more quickly mixed with the ICM. The choice of the jet velocity input $\| \mathbf{u_{\rm J}} \| \simeq 10^4 \, \rm km.s^{-1}$ (or equivalently the mass loading factor $\eta=100$) is particularly arbitrary but based on earlier works from \cite{ommaetal04}, \cite{cattaneo&teyssier07} and \cite{duboisetal09}. The same simulation performed with $\| \mathbf{u_{\rm J}} \| \simeq 3.10^4 \, \rm km.s^{-1}$ produces results in very strong agreements with the ones presented here (mass of the most massive BH and stellar mass of the central galaxy agrees within 1\%) suggesting that even strong variations of $\| \mathbf{u_{\rm J}} \|$ keep our results unchanged.

We set $r_{\rm J}$ and $h_{\rm J}$ equal to $\Delta x$, and the energy efficiency $\epsilon_{\rm f} = 1$ so as to reproduce the $M_{\rm BH}$--$M_*$ and $M_{\rm BH}$--$\sigma_*$ observational relations.
Larger values of $r_{\rm J}$ and $h_{\rm J}$ have been tested at the same resolution $\Delta x \simeq 1$ kpc in a cosmological simulation (as opposed to a resimulation like the one presented in this paper) 
and they produce BHs which are too massive with respect to their host galaxy. Note that our total efficiency $\epsilon_{\rm r}\epsilon_{\rm f} = 0.1$ is also in good agreement with the average value
 obtained by general relativistic magneto-hydrodynamics numerical simulations of the accretion-ejection mechanism in accretion discs around spinning BHs (e.g. \citealp{devilliersetal05}, 
\citealp{hawley&krolik06}, or \citealp{benson&babul09} and references therein). Lower $\epsilon_{\rm f}$ values again cause black holes to become more massive, overshooting the $M_{\rm BH}$--$M_*$ observational relation.

\section{Simulation Set-up}
\label{IC}

The simulations are run with the Adaptive Mesh Refinement (AMR) code {\sc ramses} \citep{teyssier02}. The evolution of the gas is followed using a second-order unsplit Godunov scheme
for the Euler equations. The Riemann solver used to compute the flux at a cell interface is the acoustic solver using a first-order MinMod Total variation diminishing scheme to
reconstruct the interpolated variables from their cell-centered values. Collisonless particles (dark matter, stars and sink particles) are evolved using a particle-mesh solver
with Cloud-in-Cell (CIC) interpolation.

We assume a flat $\Lambda$CDM cosmology with total matter density $\Omega_{m}=0.3$, baryon density $\Omega_b=0.045$, dark energy density $\Omega_{\Lambda}=0.7$, fluctuation
amplitude at $8 \, h^{-1}.\rm Mpc$ $\sigma_8=0.90$ and Hubble constant $H_0=70\, \rm km.s^{-1}.Mpc^{-1}$ that corresponds to the Wilkinson Microwave Anisotropies Probe (WMAP) 1 year best-fitting cosmology \citep{spergeletal03}. The simulations are performed using a
resimulation (zoom) technique:
the coarse region is a $128^3$ grid with $M_{\rm DM}=2.9\times10^{10} \, \rm M_{\odot}$ DM resolution in a $80\,\rm h^{-1} Mpc$ simulation box. This region contains a smaller $256^3$
equivalent grid in a sphere of radius $20\,\rm h^{-1} Mpc$ with $M_{\rm DM}=3.6\times10^9 \, \rm M_{\odot}$ DM resolution, which in turn encloses the final high resolution sphere with
radius $6\,\rm h^{-1} Mpc$, $512^3$ equivalent grid and $M_{\rm DM}=4.5\times10^8 \, \rm M_{\odot}$ DM resolution. Figure~\ref{nice_dm_stars} shows the distribution of DM and
the distribution of stars in the zoom region and in the galaxy cluster at $z=0$.

The smallest region is the resimulation zone where cells may be refined up to $\ell_{\rm max}=16$ levels of refinement, reaching $1.19 \, \rm h^{-1}.kpc$, following a
quasi--Lagragian criterion: if more than 8 dark matter particles lie in a cell, or if the baryon mass exceeds 8 times the initial dark matter mass resolution, the cell
is refined. This strategy allows  AMR codes, such as {\sc ramses}, which use CIC interpolation in their gravity solver, to avoid propagating discreetness noise from
small scales \citep{romeoetal08}. A Jeans length criterion is also added to ensure the numerical stability of the scheme on all levels $\ell < \ell_{\rm max}$
\citep{trueloveetal97}, and where $\delta \rho=\rho/\bar \rho > 10^5$: the cells fulfilling these conditions must sample the local Jeans length with more than
4 cells. We point out that the $\ell_{\rm max}=16$ level of refinement is only reached at $a_{\rm exp}=(1+z)^{-1}=0.8$, and that the actual maximum level of
refinement for a given redshift is increased as the expansion factor grows with time, i.e. $\ell_{\rm max}=15$ at $a_{\rm exp}=0.4$, $\ell_{\rm max}=14$ at $a_{\rm exp}=0.2$, etc.
This allows us to resolve the smallest scales with a roughly constant physical size ($0.95<\Delta x<1.9\, \rm h^{-1}$.kpc), rather than a constant comoving size.

\begin{figure}
  \centering{\resizebox*{!}{12cm}{\includegraphics{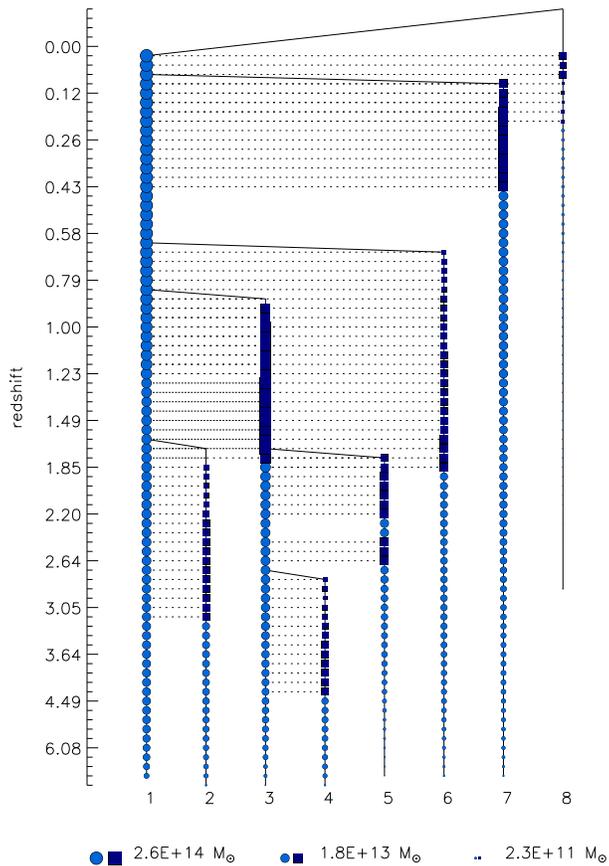}}}
  \caption{Dark matter halo merger history of the resimulated galaxy cluster. Only the 8 most massive branches and sub-branches are shown. Halos are designated with light blue circles 
and sub-halos with dark blue squares. The final merger between a sub-halo and its host halo is represented by a solid line connecting both objects. Note that the cluster main branch 1
experiences two major mergers, one with branch 2 at $z \sim 3.1$ and one with branch 3 at $z \sim 1.7$. However, these mergers are completed at $z \sim 1.7$ and $z \sim 0.8$ respectively,
epochs which coincide with the merger of the central galaxies hosted in these halos. Central BHs merge later (see Fig~\ref{mbhvsredshift}).}
    \label{mergertree_DM}
\end{figure}

The resimulated region tracks the formation of a galaxy cluster with a 1:1 major merger occurring at $z=0.8$ .
Indeed throughout its formation, the cluster chosen for resimulation passes through different dynamical stages: both a morphologically perturbed epoch occurring at
half the age of the Universe, and a relaxed state at late times, which permits us to study the different associated states of the BH self-regulated growth.
Figure~\ref{mergertree_DM} shows the dark matter halo merger tree history for this cluster. 
Halos and sub-halos are identified and followed using the Most massive Sub-node Method (MSM) algorithm described in~\cite{tweedetal09}, which isolates
bound substructures. The cluster experiences a major halo merger at $z\simeq 1.7$ and two proto-clusters progenitors merge together
earlier at $z\simeq 3.1$. These mergers end at $z \sim 0.8$ and $z \sim 1.7$ respectively, when 
the central galaxies hosted in the (sub)halos merge. Central BHs hosted by these galaxies merge later, as shown in Fig~\ref{mbhvsredshift}.
At later times, most of the mass growth of the cluster occurs through diffuse accretion or minor mergers. In the following, we discuss how
such events might trigger or halt the AGN activity of the central (most massive) BH.

\section{Growth of  SMBHs and their activity}
\label{BHgrowth}

\begin{figure}
  \centering{\resizebox*{!}{8cm}{\includegraphics{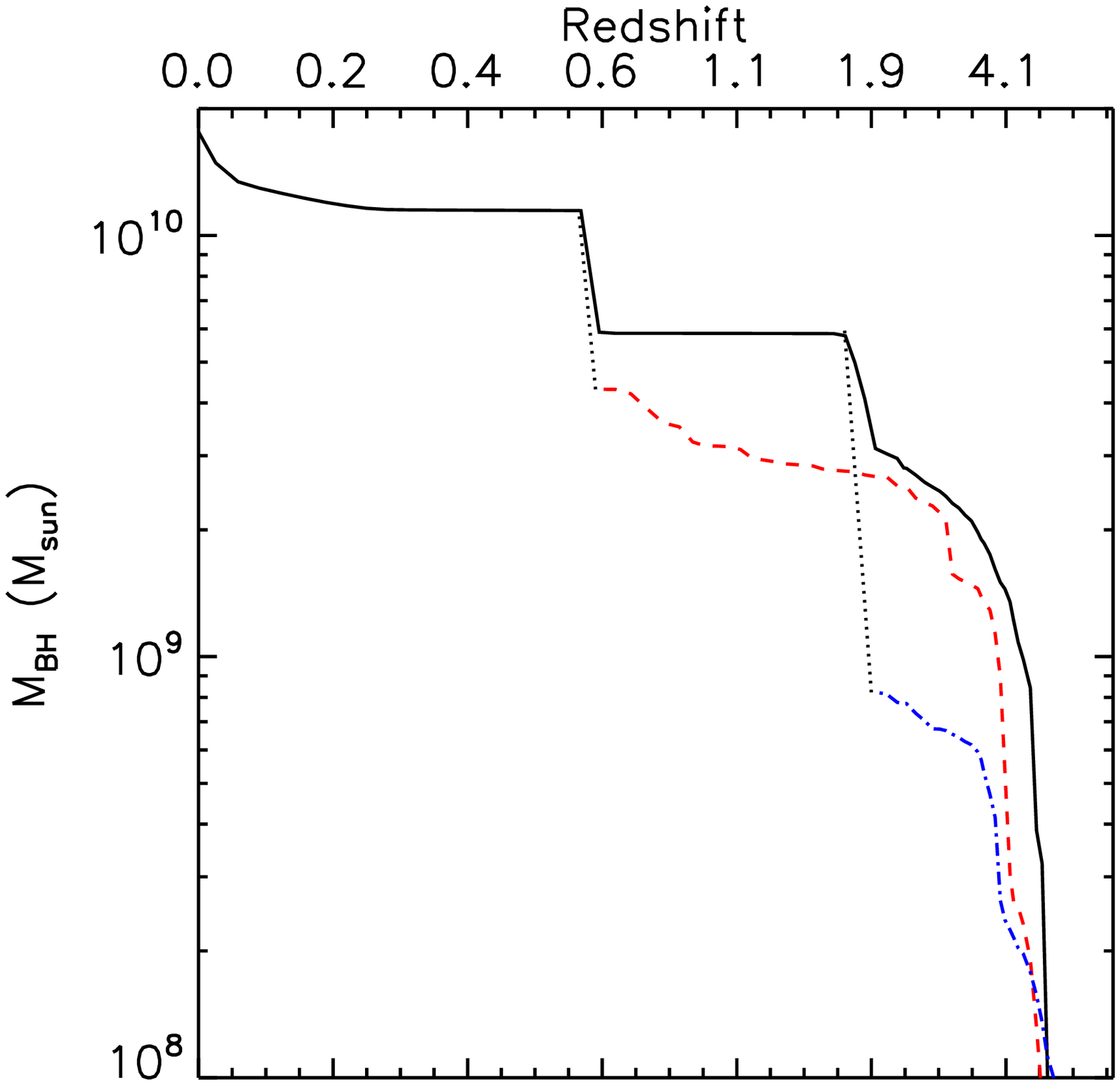}}\vspace{-2.0cm}}
  \centering{\resizebox*{!}{8cm}{\includegraphics{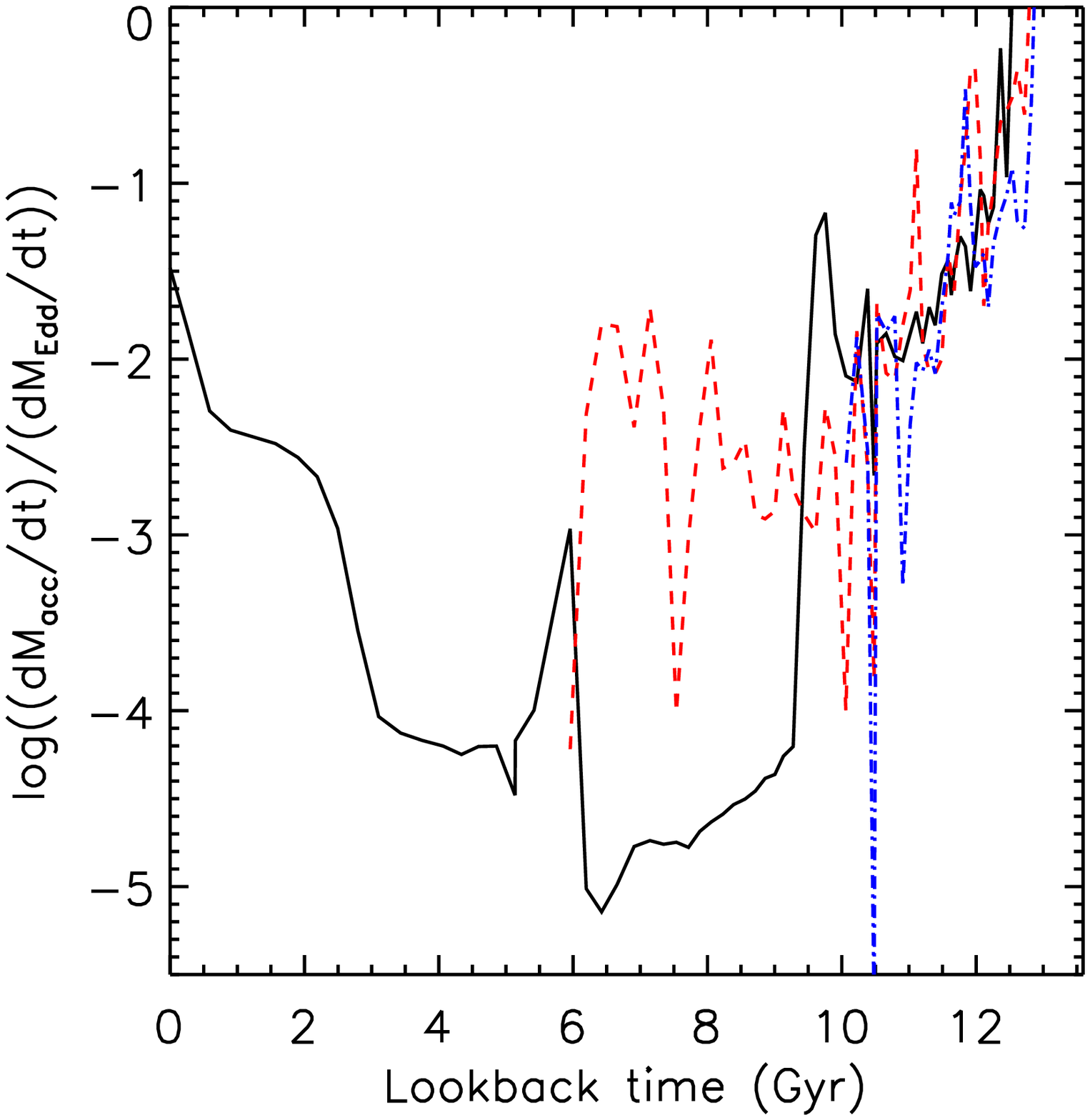}}}
  \caption{Mass evolution (upper panel) and  accretion rate relative to the Eddington limit (bottom panel) of the most massive BH (black solid lines), and two of its most massive progenitors involved 
in the episodes illustrated in figure~\ref{nice_jet} and~\ref{nice_merger}, i.e. central BH of branch 2 (blue dot-dashed lines) and branch 3 (red dashed lines) of the merger tree (figure~\ref{mergertree_DM}).
Black vertical dotted lines in the upper panel mark the mergers of the progenitors with the most massive BH.}
    \label{mbhvsredshift}
\end{figure}

The growth of a BH is tightly linked to the accretion history of its host halo (c.f. coeval growth scenario advocated by \citealp{milleretal06} and \citealp{hopkinsetal07}). In principle, cold gas which flows directly into 
the central nucleus in a free-fall time will very efficiently grow BHs. However this rapid growth might be substantially reduced by AGN activity that could expel both energy and material in the 
vicinity of the BH. In this case, self-regulation of the BH growth possibly drives the relations observed between BH mass and 
their host galaxy properties \citep{magorrianetal98, tremaineetal02, haring&rix04}. 

At high redshift ($z>2$), most galaxies seem to harbor a massive cold gas disc component, both in cosmological hydrodynamics simulations \citep{ocvirketal08} and
 in the observations \citep{shapiroetal08}, which can be tapped to fuel rapid BH growth. Accordingly, 
in the cosmological re-simulation of a cluster presented here, the central BH reaches a few tenths of its final mass when the Universe is less than 2 Gyr old, accreting 
at a rate above $\sim$1 \% of its Eddington limit (figure~\ref{mbhvsredshift}). As the initial seed mass of the black hole is $10^{5}\, \rm M_{\odot}$, this means that 
its mass increases by a factor $10^{4}$ in a tenth of the Hubble time. 
On Fig.~\ref{mbhvsredshift} the growth of two of its most massive BH progenitors (indicated by red dashed and blue dot-dashed lines on the figure) is also displayed until 
they merge with the final BH ($M_{\rm BH}=1.7\,10^{10}\, \rm M_{\odot}$ at $z=0$). These mergers (vertical dotted lines in upper panel of Fig~\ref{mbhvsredshift})
coincide with two important halo mergers in the history of the cluster (branches labelled 2 and 3 on Fig.~\ref{mergertree_DM}). 
At high redshift ($z>2$), these BH progenitors behave like the main one: they accrete gas at a fraction of their Eddington rate and this fraction steadily decreases 
with time from $z=4$ to $z=2$. We know \citep{birnboim&dekel03, keresetal05, dekel&birnboim06, ocvirketal08} that for the most massive halos 
(those with masses $M_{\rm stream} \geq 6 \times 10^7 (1+z)^8 \,\rm M_{\odot}$ at $z \geq 2$), cold accretion of gas from the IGM is efficiently thermalized at a few virial radii by 
an accretion shock, and as a result we expect the accretion rate in the centre of the halo to drop. From Fig.~\ref{mergertree_DM} one can see that the DM halos of branches 2 and 
3 have masses $ \approx 10^{13}\,\rm M_{\odot}$ at redshift $z=3$ and therefore satisfy the $M_{\rm stream}$ criterion. On that account we claim that this explains, in part, the decrease  
of the BH accretion rate relative to its Eddington rate. 

Mergers are a non-negligible growing mode at intermediate and lower redshift, as there is less cold gas to feed the BH in the massive cluster. Indeed fig.~\ref{mbhvsredshift} 
shows that the BH doubles its mass at redshift $\simeq1.6$ when two BHs of comparable mass ($M_{\rm BH}=3.1 \, 10^{9}\,\rm M_{\odot}$ and 
$M_{\rm BH}=8.2 \, 10^{8}\,\rm M_{\odot}$) coalesce. The extra amount of mass ($\simeq 2 \, 10^9 \, \rm M_\odot$) comes from the fast accretion of material brought in by the 
galaxy major merger ($M_{*}=1.3 \, 10^{12}\,\rm M_{\odot}$ and $M_{*}=6.1 \, 10^{11}\,\rm M_{\odot}$). This merger appears in the DM merger tree (figure~\ref{mergertree_DM}) when
branches 1 and 2 join at redshift $1.6$. It is interesting to note that the first BH that forms is not necessarily the most massive one at late times (in our case the BH hosted by
branch 2 halo forms first), as already pointed out by \cite{dimatteoetal08}. 

\begin{figure*}
  \centering{\resizebox*{!}{5.5cm}{\includegraphics{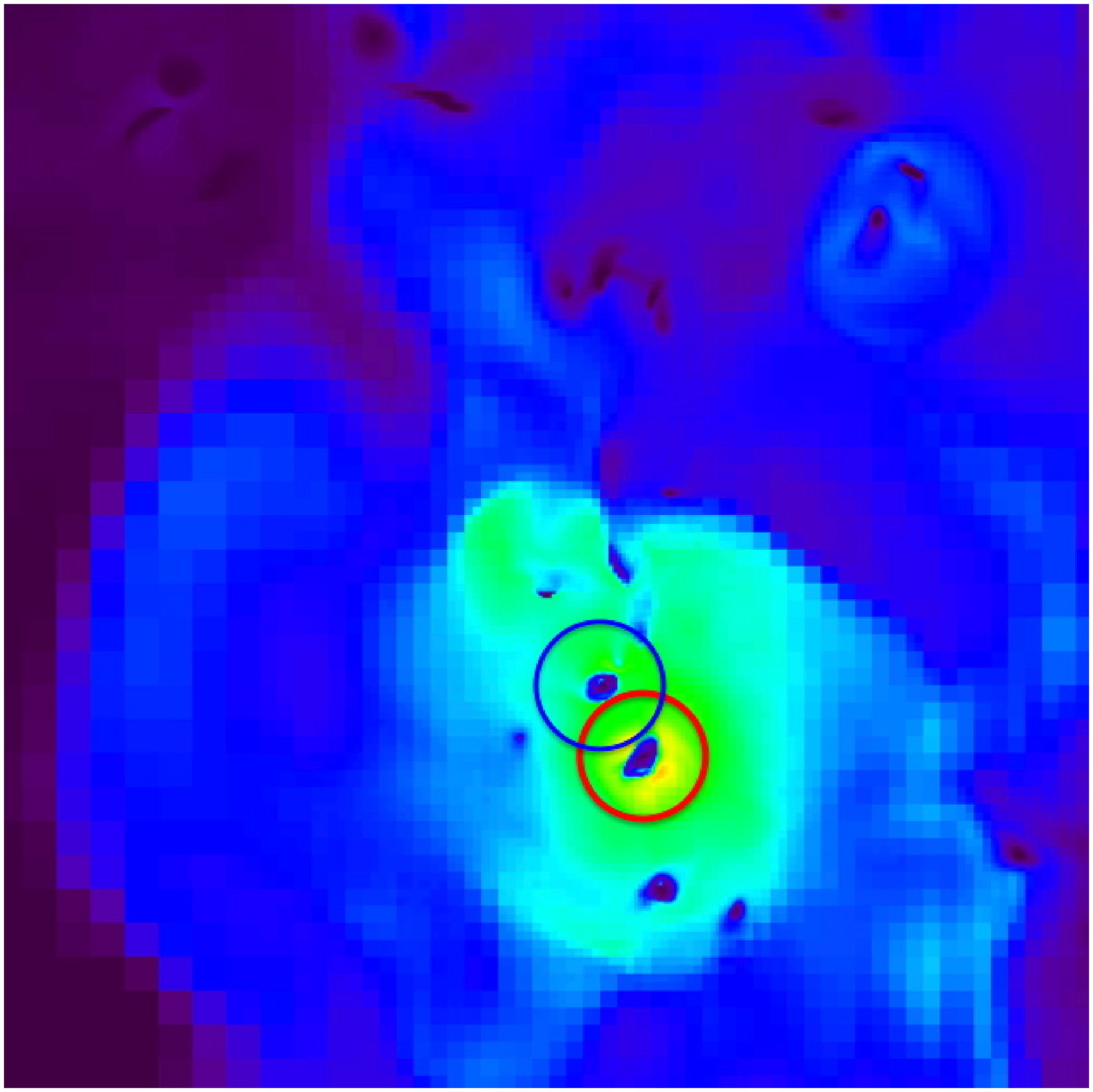}}}
  \centering{\resizebox*{!}{5.5cm}{\includegraphics{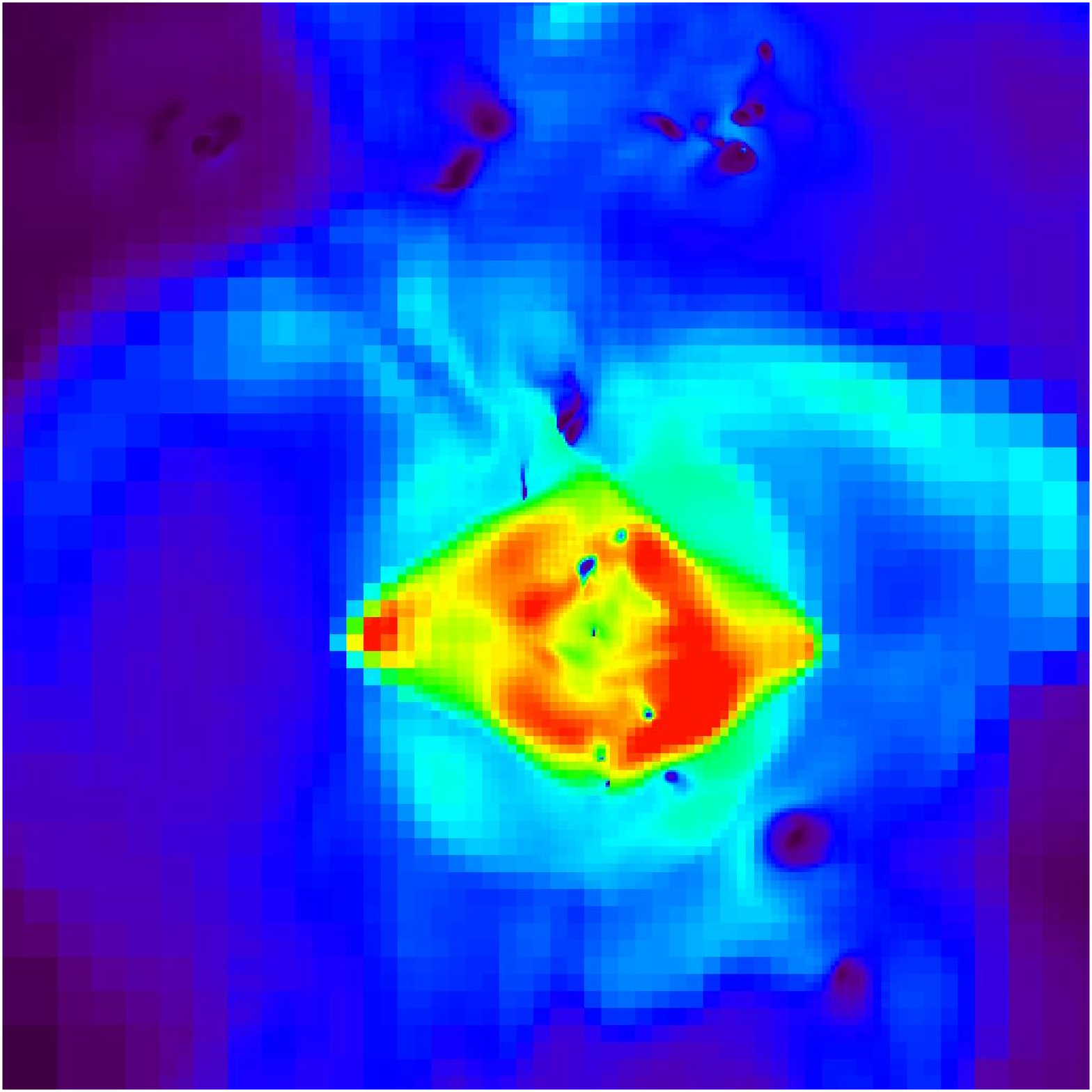}}}
  \centering{\resizebox*{!}{5.5cm}{\includegraphics{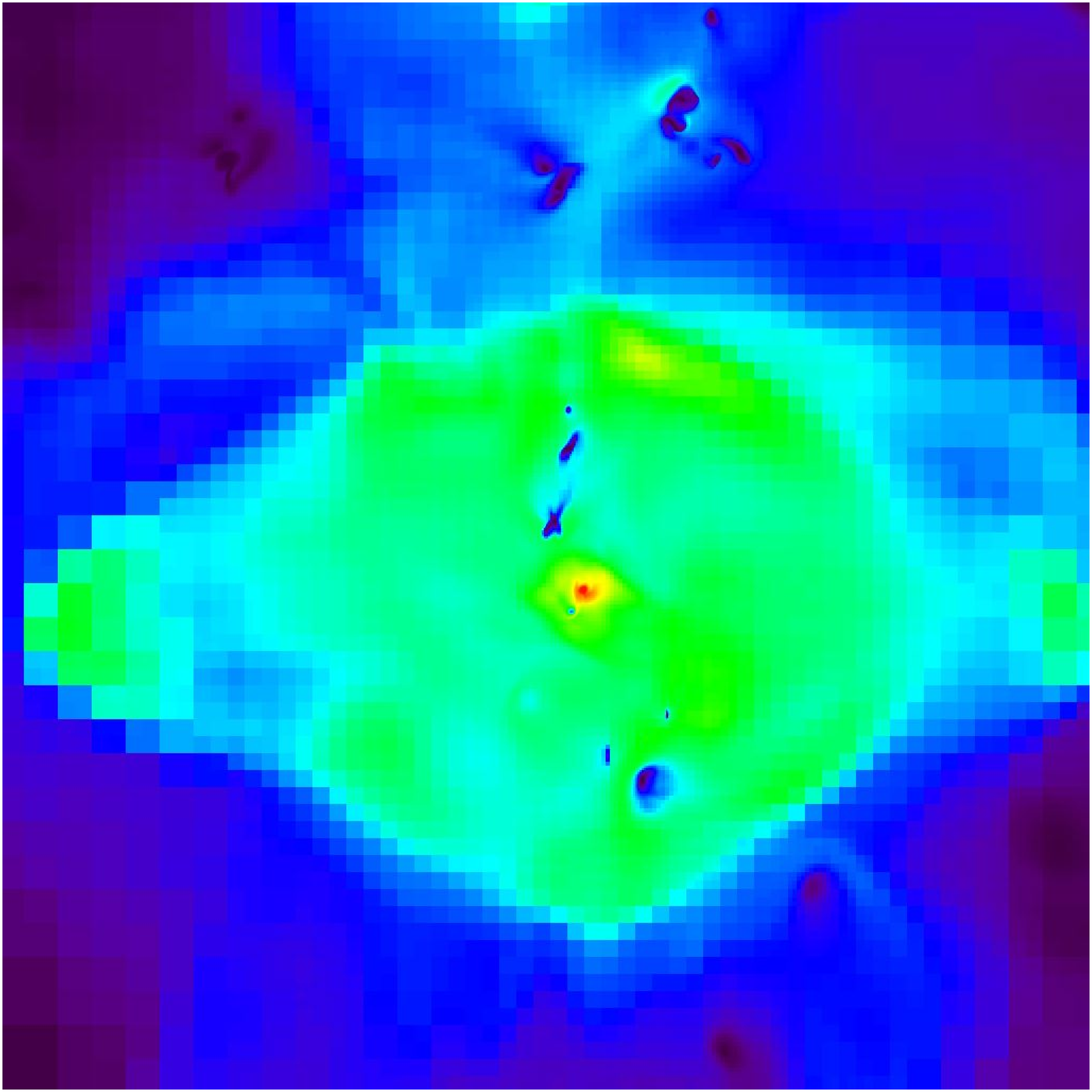}}}
  \centering{\resizebox*{!}{5.5cm}{\includegraphics{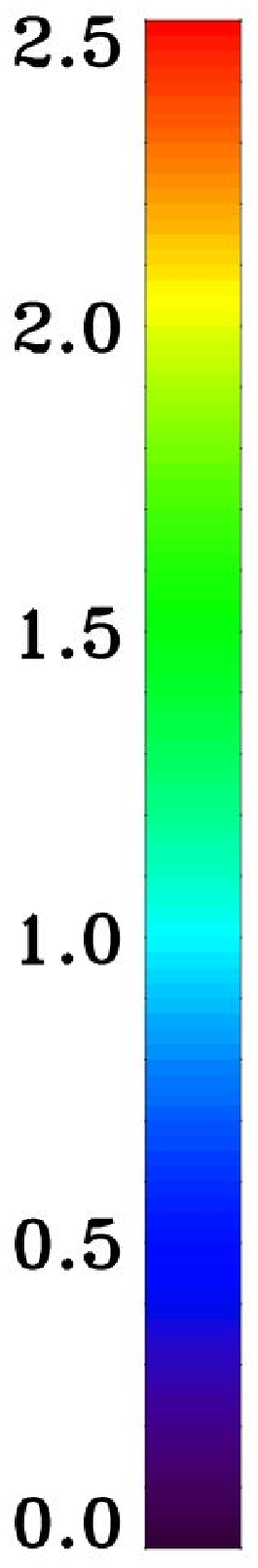}}}
  \caption{Projected temperature in linear scale along the z direction at $z=1.76$, $z=1.56$ and $z=1.49$ from left to right of the most massive cluster progenitors. Branch 1 of the merger tree corresponds 
to the red circle and branch 2 to the blue circle (see Fig~\ref{mergertree_DM}). This redshift sequence corresponds to pre-central galaxy merger, central galaxy and BH mergers and post-central galaxy merger respectively.
The red saturated region corresponds to a temperature $T\simeq2.5-3$ keV. The size of the images is $3.6$ Mpc in comoving units. }
    \label{nice_jet}
\end{figure*}

\begin{figure*}
  \centering{\resizebox*{!}{5.5cm}{\includegraphics{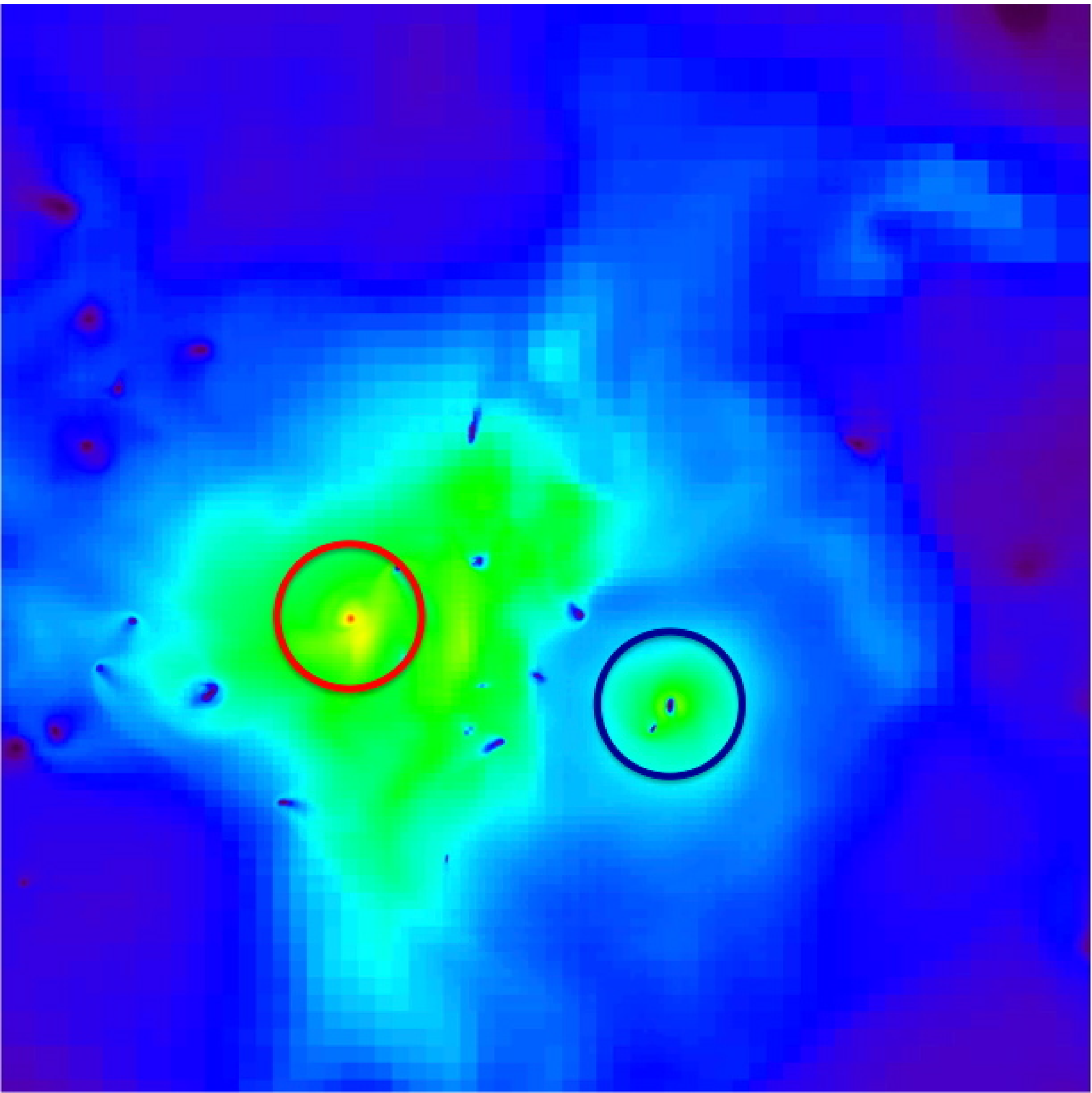}}} \hspace{-0.15cm}
  \centering{\resizebox*{!}{5.5cm}{\includegraphics{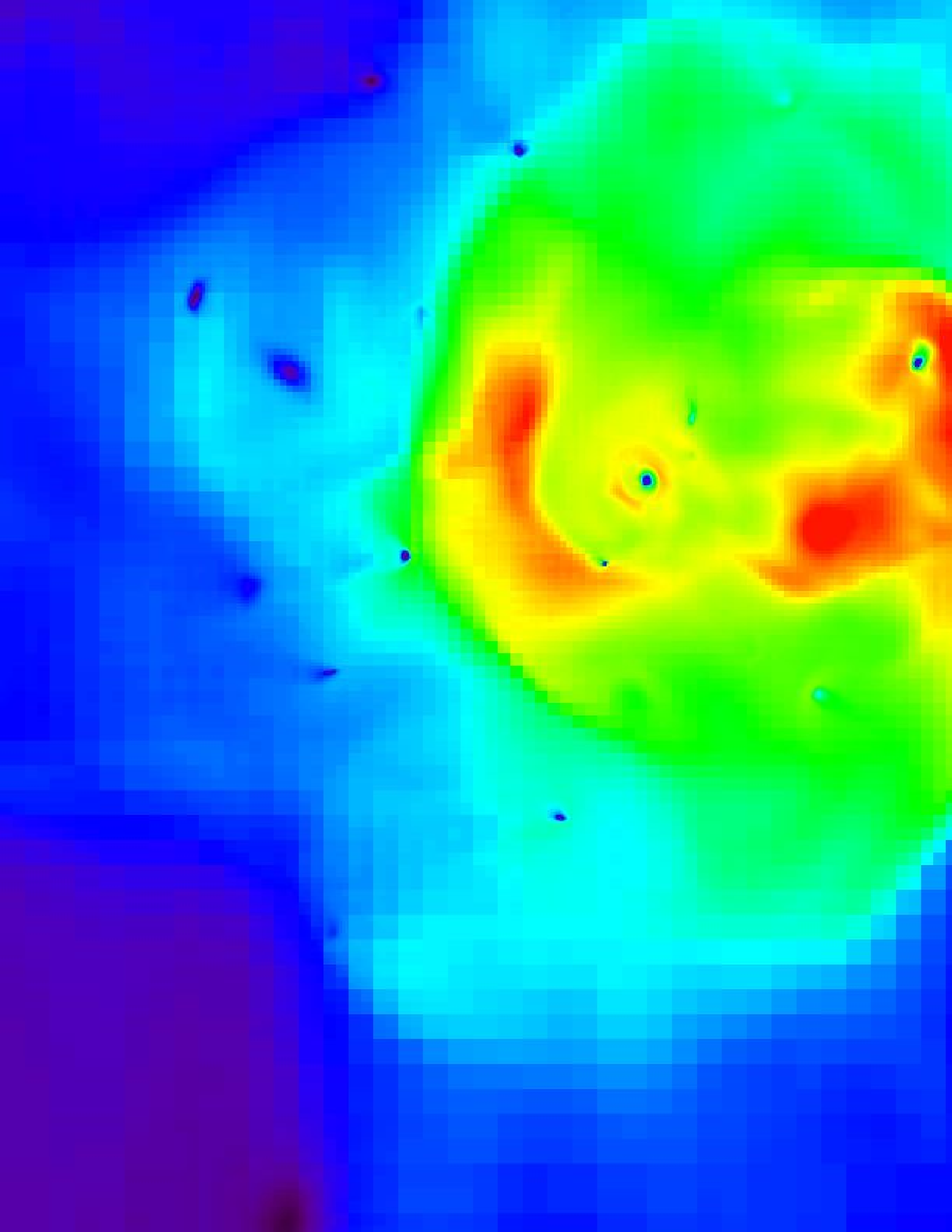}}}
  \centering{\resizebox*{!}{5.5cm}{\includegraphics{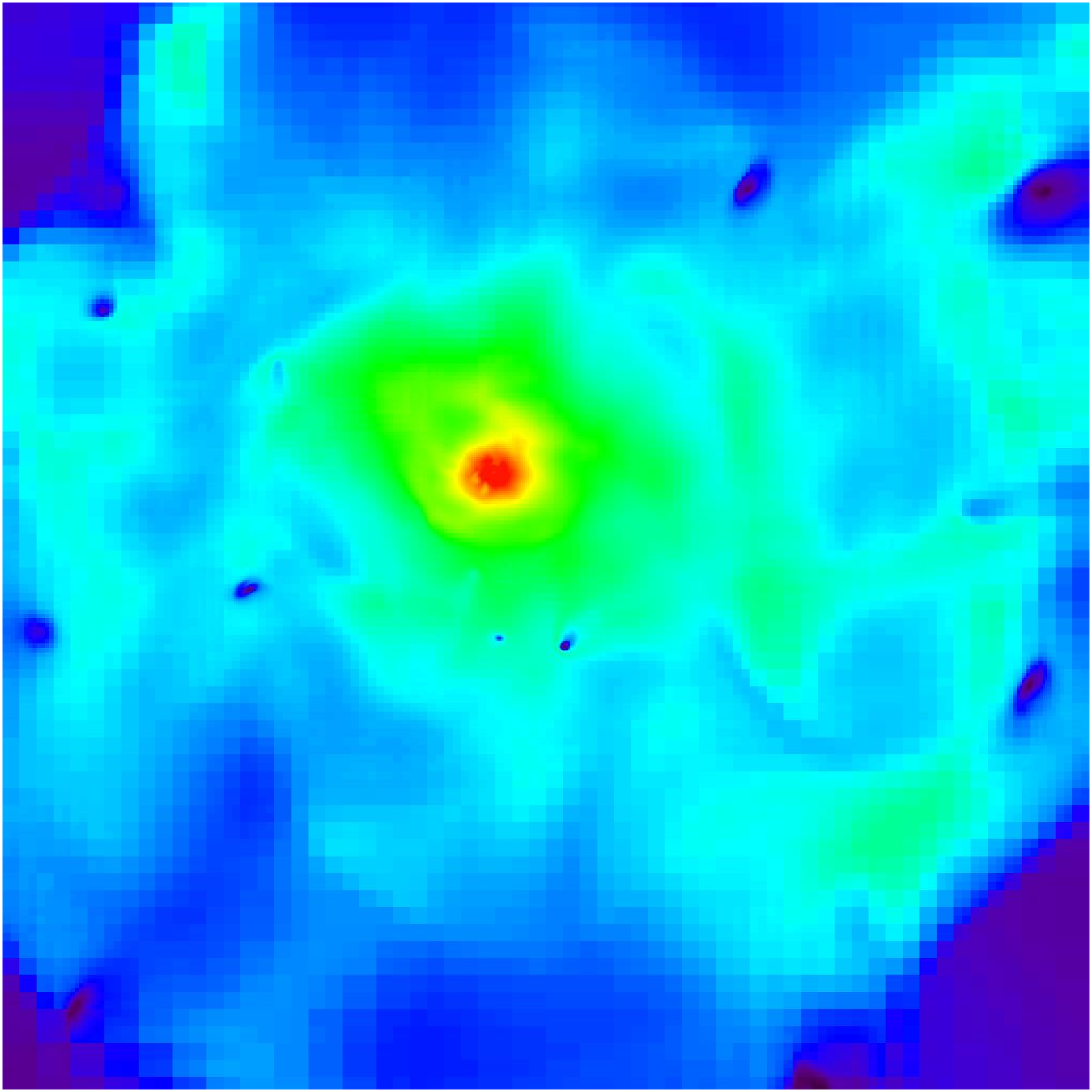}}}
  \centering{\resizebox*{!}{5.5cm}{\includegraphics{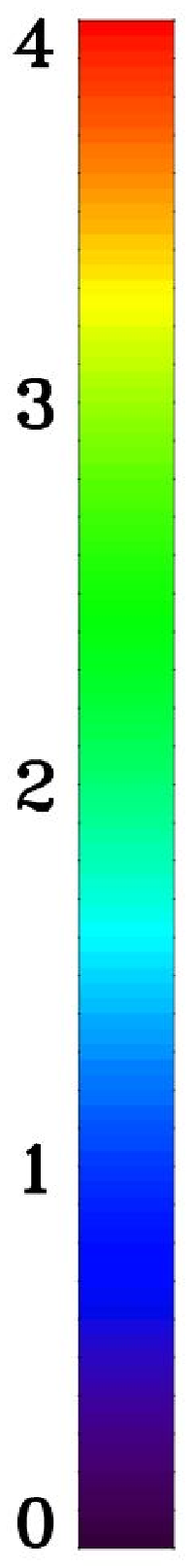}}}
  \caption{Projected temperature in linear scale along the x direction at $z=0.88$, $z=0.74$ and $z=0.58$ from left to right of the cluster during its merging phase. Branch 1 of the merger tree corresponds 
to the red circle and branch 3 to the blue circle (see Fig~\ref{mergertree_DM}).  This redshift sequence corresponds to 
pre-central galaxy merger, slightly post-central galaxy merger and BH merger respectively.The red saturated region corresponds to a temperature $T\sim4-6$ keV. The size of the images is $3.6$ Mpc in comoving units. }
    \label{nice_merger}
\end{figure*}

\begin{figure*}
  \centering{\resizebox*{!}{5.5cm}{\includegraphics{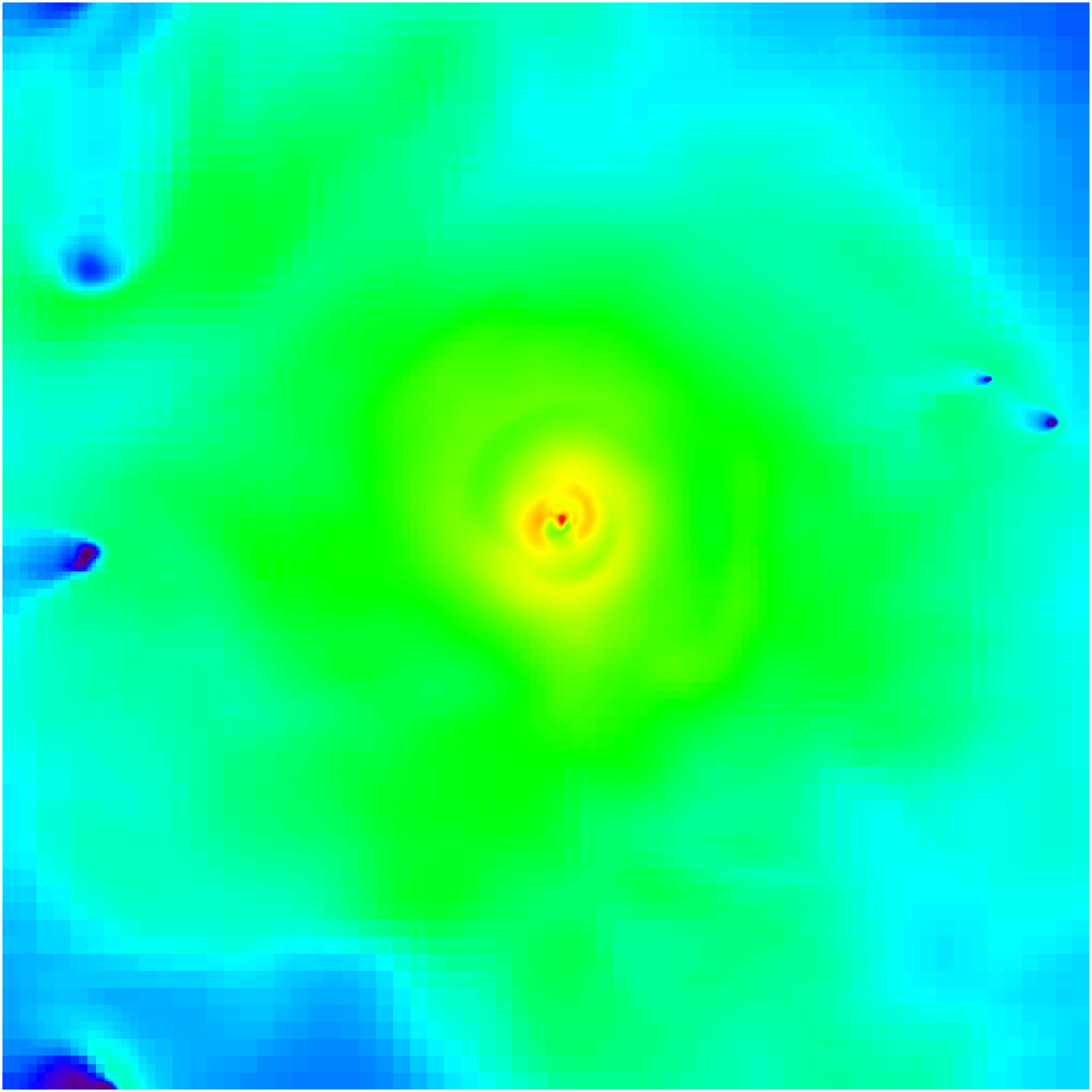}}}
  \centering{\resizebox*{!}{5.5cm}{\includegraphics{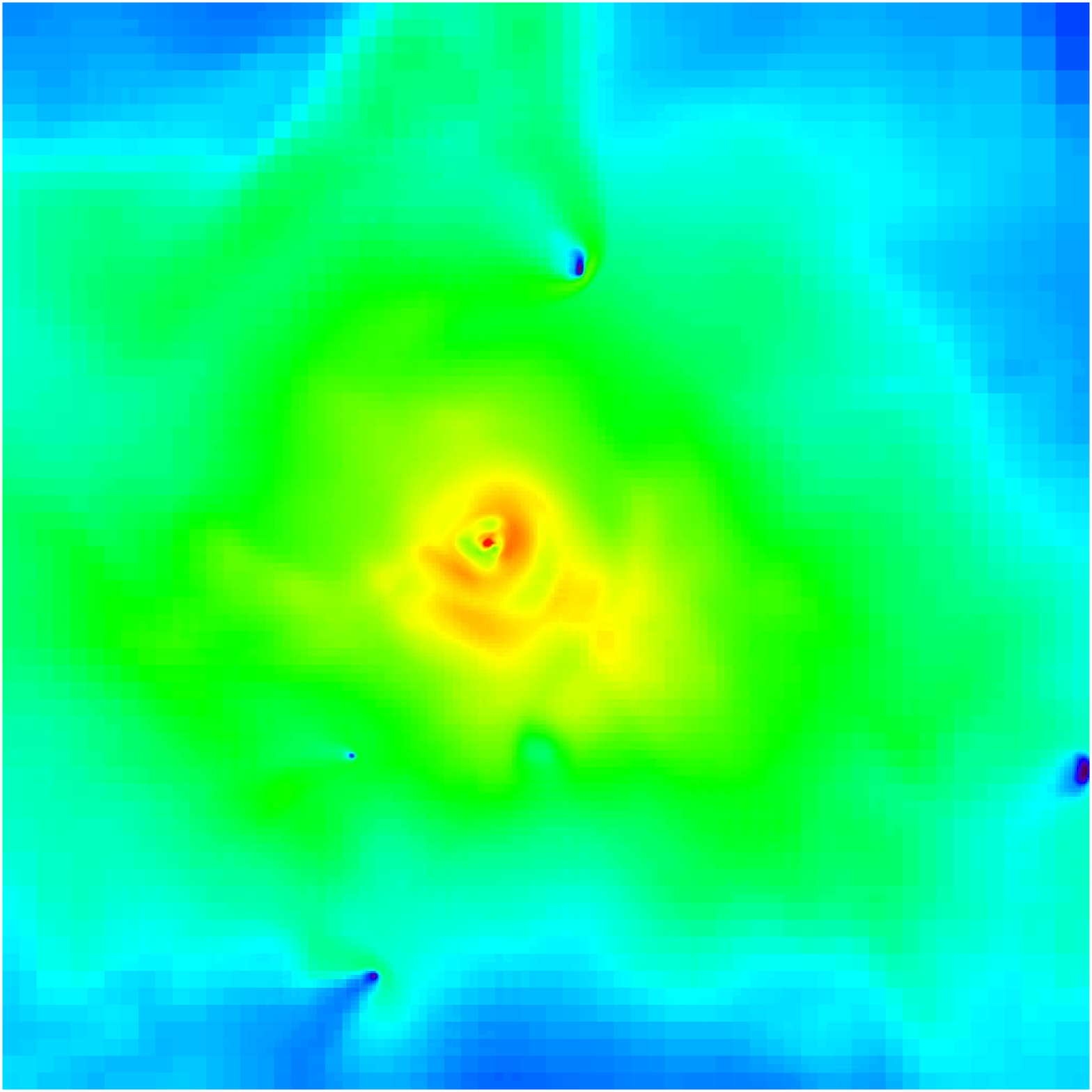}}}
  \centering{\resizebox*{!}{5.5cm}{\includegraphics{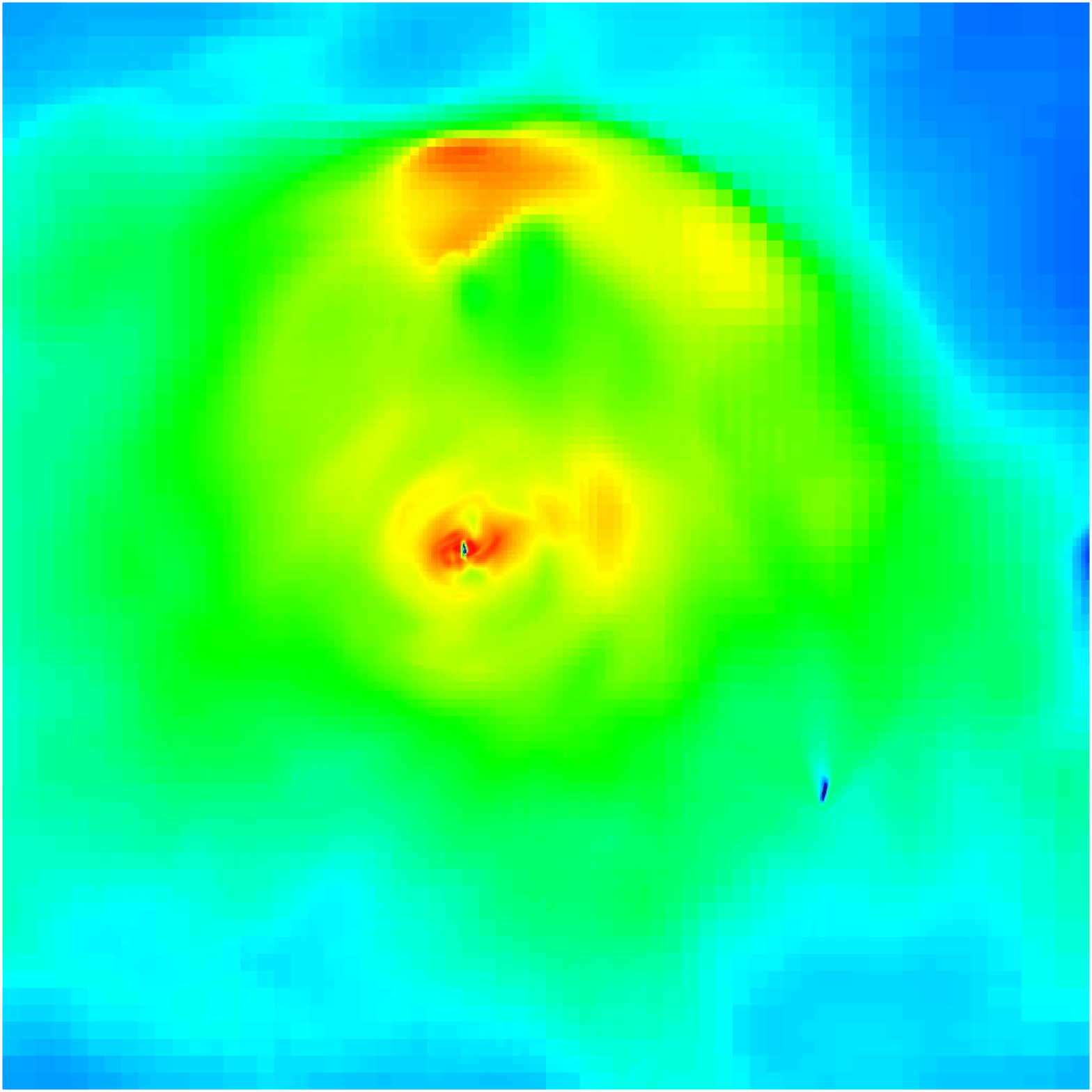}}}
  \centering{\resizebox*{!}{5.5cm}{\includegraphics{./colorbar_0-4keV.ps}}}
  \caption{Projected temperature in linear scale along the z direction at $z=0.09$, $z=0.04$ and $z=0$ from left to right of the cluster during its relaxed phase. The size of the images is $1.8$ Mpc in comoving units. }
    \label{nice_radio}
\end{figure*}

Fig.~\ref{nice_jet},~\ref{nice_merger} and~\ref{nice_radio} show three different episodes of the formation and evolution of the cluster, respectively a high-redshift major merger between two gas-rich galaxies (Fig.~\ref{nice_jet}), the major merger of the two clusters (Fig.~\ref{nice_merger}), and the relaxation of the cluster at late times (Fig~\ref{nice_radio}).

The halo merger between branches 1 and 2 at $z=3.1$ results in a cataclysmic episode for its host galaxies at $z\simeq 1.6$: a large amount of gas is expelled far from the core of the halo, 
reaching the virial radius, and the resulting disc of gas from the two merging galaxies is almost completely disrupted. This sequence of events is illustrated in figure~\ref{nice_jet}. 
On the left panel, we observe the encounter of the two gas-rich galaxies before they merge. Shortly after they merge (fig.~\ref{nice_jet} middle panel), their respective BHs do so as well 
which results in a strong jet that disrupts most of the cold baryon content in the galaxy and shock heats the ambient medium to high temperature. The jet propagates supersonically at Mach $3$ 
($u_{\rm jet}\simeq 3000\, \rm km.s^{-1}$ and $c_{\rm s}\simeq 1000 \, \rm km.s^{-1}$) before being stopped by the intergalactic 
medium at $r\simeq1.2$ Mpc, which corresponds to about $3 \, r_{\rm vir}$ at this redshift (fig~\ref{nice_jet} right panel).

The disruption of cold material by AGN feedback has already been noted by \cite{dimatteoetal05} in idealized simulations of a gas rich merger. 
It is comforting to confirm their results within a cosmological setting. \cite{khalatyanetal08} have also pointed out that mergers could trigger a high level of AGN activity during the formation of a small galaxy group. Finally, we see two hotspots during the jet propagation, that look like radio lobes 
(fig.~\ref{nice_jet} right panel).  Such events (strong jets following a merger) become rarer as time goes on since the combined action of star formation and early 
AGN activity strongly diminishes the cold and dense gas content in massive halos. 

\begin{figure}
  \centering{\resizebox*{!}{8cm}{\includegraphics{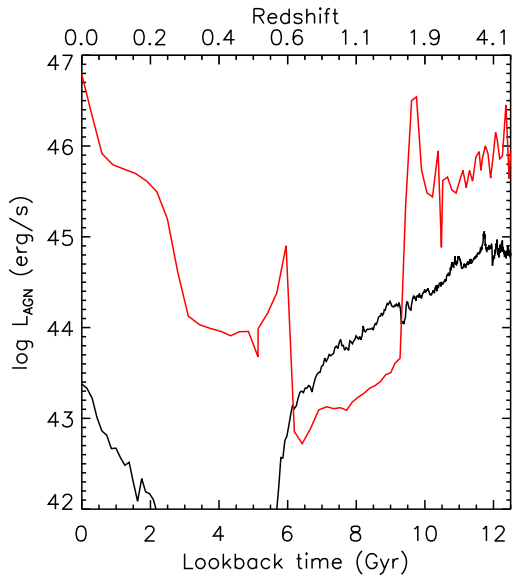}}}
  \caption{AGN luminosity of the most massive BH at $z=0$ as a function of time (red curve). For comparison,
a simple estimate of the contribution from supernova activity based on star formation rate is given (black curve).
Note however that this contribution is negligible compared to that of the AGN and is not included in the simulation.}
    \label{lagnvsredshift}
\end{figure}

To check how this striking result is affected by our limited spatial resolution, we performed the same simulation with one more level of 
refinement ($\Delta x=0.6\, \rm h^{-1}.kpc$). The same burst appears at the same redshift but its power is slightly lower, because more gas has been 
pre-heated by a strong AGN activity in a previous merger taking place at higher redshift $z\simeq4$. This is not a very surprising effect: with more resolution, the density contrast is more 
pronounced especially in poorly resolved high-redshift galaxies, which, in turn, leads to a faster accretion rate at early times. As in our standard run, the cold gas 
in the core of the halo is strongly disrupted by the AGN activity triggered by the wet merger occurring at $z=1.6$, so that BH growth and AGN luminosity 
are suppressed for 3 Gyrs (see figure~\ref{lagnvsredshift} from $z=1.6$ to $z=0.6$). Therefore we can reasonably claim that most of the important features 
describing the BH growth are already captured by our standard resolution run.

Figure~\ref{lagnvsredshift} also shows the energy that would be released by supernova feedback if it were implemented in our simulation. 
To estimate this energy release, we assume that stars are distributed according to a \cite{salpeter55} Initial Mass Function, for which each massive star ($M_{\star} > 8 \, \rm M_{\odot}$) deposits $10^{51}$ erg per $10\, \rm M_{\odot}$ into the ISM.
 We see that the energy from this form of feedback is always lower than that from the AGN at all times, except during the post-merger phase from $z=1.6$ to $z=0.6$. However, in this post-merger phase the high level of supernova 
feedback is an artifact of the way star formation is computed: we trace back the star formation history (SFH) of the central galaxy using all the stars which belong to it at $z=0$, so 
that the star formation and hence the supernova rate we derive from it includes that of its accreted satellites. In this redshift range ($0.6<z<1.6$), the star formation rate is dominated 
by the galaxy progenitor that has not undergone the cataclysmic quasar phase (hosted by the branch 3 halo on Fig~\ref{mergertree_DM}), whilst the star formation activity in the 
quasar galaxy progenitor (hosted by the branch 1 halo) is completely suppressed. In light of this, it is a fair approximation to neglect the feedback from supernovae on the evolution 
of this galaxy cluster.

At $z\simeq 1.7$, another halo major merger (1:1) occurs (branches 1 and 3 in the merger tree (figure~\ref{mergertree_DM})). However, the most 
massive ($M_{\rm BH}=5.9 \, 10^{9}\,\rm M_{\odot}$) BH only merges with the ($M_{\rm BH}=4.3 \, 10^{9}\,\rm M_{\odot}$) BH of its cluster companion at a much later epoch ($z=0.58$
right panel of Fig~\ref{nice_merger}). Note that, in this case, the BH merger also takes place quite a long time after the central galaxies hosting the BHs merge together. 
As a matter of fact, the $z=0.8$ major galaxy merger drives the cluster gas to temperatures twice the virial temperature thanks to a violent shock wave (middle panel of Fig~\ref{nice_merger}). 
However, during this galaxy merger phase, the accretion rate onto the most massive BH drops to negligible 
values ($\sim 10^{-5}\dot M_{\rm Edd}$ fig.~\ref{mbhvsredshift} solid black line), and only its companion continues to accrete at a moderate rate 
(a few $\sim 10^{-3}\dot M_{\rm Edd}$, fig.~\ref{mbhvsredshift} red dashed line). The resulting AGN activity, even when boosted by the final BH merger at $z=0.58$ does not seriously impact the highly 
pressurized ICM gas which confines the jet energy to the very central parts of the cluster (right panel of Fig~\ref{nice_merger}). After the major BH merger, the accretion rate onto 
the central BH becomes extremely small ($10^{-4} \dot M_{\rm Edd}$) due to the complete evaporation of leftover cold material by the final outburst of AGN activity. Subsequently the BH stays in this almost-dead phase for 2 Gyr.

It is striking that the less massive BH progenitor of the latter BH merger (at $z=0.58$) accretes gas at $0.01\,\dot M_{\rm Edd}$ before the merger, 
whereas accretion onto the most massive BH is negligible. The reason for this different behavior can be understood by looking at the temperature maps of both 
cluster progenitors (left panel of figure~\ref{nice_merger}): the most massive cluster (branch 1, red circle) is slightly warmer than 
its companion (branch 3, blue circle), as it has been pre-heated by important quasar activity at earlier redshifts (z$\sim1.56$). On the other hand, the less massive progenitor  
did not experience such strong pre-heating, and as a result has a lower gas temperature, and therefore a higher accretion rate during the pre-merger phase.   

Finally, the cluster relaxes and the inner halo cold gas reservoir gets replenished, fueling a faint accretion onto the BH. This translates into low 
AGN activity. Episodically, stronger jets are produced by the AGN in this phase which yield small perturbations of the ICM temperature in the form of sound waves 
(figure~\ref{nice_radio}).  These jets are roughly sonic ($u_{\rm jet}\simeq 1000\, \rm km/s$ and $c_s \simeq 1300 \, \rm km.s^{-1}$), and are efficiently thermalized by the ICM.
As a result they do not propagate farther than a few $10$ kpc.

\begin{figure}
  \centering{\resizebox*{!}{8cm}{\includegraphics{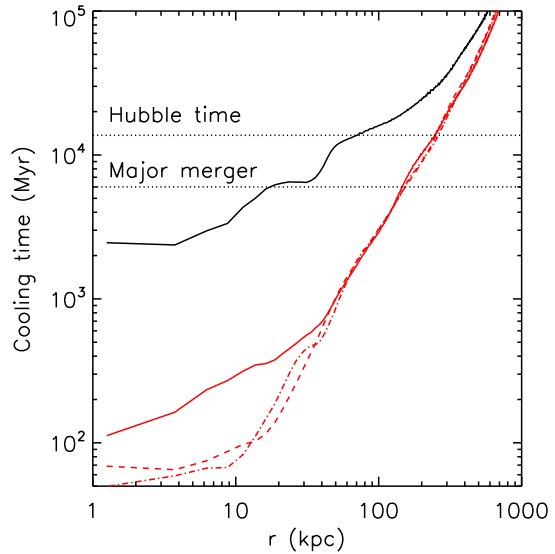}}}
  \caption{Volume-weighted cooling time for the run without AGN (black) at $z=0$ and with AGN (red) at $z=0$ (solid), $z=0.04$ (dashed) and $z=0.09$ (dot-dashed). The horizontal dotted lines are the Hubble time and the time since the major merger of the two 
central galaxies at $z\simeq0.8$.}
    \label{ctimevsr}
\end{figure}

Nevertheless, the part of the jet energy which is carried by these sound waves limits the cooling flow in the cluster core. However the cooling time in 
the core ($<100$ kpc) is extremely short (smaller than a Gyr, see figure~\ref{ctimevsr}), so the cooling flow eventually develops again and feeds the BH afresh.
As a result, the BH accretion rate increases from $10^{-4}\, \dot M_{\rm Edd}$ to a few $10^{-2}\, \dot M_{\rm Edd}$ at z$\sim0$, giving rise to late-time AGN activity.

\begin{figure}
  \centering{\resizebox*{!}{5.4cm}{\includegraphics{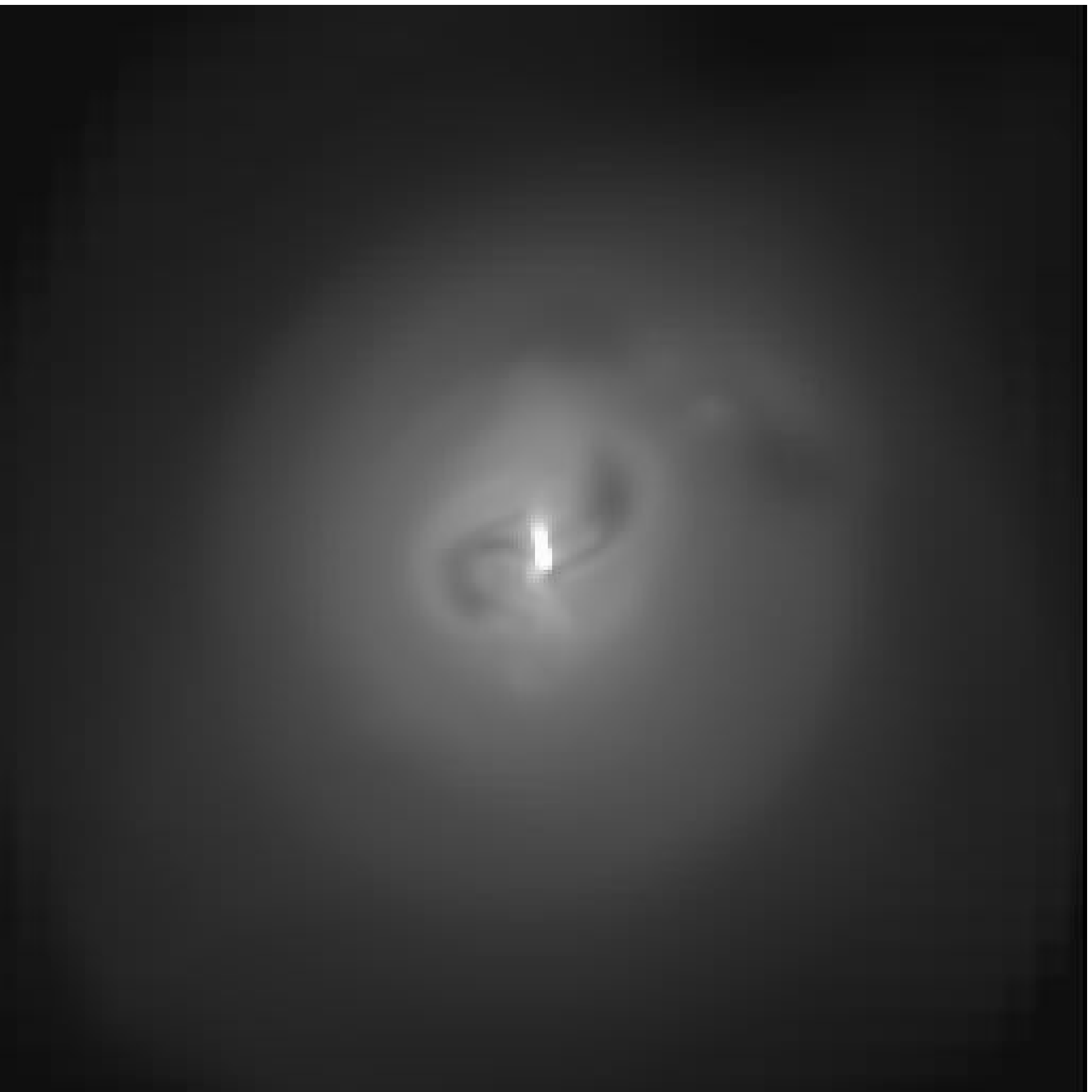}}}
  \centering{\resizebox*{!}{5.4cm}{\includegraphics{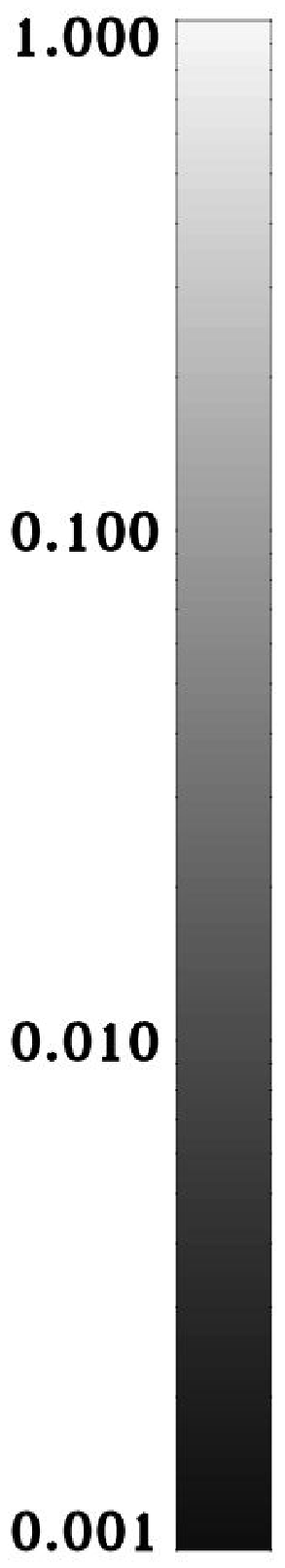}}}
  \centering{\resizebox*{!}{5.4cm}{\includegraphics{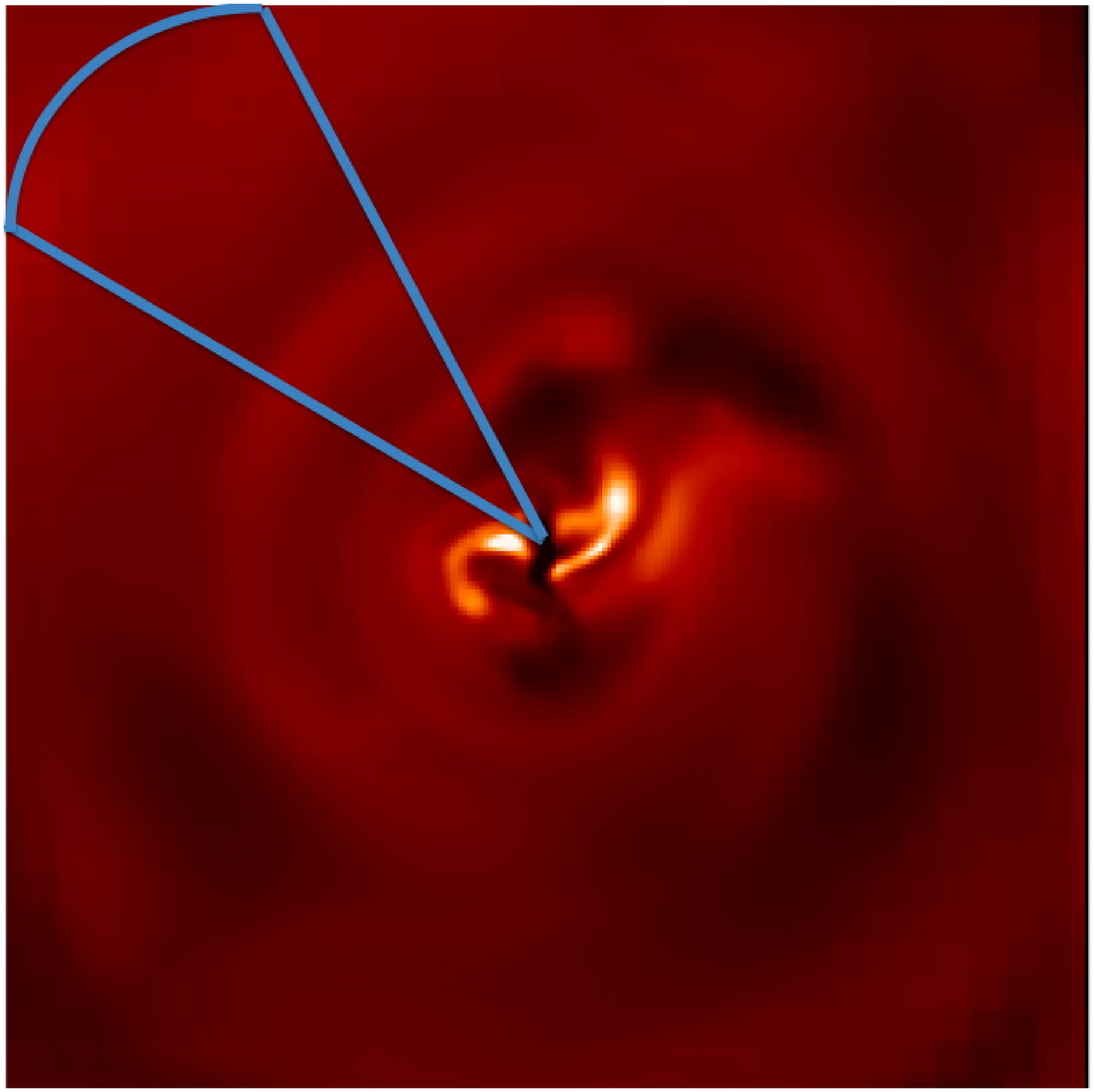}}}
  \centering{\resizebox*{!}{5.4cm}{\includegraphics{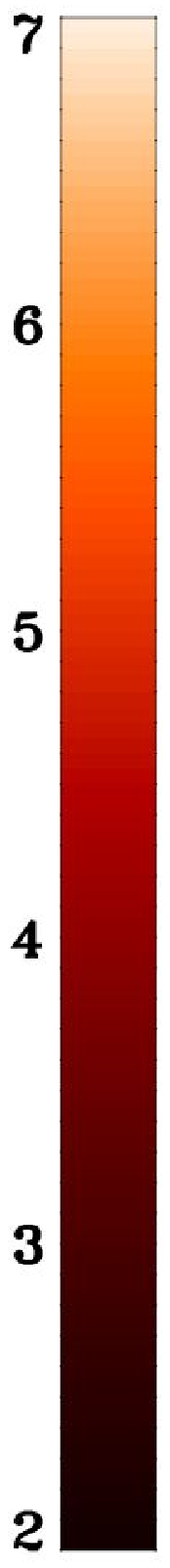}}}
  \centering{\resizebox*{!}{5.4cm}{\includegraphics{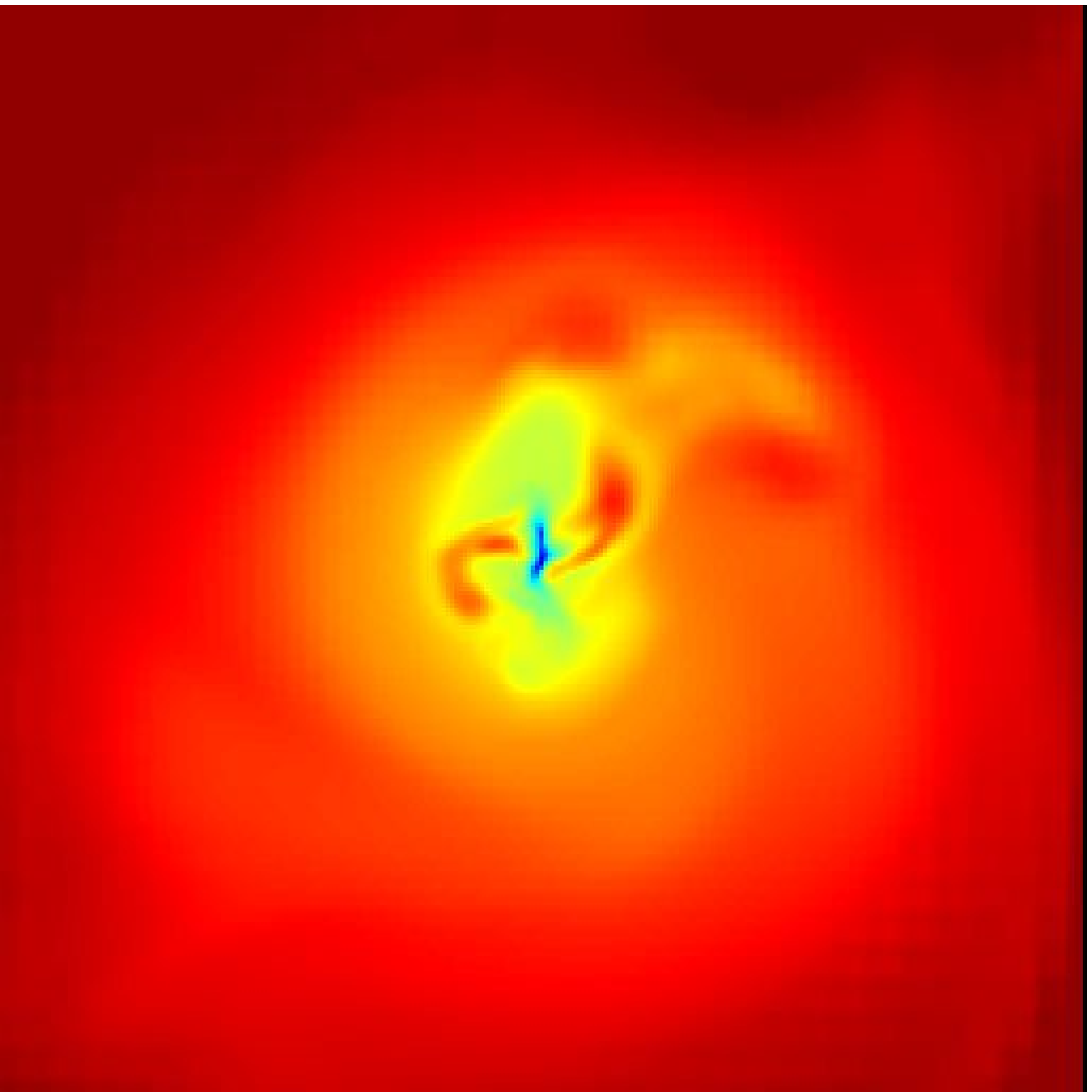}}}
  \centering{\resizebox*{!}{5.4cm}{\includegraphics{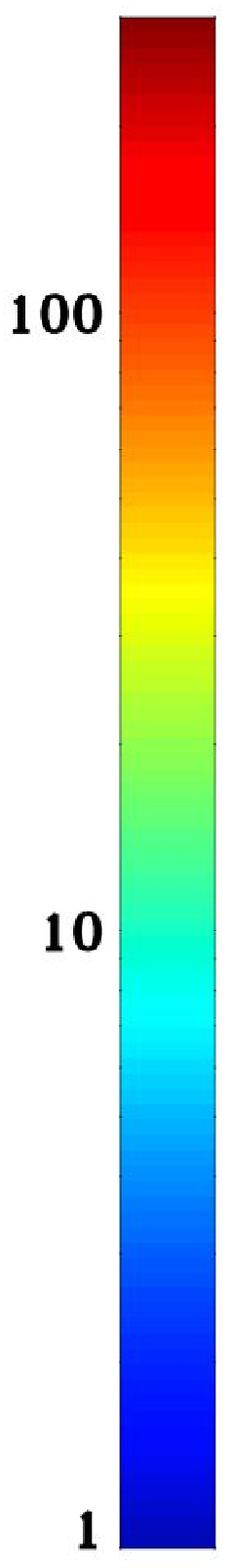}}}
  \centering{\resizebox*{!}{5.4cm}{\includegraphics{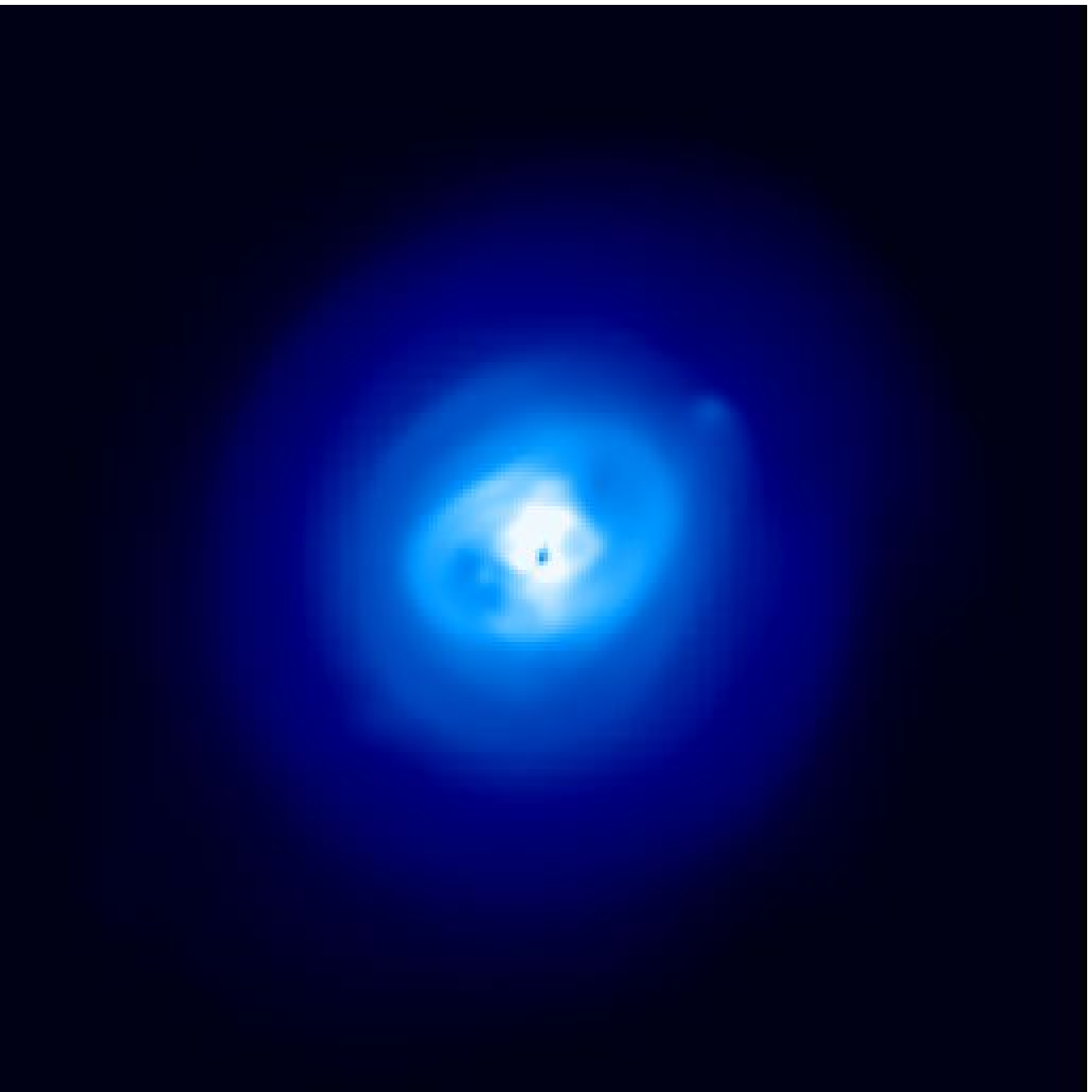}}}
  \centering{\resizebox*{!}{5.4cm}{\includegraphics{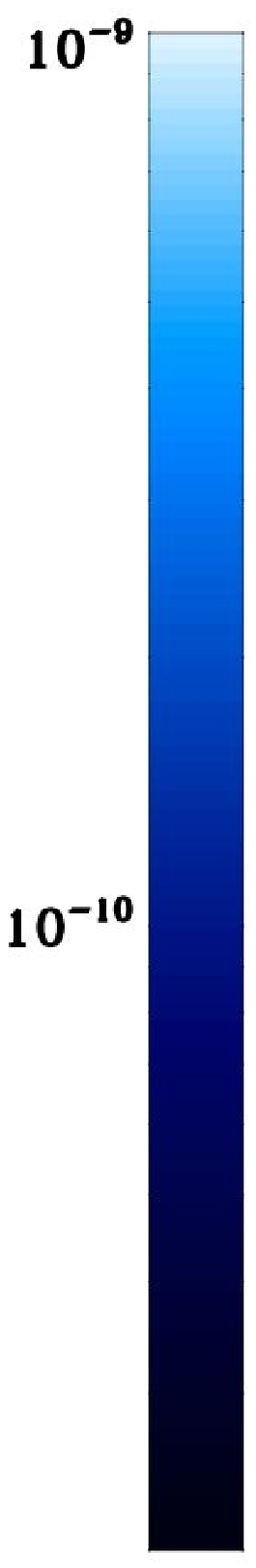}}}
  \caption{Density (uppermost panel), temperature (second from top), entropy (third from top) and pressure (bottom panel) slices through the central AGN at $z=0$. Color bars on the side of each panel indicate units in H.cm$^{-3}$, keV, keV.cm$^{-2}$, and erg.cm$^{-3}$
from top to bottom. Images are 447 kpc on a side. The cone shows the solid angle used to compute the averaged quantities in figure~\ref{profiles_pencil}.}
    \label{jet_slice}
\end{figure}

\begin{figure}
  \centering{\resizebox*{!}{6cm}{\includegraphics{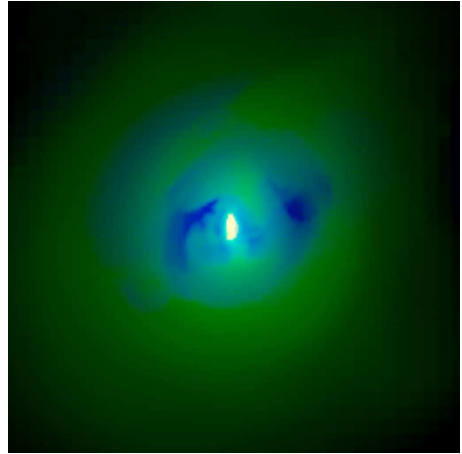}}}
  \caption{Composite RGB image of the simulated Xray emission $0.3$--$1$ (red), $1$--$3.5$ (green), $3.5$--$10$ keV (blue) bands for the AGN run at $z=0$. The image size is 447 kpc on a side.  }
    \label{nice_Xray}
\end{figure}

\begin{figure}
  \centering{\resizebox*{!}{6cm}{\includegraphics{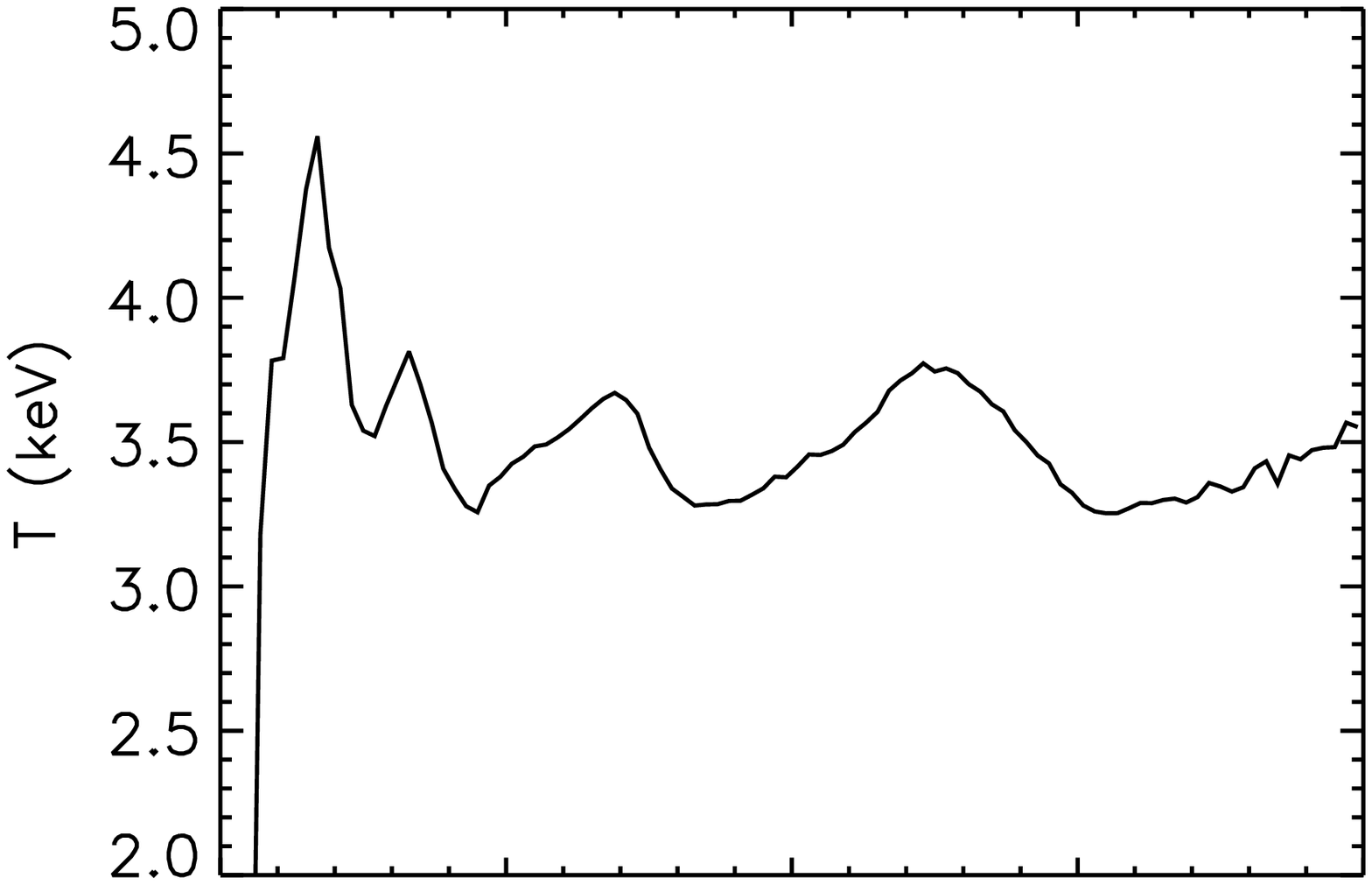}}\vspace{-1.7cm}}
  \centering{\vspace{-1.7cm}\resizebox*{!}{6cm}{\includegraphics{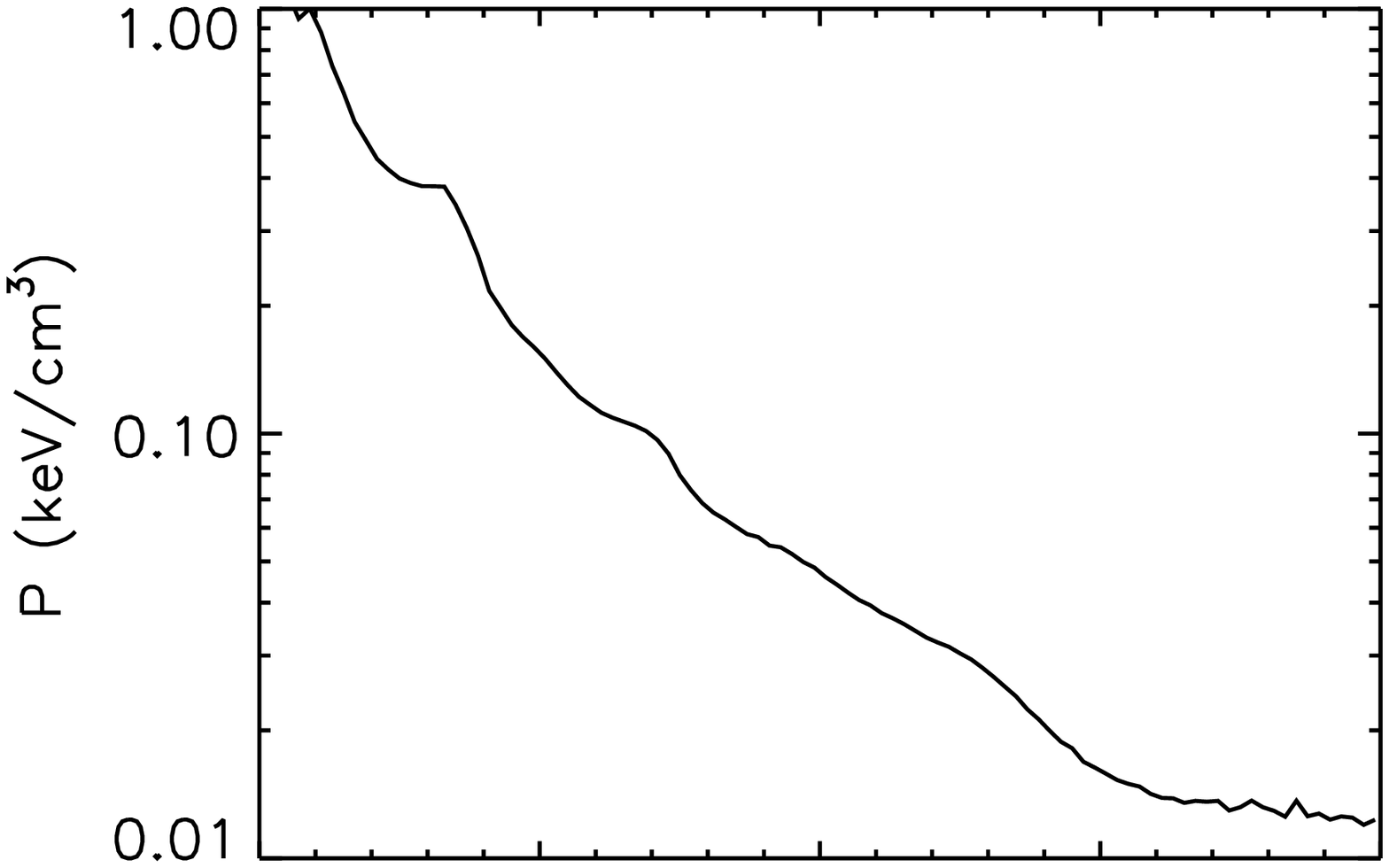}}}
  \centering{\vspace{-1.7cm}\resizebox*{!}{6cm}{\includegraphics{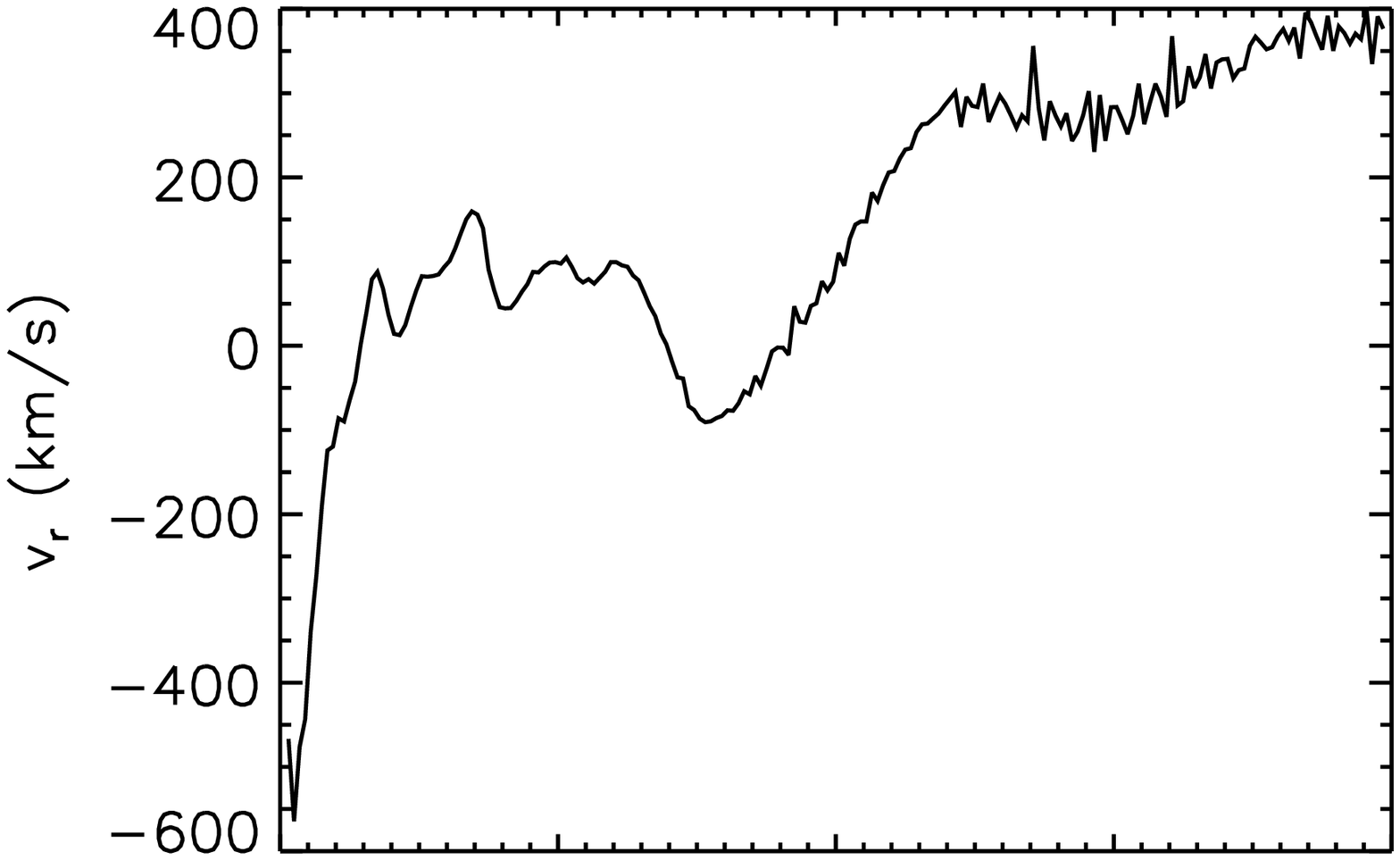}}}
  \centering{\resizebox*{!}{6cm}{\includegraphics{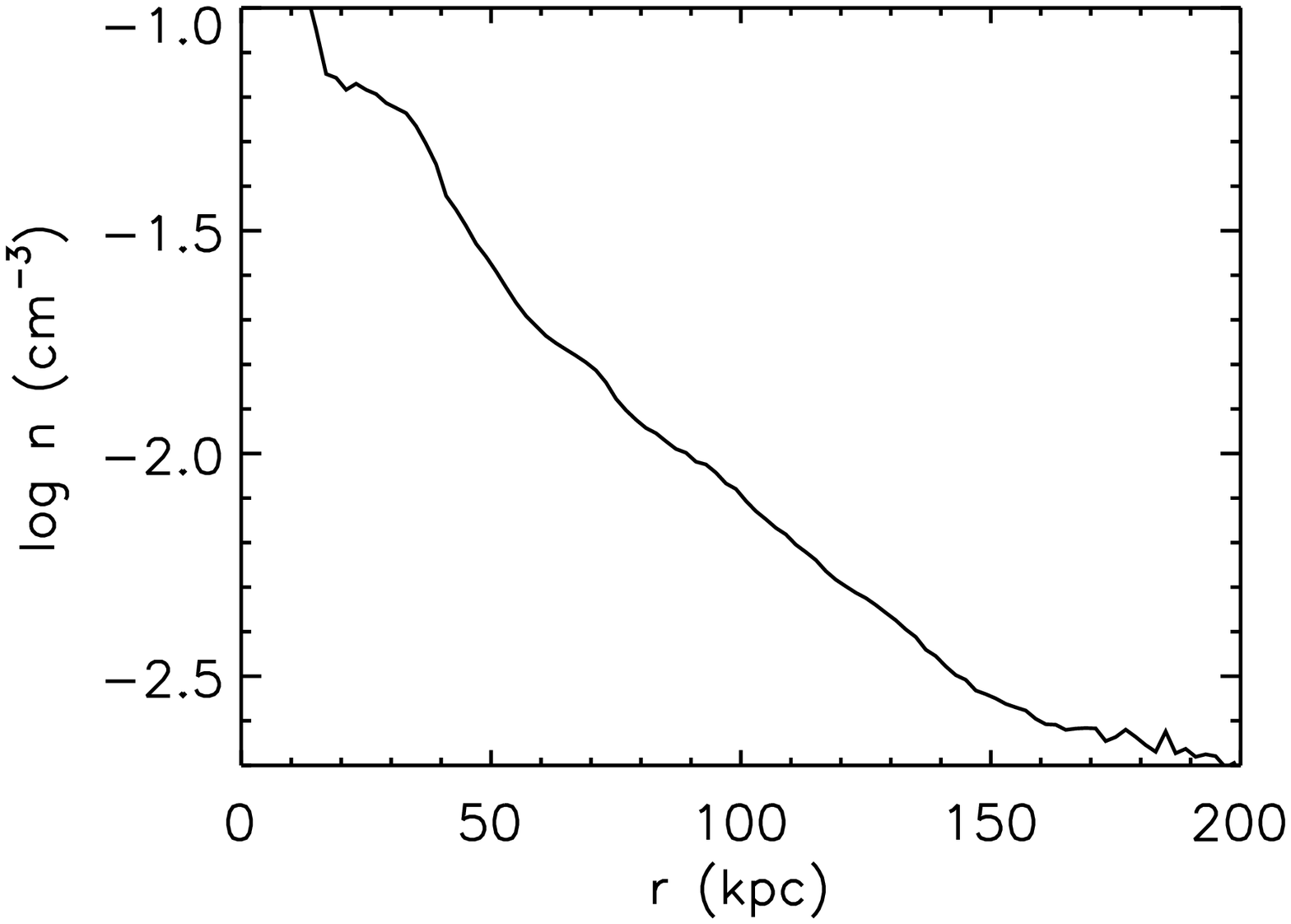}}}
  \caption{From top to bottom: volume-weighted average temperature, pressure, radial velocity and density as a function of radius in the solid angle seen in figure~\ref{jet_slice} at $z=0$ for the AGN run.}
    \label{profiles_pencil}
\end{figure}

Figure~\ref{jet_slice} shows different physical properties in a slice of gas cut through the AGN jet at $z=0$. They show that the jet is at low-density and high 
temperature in good agreement with high resolution simulations of jet-formation (see \citealp{heinzetal06, simionescuetal09} for example). A remarkable feature is 
the formation of two under-dense cavities filled with sonic-jet material. We have computed a simulated X-ray map of these cavities in 3 different temperature bands
(see figure~\ref{nice_Xray}). These cavities are reminiscent of the ones observed in Perseus A by \cite{fabianetal06} in which a strong cooling core is also visible. 
In our simulation, an extended cooling-core  (a few $10$ kpc across) is absent, but the cooling flow which gives rise to late-time AGN activity is clearly present. As in 
the \cite{fabianetal06} observations, we interpret the ripples induced by our jet modelling as sound waves. This can be seen in figure~\ref{profiles_pencil} where radial velocities
are always sub-sonic. These sound waves, provided one can dissipate them viscously \citep{fabianetal03, ruszkowskietal04b} can reheat the ICM at distances larger than the scalelength of the jet. 
In our simulation, no explicit viscosity is included, but these spherical sound waves do not appear at radii larger than $r_{500\rho_{\rm c}}$, suggesting that they have been dissipated by numerical 
viscosity on these scales.

\section{Regulation of the cooling catastrophe}
\label{Cooling_catastrophe}

In the absence of strong feedback processes to offset the cooling of gas in the potential wells of massive DM halos, too many massive galaxies are formed
both in CDM cosmological hydrodynamical simulations and semi-analytic models of galaxy formation and evolution. In our cluster zoom simulation, when no AGN feedback is considered, 
the final mass  of stars in the central cD galaxy is very high, $M_{*}\simeq1.7\, 10^{13} \, \rm M_{\odot}$,  for a $M_{\rm 500}=2.4\, 10^{14}\, \rm M_{\odot}$ ($M_{\rm 200}=2.9\, 10^{14}\, \rm M_{\odot}$) dark matter halo with radius $r_{\rm 500}=940$ kpc 
(resp. $r_{\rm 200}=1370$ kpc, see figure~\ref{nice_dm_stars})\footnote{All quantities with subscripts 200 or 500 refer to regions with overdensities 200 or 500 times 
larger than critical ($\rho_c=3H^2/(8\pi G)$). }. In the presence of stirring from AGN feedback, the total stellar mass is reduced to $M_{*}\simeq5.6 \,10^{12} \, \rm M_{\odot}$, i.e. by more than a factor of 3. To compute the stellar mass content, we use the same
 MSM algorithm \citep{tweedetal09} as for the dark matter but with different parameters, since stars are more clustered than DM particles. This tool efficiently separates one galaxy from another, especially the central galaxy from its satellite galaxies (see figure~\ref{nice_dm_stars}). However, the algorithm 
used in this method attributes all the stars present in the dark matter halo and not part of satellite galaxies to the central one. As a result, a non-negligible part of the stellar mass of the central galaxy resides in the intra-cluster stellar halo (composed of all the stars 
stripped from satellite galaxies of the cluster), which has a very large extent (up to $\sim 400$ kpc). This caveat must be borne in mind when comparing the stellar mass of the central object with observations: our estimate only provides an upper limit of the stellar mass content 
of the central galaxy.

\begin{figure}
  \centering{\resizebox*{!}{8cm}{\includegraphics{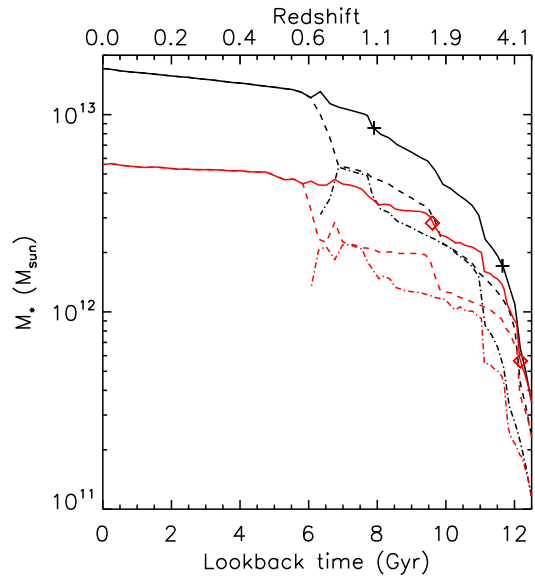}}}
  \caption{Stellar mass evolution of the two galactic progenitors (dashed and dot-dashed lines) involved in the major merger of the cluster for the run without AGN (black) and with AGN (red).
 The solid lines show the cumulative stellar mass of these two progenitors. Crosses and diamonds indicate the $0.5 \, M_{*}(z=0)$, and $0.1 \,M_{*}(z=0)$ epoch respectively.  }
    \label{mgalvsredshift}
\end{figure}

\begin{figure}
  \centering{\resizebox*{!}{8cm}{\includegraphics{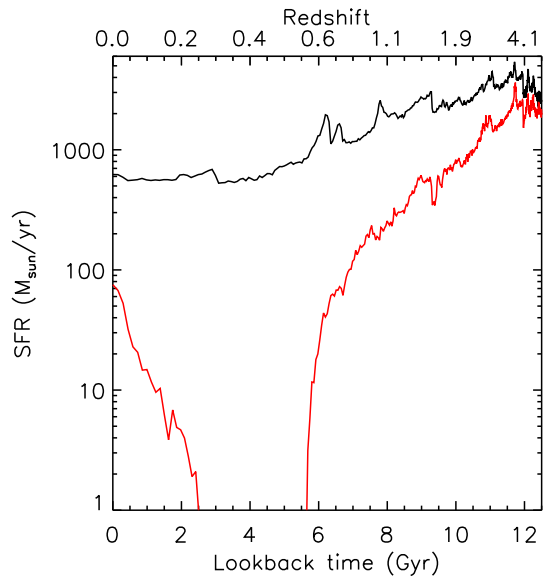}}}
  \caption{Star formation rate as a function of the lookback time for the most massive galaxy at $z=0$ for the no-AGN (black) and AGN run (red).}
    \label{sfr_mygal}
\end{figure}

Figure~\ref{mgalvsredshift} compares the stellar mass evolution as a function of redshift for the most massive galaxy at $z=0$ (solid line) and its two most massive progenitors
(dashed and dot-dashed lines, central galaxies of DM halos identified as branch 1 and 3 in figure~\ref{mergertree_DM}) in the AGN and no-AGN runs (red and black sets of curves respectively). We observe that the reduction of the stellar mass is a continuous process which begins at an early stage 
(around $z\sim5$) but gets amplified as time goes on to reach a factor 3 at $z\sim1$. Both progenitors seems to follow the same reduction of their stellar content which suggests that even the branch 3 cluster, which does not exhibit 
any strong quasar activity at high redshift, is able to prevent some gas from falling onto the central galaxy. The process is more quiescent as can be seen in its 
BH accretion rate (red dashed curve in figure~\ref{mbhvsredshift}), but even this continuous and moderate AGN activity can efficiently reduce the star formation.

We evaluate the bulge to disc mass ratio of the central galaxy by using the kinematics of its star particles. First, we identify the rotation axis of the galaxy to define
the correct cylindrical reference frame in which we project the velocity components of each star particle. A particle belongs to the bulge if its circular velocity is 
lower than half its total velocity. With that definition, the bulge-to-total mass ratio of the central galaxy is $0.75$ for the simulation with AGN, and $0.80$ 
for the simulation without AGN. Thus it appears that although AGN dramatically change the SFH, they have a much less significant impact on the morphology of a galaxy.

Figure~\ref{sfr_mygal} shows the star formation rate for the central galaxy as a function of time for the two runs. To compute the time evolution of the SFR, 
we simply have identified the stars belonging to the central galaxy at $z=0$, and traced them back in time. Thus it is the SFH summed over all the stars 
of all the satellite galaxies that have been accreted onto the central galaxy throughout its evolution. The SFR continuously decreases with time due to early 
AGN activity (z$\sim4$), but the dramatic decline in SFR occurs around $z=0.6$ when the BHs hosted by the central galaxy merger remnant of branch 1 and 3 halos finally 
merge. The vast majority of the cold gas is heated up by the AGN during this violent merger. In contrast, the cold gas in the no-AGN case is simply compressed in the galaxy 
merger which results in a double small star formation peak around the same redshift ($z\sim0.6$).
The latter effect is a well-known property of merging galaxies without AGN \citep{mihos&hernquist96}. However, mergers of galaxies containing BHs
boost the accretion of gas onto the BH fueling strong AGN activity and produce a dip in the SFR by reducing the cold-gas content.
These features are clearly seen at redshift $z=1.6$ and $z=0.6$ in figure~\ref{sfr_mygal}. 
Such a behavior has already been analyzed in detail in idealized (as opposed to cosmological) simulations of galaxy merger \citep{springeletal05, dimatteoetal05}.
Our work confirms that it also occurs in more realistic cosmological configurations.

\begin{figure}
  \centering{\resizebox*{!}{8cm}{\includegraphics{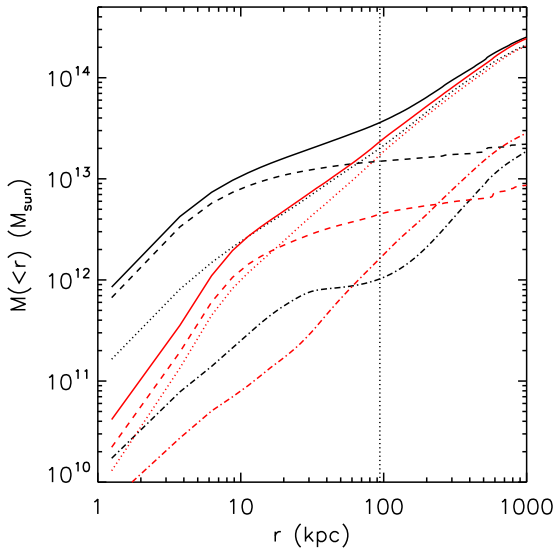}}}
  \caption{Integrated total mass (solid lines), dark matter mass (dotted lines), stellar mass (dashed lines) and gas mass (dot-dashed lines) for the run without AGN (black) and with AGN (red). The dotted vertical line where $r=0.1\times r_{\rm 500}$.}
    \label{massvsr}
\end{figure}

We have measured the cumulative mass profiles of the different components (gas, stars, dark matter) as a function of radius for the cluster at $z=0$ (figure~\ref{massvsr}).
In the absence of feedback, most of the cold baryons are concentrated within the galaxy (in the inner 10 kpc): i.e. the cooling catastrophe has occurred. There 
is a strong difference in the cumulative profiles between the runs with and without AGN activity, especially  in the central region of the cluster. 
They differ by a factor 5 in total mass at $r=10$ kpc and by a factor 1.5 at $r=100$ kpc. Without AGN, the gravitational potential is steeper in the centre of the cluster, baryons accumulate and gravitationally 
pull DM along with them. This is predicted to have severe consequences when simulating the gravitational lensing effect of such structures \citep{peiranietal08, meneghettietal10}. 

\begin{figure}
  \centering{\resizebox*{!}{8cm}{\includegraphics{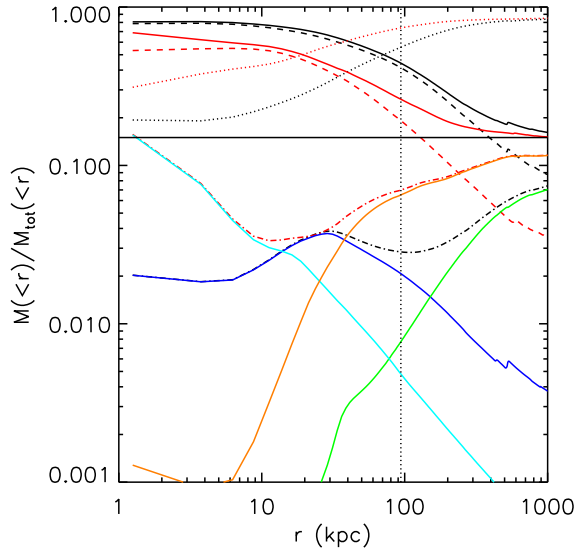}}}
  \caption{Baryon (solid lines), dark matter (dotted lines), stellar (dashed lines) and gas (dot-dashed lines) cumulated mass fractions for the run without AGN (black) and the run with AGN (red). Dark blue is the cold gas component and green the hot gas component for the no-AGN run. Light blue is the cold gas component and orange the hot gas component for the AGN run. The dotted vertical line where $r=0.1\times r_{\rm 500}$. The horizontal line indicates the Universal baryon fraction $f_{\rm b}=\Omega_{\rm b}/\Omega_{\rm m}=0.15$. }
    \label{mfracvsr}
\end{figure}

Baryon fraction as a function of radius provides us with a good benchmark to quantify the influence of feedback processes. We have computed the cumulative fractions 
of stars, gas and dark matter for both runs (figure~\ref{mfracvsr}). We observe a very small difference in the total baryon fraction at $r_{200}$ and $r_{500}$ which means that this 
quantity is relatively immune to the presence of feedback. In contrast, the baryon fraction 
has decreased by a factor 2 at $r=0.1 r_{500}$. This means that half of the baryons that were concentrated in the 
inner parts of the halo have been redistributed by the AGN feedback at larger radii. However, as the baryon fraction seems relatively independent of the presence 
of the AGN at very large radii ($>r_{500}$), we can conclude that only a modest fraction of the gas is expelled by the AGN outside of the cluster.

Stellar and gas fractions yield more clues as to the impact of AGN feedback on the cluster history. There is a strong decrease in the stellar fraction, even at 
large radii ($r>r_{500}$), meaning that star formation has been efficiently suppressed by the AGN activity. We see that the stellar fraction at $r_{500}$ 
has been lowered by more than a factor 2, which is comparable with the results obtained by~\cite{duffyetal10}. 
Table~\ref{tabfrac} show the stellar fractions $f_{500}^{\rm star}$, gas fractions $f_{500}^{\rm gas}$ and baryon fractions (gas and stars) $f_{500}^{\rm b}$ measured 
at $r_{500}$ in our simulations and compared to X-ray or near infrared measurements made by various groups \citep{linetal03, gastaldelloetal06, vikhlininetal06, arnaudetal07, gonzalezetal07, giodinietal09, sunetal09}. 
Observational data values are obtained using the best fit these authors provide for $f_{500}^{\rm star}$, $f_{500}^{\rm gas}$ and $f_{500}^{\rm b}$ as a function of $M_{500}$. From this comparison with observations, the simulation 
which does not include AGN feedback is clearly ruled out, as the stellar fraction is too high by a factor $\sim 3-4$ and the gas fraction too low by 25\% in the most favorable case. 
This clearly indicates that the ICM has undergone 
 a cooling catastrophe: too much gas has been depleted and transformed into stars. On the other hand, the simulation with AGN feedback shows reasonable agreement with the stellar fraction 
estimated from X-ray measurements by \cite{gonzalezetal07} 
and \cite{giodinietal09}, but still overestimates the stellar fraction from near infrared data by \cite{linetal03} by about a factor 2. Gas fraction in the simulation with AGN, is also within the range of values inferred from X-ray gas emission, albeit on the high side. 
The major flaw of our simulations is their inability to match the lower than Universal baryon fraction observed in small galaxy clusters ($M_{500}< 10^{15} \, \rm M_{\odot}$). We note that the discrepancy would be even more blatant
 had we used WMAP 5 year parameters \citep{dunkleyetal09} since 
the Universal baryon fraction goes up to 18\% for this cosmology. In the case of our AGN simulation, the cluster baryon fraction is close to the Universal baryon fraction $\Omega_{\rm b}/\Omega_{\rm m}=0.15$,
 but we fail to push gas far enough out 
of the cluster potential well to match lower observational values sitting around $\sim0.13$.

\begin{table}
\caption{Comparison of the stellar, gas and baryon fractions at $r_{500}$ in our simulations with the observational data.}
\label{tabfrac}
\begin{tabular}{@{}|l|c|c|c|c|c|}
  \hline
  & no AGN & AGN & L03$^a$ & G07$^b$ & G09$^c$ \\
  \hline
  $f_{500}^{\rm star}$ & 0.090 & 0.036 & 0.019 & 0.023 & 0.032\\
  $f_{500}^{\rm gas}$ & 0.073 & 0.116 & 0.117 & 0.101 & 0.090\\
  $f_{500}^{\rm b}$ & 0.163 & 0.152 & 0.136 & 0.124 & 0.122\\
  \hline
\end{tabular}

\medskip
$^a$ Observational data from \cite{linetal03}\\
$^b$ Observational data from \cite{gonzalezetal07}. Their gas fraction is the best fit to data from \cite{vikhlininetal06} and \cite{gastaldelloetal06}. \\
$^c$ Observational data from \cite{giodinietal09}. Their stellar fraction are the best fit of their data combined with data from \cite{linetal03}. Their gas fraction is the best fit to data from \cite{vikhlininetal06}, \cite{arnaudetal07} and \cite{sunetal09}
\end{table}

The gas fraction behavior is somewhat counter-intuitive: it is larger in the run with AGN feedback, whatever the radius is. This is explained by the fact that AGN 
removes gas from the central regions of the cluster to replenish its outer parts. AGN feedback thus transforms cold gas contained in the central galactic disc into 
hot and diffuse halo gas. Moreover, the gas fraction has a remarkable feature in the form of a pronounced dip at intermediate radius (15 kpc for the no-AGN and 
100 kpc for the AGN case) which marks the transition between the cold/dense phase ($n >0.1 \, \rm cm^{-3}$), and the hot/diffuse component. 
Such dips in the gas fraction also are commonplace in X-ray cluster surveys \citep{vikhlininetal06}.
Another interesting result from Fig.~\ref{mfracvsr} is that the dark matter to total mass fraction in the cluster core ($r < 10$kpc) is higher in the case of AGN feedback, 
even though the total amount of dark matter is lower in this case. We attribute this to the domination of the mass budget by the stellar component which pulls
DM along with it through adiabatic contraction \citep{blumenthaletal86}. 

\begin{figure}
  \centering{\resizebox*{!}{6cm}{\includegraphics{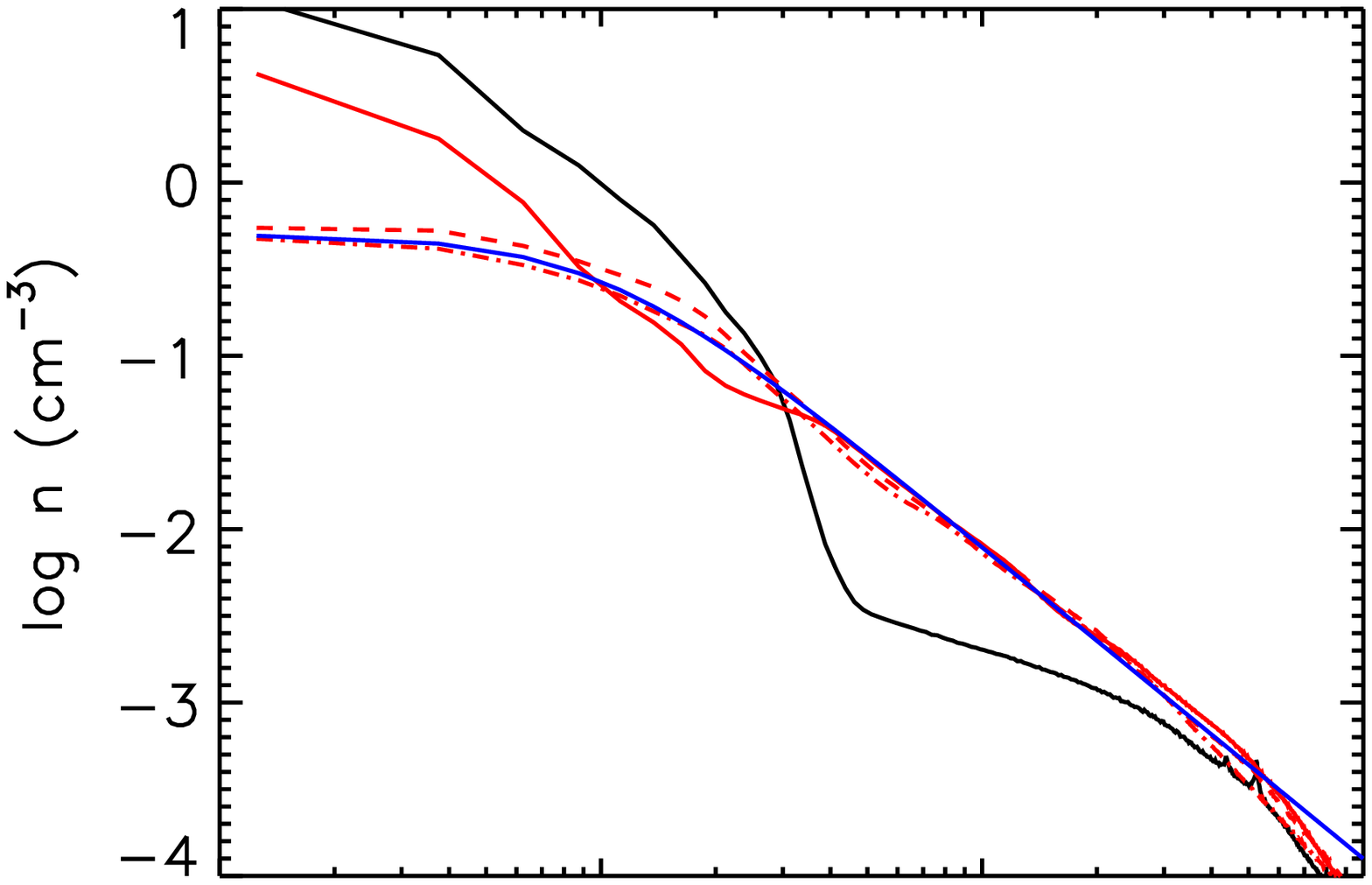}}\vspace{-1.7cm}}
  \centering{\vspace{-1.7cm}\resizebox*{!}{6cm}{\includegraphics{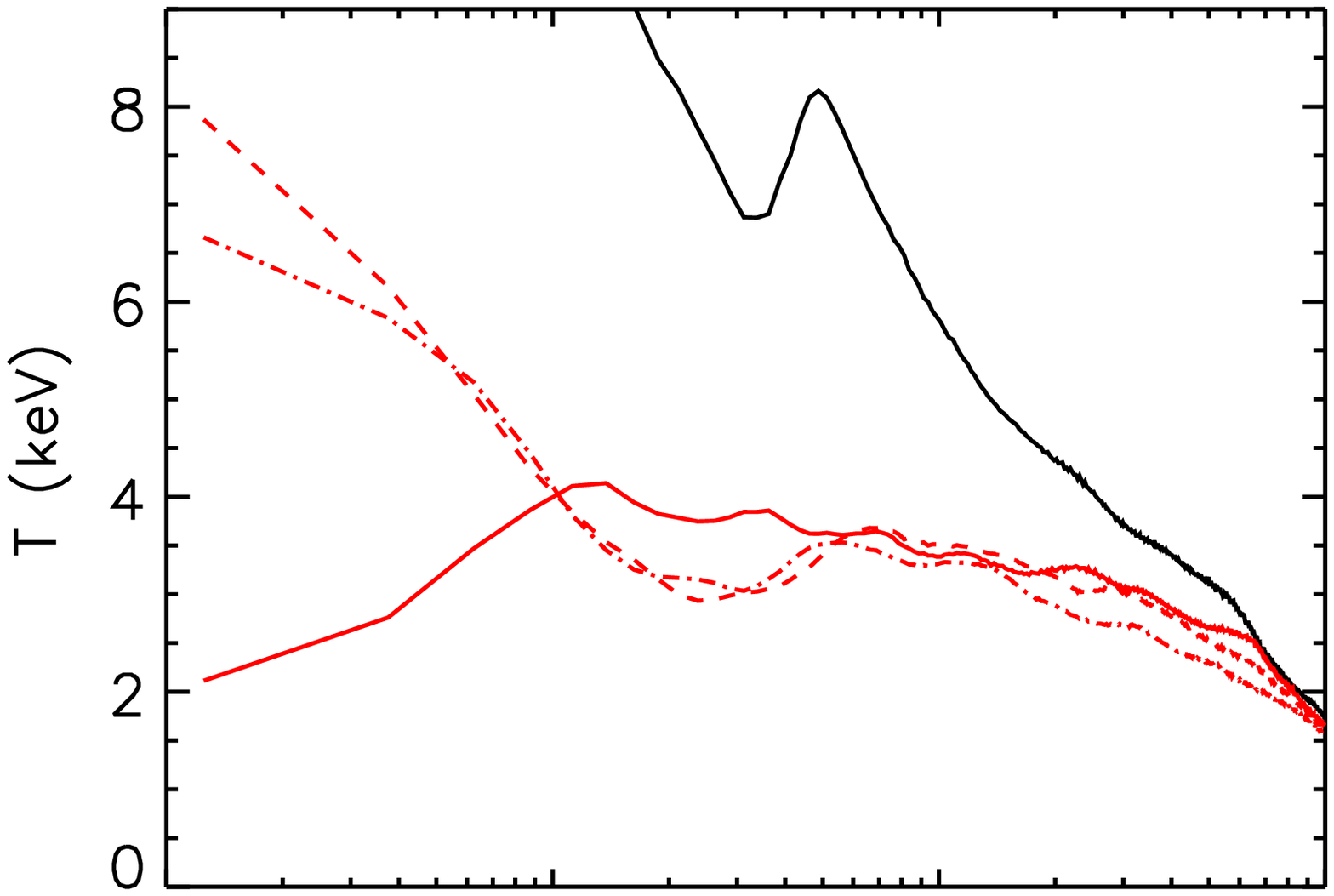}}}
  \centering{\vspace{-1.7cm}\resizebox*{!}{6cm}{\includegraphics{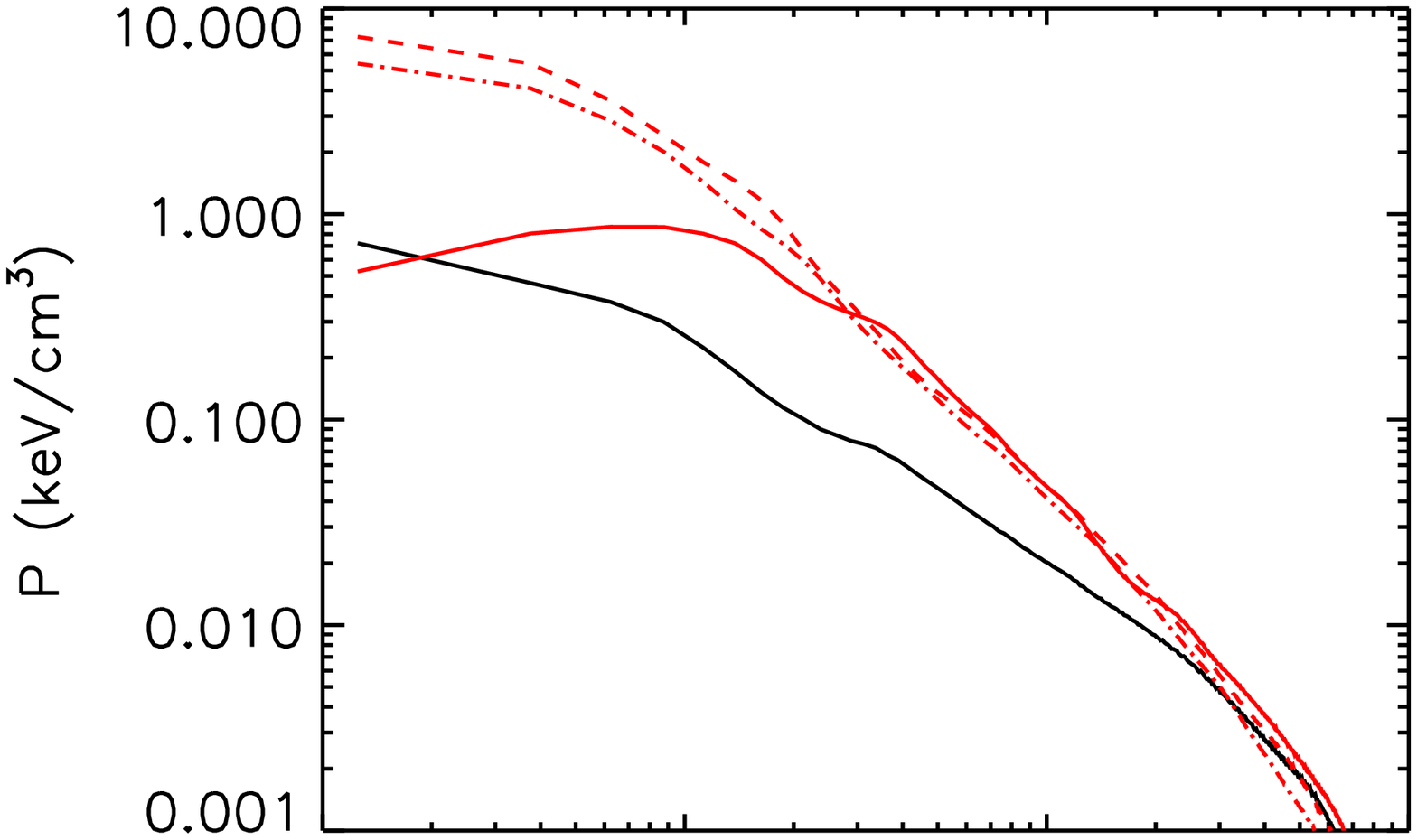}}}
  \centering{\resizebox*{!}{6cm}{\includegraphics{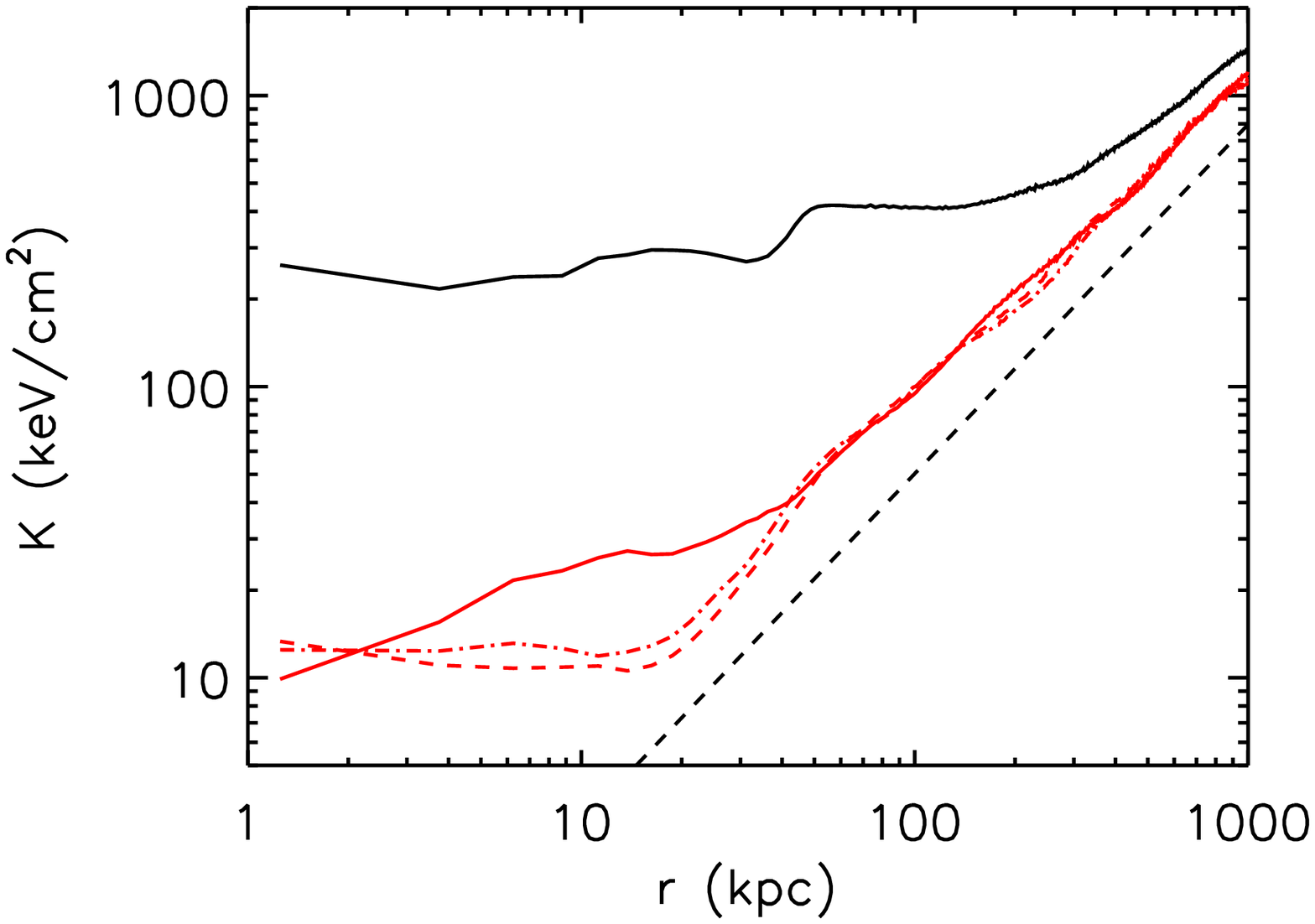}}}
  \caption{From top to bottom: volume-weighted average number density, temperature, pressure and entropy as a function of radius for the run without AGN (black solid) at $z=0$, and the run with AGN at $z=0$ (red solid), $z=0.04$ (red dashed) and $z=0.09$ (red dot-dashed). The blue line in the number density plot is a $\beta$ profile fit with $\beta=0.6$. 
The dashed black line in the entropy plot describes a $K\propto r^{1.2}$ power-law. }
    \label{gasprofiles}
\end{figure}

Finally, we compare the thermodynamical properties of the gas in the two runs in figure~\ref{gasprofiles}. We have fitted the gas density profile in the AGN run with a $\beta$ 
profile of the form 
\begin{equation}
\rho=\rho_{\rm s}  \left (1+(r/r_{\rm c})^2\right)^{-3\beta/2}\, ,
\end{equation}
where $\rho_{\rm s}=0.5\, \rm cm^{-3}$, $r_{\rm c}=10$ kpc and $\beta=0.6$. This profile matches the density profile of the relaxed cluster at different times ($z=0$, $z=0.04$ and $z=0.09$, 
which are separated by $500$ Myr) in the intermediate radius range $0.05$-$1$ $r_{500}$. When the cluster is relaxed, the same analytic profile extends to the core of the cluster, but as soon as a 
cooling flow develops it fails to describe the numerical gas density accurately. Indeed, the core of the cluster shrinks as gas flows in, so
$r_{\rm c}$ drops, but the $\beta$ index stays identical as the profile remains unaffected by the cooling flow on large scales. By contrast, fitting the gas density profile with 
a $\beta$ profile for the simulation where no AGN is present turns out to be an impossible task, since the core radius becomes smaller than the spatial resolution in that case.

\begin{figure}
  \centering{\resizebox*{!}{6cm}{\includegraphics{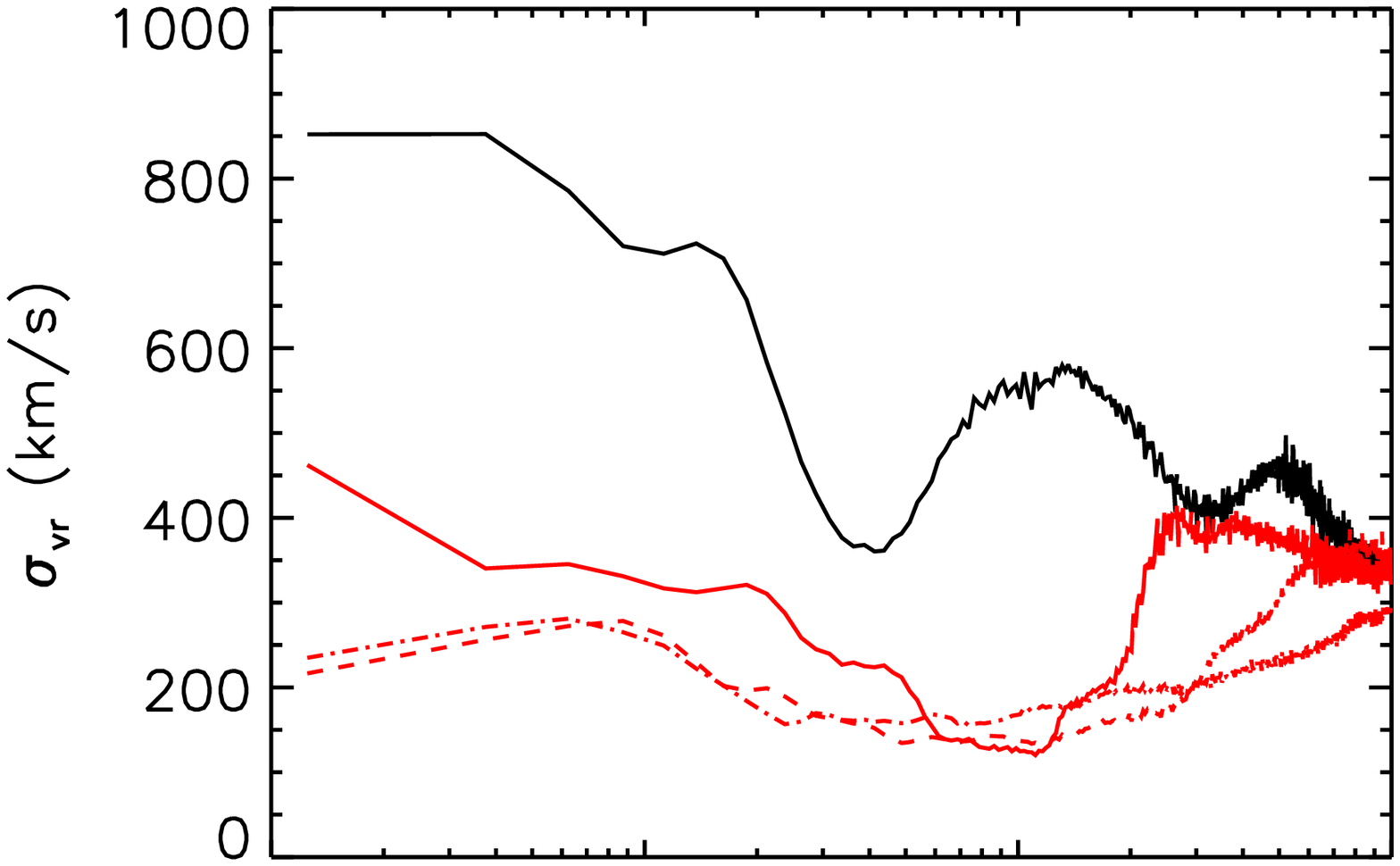}}\vspace{-1.7cm}}
  \centering{\resizebox*{!}{6cm}{\includegraphics{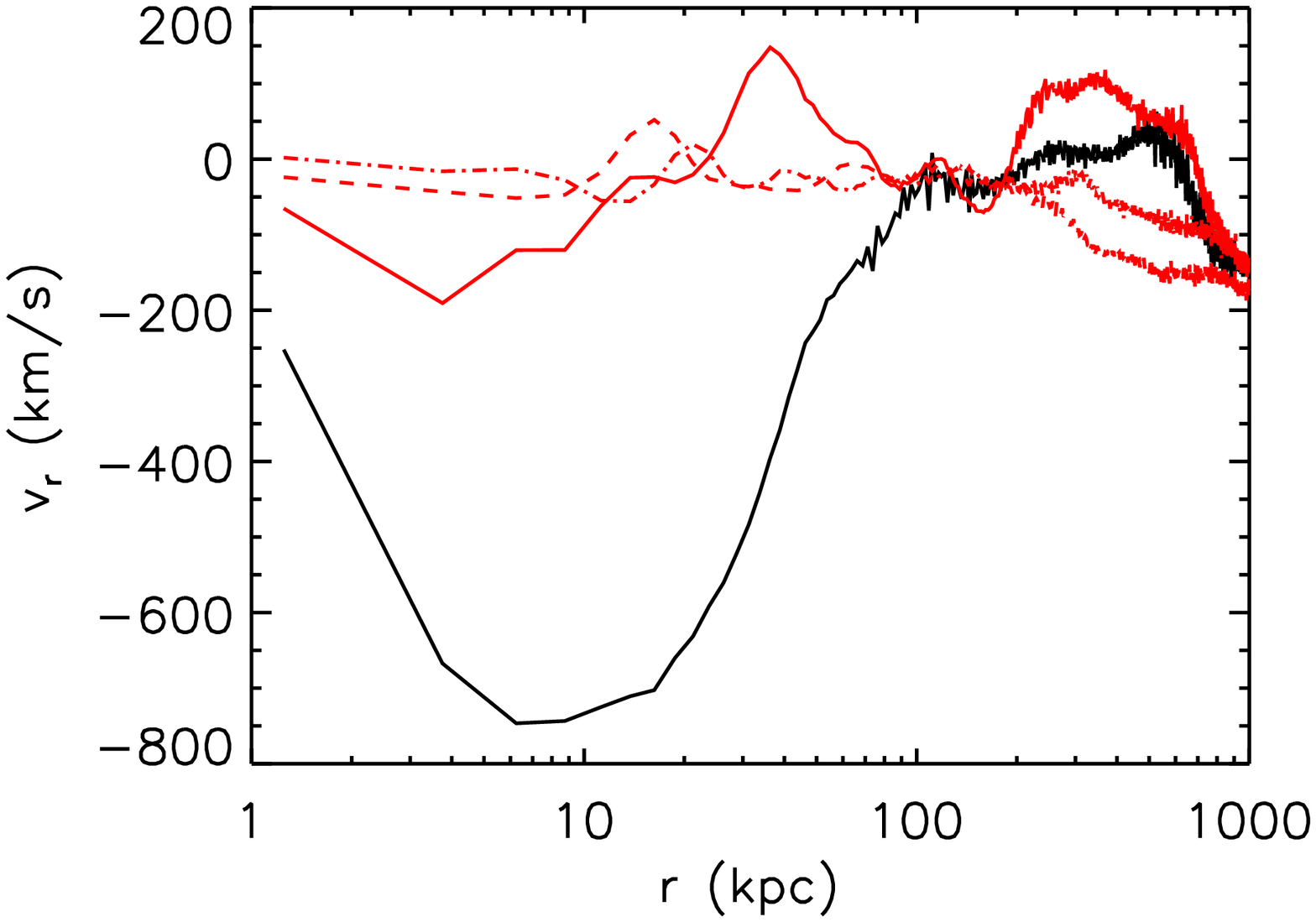}}}
  \caption{Same as figure~\ref{gasprofiles} for the volume-weighted average radial velocity dispersion (upper panel) and radial velocity (bottom panel) of the gas.}
    \label{velprofiles}
\end{figure}

Surprisingly, the simulation without the AGN, which has endured a cooling catastrophe for Gyrs, exhibits a very hot gas core (temperature in excess of 9 keV) with a very steep
profile (see second panel from the top on Fig~\ref{gasprofiles}). Actually a massive central cold disc component is also present in this run and would appear on
Fig~\ref{gasprofiles} if we were measuring mass-weighted instead of volume-weighted quantities, but the properties of the gas at the center of the cluster would then reflect the properties of the ISM instead of the diffuse ICM. 
This very hot thermal part, in the no-AGN run, arises from the cluster need for more thermal energy to support the gas against the extra gravitational compression generated by adiabatic contraction. 
Due to the very high temperature and a lack of diffuse gas around the post-merger galaxy this energy is not easily radiated away (the cooling time is greater than 2 Gyr as shown on Fig~\ref{ctimevsr}). 
However we can clearly see in figure~\ref{velprofiles} that the gas, close to the galactic disc ($r< 50$ kpc), still collapses onto that galaxy due to the lack of pressure support (figure~\ref{gasprofiles}),
which explains the depletion of the gas component at $r\simeq100$ kpc.

On the other hand, when the AGN is active the temperature profile is stabilized and looks quasi-isothermal in the range $0.05$-$1$ $r_{500}$ (second panel of Fig~\ref{gasprofiles}).
Before $z=0$, the temperature is a factor 2 to 4 higher in the inner 10 kpc, due to heating from the jet, which remains confined in that region. As the gas radiates away the jet energy,
its temperature drops and a cooling flow develops because of a lack of pressure support in the core (second panel from the bottom on Fig~\ref{gasprofiles}). Small variations of
temperature with radius in the form of wiggles can be observed in Fig~\ref{gasprofiles} (second panel from the top, solid red curve). These correspond to the propagation of sound waves into the
intra-cluster medium. These waves contribute to reheating the cooling plasma in the cluster as a whole by propagating and isotropising the energy injected by the AGN jet.
They manage to offset the extremely short cooling time within $r< 0.1\, r_{500}$ which is at least one order of magnitude shorter than the time elapsed since the last major merger
(see figure~\ref{ctimevsr}), and thus prevent most of the gas from collapsing onto the central galaxy.

The pressure cavities seen on figure~\ref{jet_slice} and in the X-ray map (figure~\ref{nice_Xray}), are visible in the pressure profile of figure~\ref{gasprofiles} (second panel from the bottom):
at $z=0$, there is a small depression in the pressure profile around $r\simeq 15$-$30$ kpc, which does not appear at earlier times when the AGN is not active enough to form these cavities. This
feature is also detectable in the radial velocity profile of the gas on figure~\ref{velprofiles} (bottom panel): there is a net radial gas outflow at $r \simeq 15-30$ kpc whose maximum corresponds
to the maximum extent of the jet and whose outwardly decreasing profile reflects the pre-shocked cocoon region. The volume-averaged velocity which we plot on this figure is however under-estimated, because the
sonic outflowing component of the jet is mixed with the quasi-steady flow or inflowing regions. The velocity inside the jet is much faster, around $1000 \, \rm km/s$.

Finally, entropy profiles (bottom panel of Fig~\ref{gasprofiles}) shows a plateau in the cluster core (at $r<20$ kpc for the AGN run, and $r< 300$ kpc for the no-AGN run) with a strong departure from the
scaling power law $K\propto r^{1.2}$. This indicates the level of turbulent mixing in the gas as explained in detail by~\cite{mitchelletal09} and is nicely illustrated by comparing entropy profiles
on figure~\ref{gasprofiles} (bottom panel) with radial velocity dispersion profiles on figure~\ref{velprofiles} (top panel). Such a comparison clearly shows that the stronger the turbulence level (or equivalently
the radial velocity dispersion), the higher the entropy. 

\section{Discussion and conclusion}
\label{discussion}

We have numerically studied how a self-regulating model of supermassive black hole growth and AGN feedback impacts the formation history of a
large cluster of galaxies. Using a resimulation technique to explicitly account for the cosmological context ($\Lambda$CDM Universe) which drives
the cluster growth and an AMR technique to solve the equations of hydrodynamics without running into entropy issues,
we find that:
\begin{enumerate}
\item[-] BHs accrete gas efficiently at high redshift ($z>2$), significantly pre-heating proto-cluster halos in the process.
\item[-] some, but not all, wet (gas-rich) mergers fuel strong episodic jet activity which transport gas from the cluster core to its intermediate/outer regions.
\item[-] reduced infall of cold gas during the more secular phase evolution of clusters produces smaller
outbursts from the central AGN which contribute to heat the whole cluster via sound waves but are inefficient at redistributing the gas
outwards.
\item[-] late-time AGN activity forms two large cavities correlated with the emergence of a small cooling flow. 
\end{enumerate}

Whilst our model for accretion onto the black hole is commonly used in simulations in the literature 
\citep{springeletal05, dimatteoetal05, sijackietal07, dimatteoetal08, booth&schaye09, teyssieretal10},
this is the first time, to the best of our knowledge, that a momentum driven jet is implemented as AGN feedback in cosmological simulations and followed in a self-consistent way.
We argue that this is one step in the correct direction
since powerful jets are observed in the Universe on scales well resolved by any of these simulations (e.g. \citealp{bridleetal94}).
Other authors have adopted a more phenomenological approach where energy is either accumulated by the BH before being released as a thermal pulse 
\citep{booth&schaye09, teyssieretal10}, or simply used as a
continuous heat source \citep{springeletal05, dimatteoetal05, dimatteoetal08}, or used both heating modes  \citep{sijackietal07}. 
Injection of non-thermal relativistic protons in rising buoyant AGN bubbles has also been explored as an alternative feedback mechanism \citep{sijackietal08}.
It is worth noting the recrudescence of efforts done to improve models of AGN feedback in idealised (non-cosmological) simulations of galaxy evolution, where kinetic energy is deposited either isotropically \citep{debuhretal10, poweretal10} or -- as in the present paper -- as a collimated jet \citep{nayakshin&power10}. These models probably capture more of the relevant jet physics that what we achieve here, and we believe that as cosmological simulation spatial resolution increases, more physical insight into the impact of AGN feedback on the ICM will be gained by coupling them to such models.
While we believe that most of the results we get are similar to those obtained with the continuous heating model and that our feedback seems less efficient at stopping the cooling catastrophe than the accumulated heating prescription, the devil certainly is in the details and we defer a more detailed comparison to a future paper (Dubois et al in prep).
It is however interesting to note that all these models, including ours, are calibrated to provide an acceptable $M_{\rm BH}$ vs $M_{\star}$ (or $M_{\rm BH}$ vs $\sigma_{\star}$) relation so that it is very unlikely that this relation can be used to constrain feedback mechanisms.

From both semi-analytic prescriptions~\citep{boweretal06, cattaneoetal06} and recent numerical simulations~\citep{ruszkowski&oh10, duffyetal10}, it has become clear that AGN feedback
must play a major role in reducing the central galaxy stellar mass in groups and clusters by a large amount. Our implementation of feedback succeeds at least partially
in reaching this goal: the stellar mass of the central object is reduced by more than a factor 3 in the run where AGN feedback stirs the gas whose properties
(density profile, temperature, radial velocity) also are in better agreement with observations. Another success of our AGN modeling is its capacity to reproduce double cavities 
separated by a cold component as seen in the X-ray emission of observed clusters (e.g. in Perseus, \citealp{fabianetal06}). Our experiments with a simple isotropic thermal input \citep{teyssieretal10} suggest that these cannot be reproduced.

One drawback of our simulation is that no supernova feedback is included. However we do not expect this feedback to be energetically relevant as a simple estimate based on the star formation 
history of our galaxies shows that it is always an order of magnitude lower than the AGN energy input (figure~\ref{lagnvsredshift}),
On the other hand, supernovae feedback releases metals into the ICM. As the cluster gas is heated to very high temperatures ($\sim 3$ keV), cooling in the ICM is primarily due to free-free collisions, 
thus we do not expect metals to strongly alter the cooling rate of the plasma. Indeed, at $T=3.5$ keV, the relative difference in the net cooling rate between a zero and a one third of solar metallicity
plasma is $0.2$ according to \cite{tozzi&norman01}. This means that, assuming a Z$=0.3\, \rm Z_{\odot}$ metallicity in the ICM, our simulation underestimates the gas cooling rate by 20 \%.

As always when analysing numerical simulations, one has to worry about spatial and mass resolution related issues. Indeed, accretion onto a SMBH happens on (sub)parsec scales, well below the 
kpc size of the smallest cell in our cluster re-simulation. Even though we use the empirically well motivated subgrid model of \cite{booth&schaye09} to account for this lack of resolution, one 
would like to test its validity by performing direct simulations. Whilst this is beyond the reach of the current generation of supercomputers, we have tried to assess both the robustness
of our subgrid and our jet implementations by performing a sub-kpc run and conclude that our results are by-and-large unchanged at this increased spatial resolution. This is also true for 
a modest increase in mass resolution. We are therefore quite confident that the conclusions we draw in this paper are robust vis-a-vis resolution issues and defer a more thorough resolution 
study to future work.

Finally, other physical mechanisms, which we do not model here, could potentially play a significant role in preventing the development of massive cooling flows.
These mechanisms involve tapping into the heat reservoir provided by the outer regions of galaxy clusters to raise the gas temperature in their core. In particular, 
recent efforts have been made to investigate the importance of anisotropic thermal conduction in idealized galaxy clusters \citep{parrish&quataert08, parrish09, bogdanovic09}. 
These studies have shown
 that the HBI can reorient the magnetic field lines in the cluster core in an azimuthal configuration that stops the inward heat flux. The most recent of these simulations 
\citep{parrishetal10} demonstrated 
that if some small turbulence is brought to break this magnetic field topology, heating is able to proceed. One has to wonder, in the context of anisotropic thermal 
conduction, whether the turbulence induced 
by small-scale and large-scale motions is able to reorder magnetic fields in cosmological simulations of galaxy clusters \citep{dolagetal05, asaietal07, dubois&teyssier08cluster, 
pfrommer&dursi09}, and, thus, 
break the shelf-shielding of the heat-flux. In particular, the presence of AGN stirring can help to disturb this magnetic equilibrium \citep{duboisetal09}, so we believe that this
 problem needs to be addressed 
with MHD cosmological simulations including both AGN stirring and anisotropic thermal conduction.  

\section*{Acknowledgment}
We thank Taysun Kimm and Stas Shabala for useful comments and discussion. YD is supported by an STFC Postdoctoral Fellowship. The simulations presented here were run on the TITANE cluster at the Centre de Calcul Recherche et Technologie in CEA Saclay on allocated resources from the GENCI grant c2009046197.

\bibliographystyle{mn2e}
\bibliography{author}

17 gid=784332012
18 uid=1861002210
20 ctime=1278672966
20 atime=1278672992
38 LIBARCHIVE.creationtime=1274973902
24 SCHILY.dev=234881027
21 SCHILY.ino=873422
18 SCHILY.nlink=1


\begin{thebibliography}{}

\bibitem[\protect\citeauthoryear{{Agertz}, {Lake}, {Teyssier}, {Moore}, {Mayer}
  \& {Romeo}}{{Agertz} et~al.}{2009}]{agertzetal09}
{Agertz} O.,  {Lake} G.,  {Teyssier} R.,  {Moore} B.,  {Mayer} L.,    {Romeo}
  A.~B.,  2009, \mnras, 392, 294

\bibitem[\protect\citeauthoryear{{Agertz}, {Moore}, {Stadel}, {Potter},
  {Miniati}, {Read}, {Mayer}, {Gawryszczak}, {Kravtsov}, {Nordlund}, {Pearce},
  {Quilis}, {Rudd}, {Springel}, {Stone}, {Tasker}, {Teyssier}, {Wadsley} \&
  {Walder}}{{Agertz} et~al.}{2007}]{agertzetal07}
{Agertz} O.,  {Moore} B.,  {Stadel} J.,  {Potter} D.,  {Miniati} F.,  {Read}
  J.,  {Mayer} L.,  {Gawryszczak} A.,  {Kravtsov} A.,  {Nordlund} {\AA}.,
  {Pearce} F.,  {Quilis} V.,  {Rudd} D.,  {Springel} V.,  {Stone} J.,  {Tasker}
  E.,  {Teyssier} R.,  {Wadsley} J.,    {Walder} R.,  2007, \mnras, 380, 963

\bibitem[\protect\citeauthoryear{{Arnaud}, {Fabian}, {Eales}, {Jones} \&
  {Forman}}{{Arnaud} et~al.}{1984}]{arnaudetal84}
{Arnaud} K.~A.,  {Fabian} A.~C.,  {Eales} S.~A.,  {Jones} C.,    {Forman} W.,
  1984, \mnras, 211, 981

\bibitem[\protect\citeauthoryear{{Arnaud}, {Pointecouteau} \& {Pratt}}{{Arnaud}
  et~al.}{2007}]{arnaudetal07}
{Arnaud} M.,  {Pointecouteau} E.,    {Pratt} G.~W.,  2007, \aap, 474, L37

\bibitem[\protect\citeauthoryear{{Asai}, {Fukuda} \& {Matsumoto}}{{Asai}
  et~al.}{2007}]{asaietal07}
{Asai} N.,  {Fukuda} N.,    {Matsumoto} R.,  2007, \apj, 663, 816

\bibitem[\protect\citeauthoryear{{Babul}, {Balogh}, {Lewis} \& {Poole}}{{Babul}
  et~al.}{2002}]{babuletal02}
{Babul} A.,  {Balogh} M.~L.,  {Lewis} G.~F.,    {Poole} G.~B.,  2002, \mnras,
  330, 329

\bibitem[\protect\citeauthoryear{{Basson} \& {Alexander}}{{Basson} \&
  {Alexander}}{2003}]{basson&alexander03}
{Basson} J.~F.,  {Alexander} P.,  2003, \mnras, 339, 353

\bibitem[\protect\citeauthoryear{{Bate}, {Bonnell} \& {Price}}{{Bate}
  et~al.}{1995}]{bateetal95}
{Bate} M.~R.,  {Bonnell} I.~A.,    {Price} N.~M.,  1995, \mnras, 277, 362

\bibitem[\protect\citeauthoryear{{Benson} \& {Babul}}{{Benson} \&
  {Babul}}{2009}]{benson&babul09}
{Benson} A.~J.,  {Babul} A.,  2009, \mnras, 397, 1302

\bibitem[\protect\citeauthoryear{{Binney} \& {Tabor}}{{Binney} \&
  {Tabor}}{1995}]{binney&tabor95}
{Binney} J.,  {Tabor} G.,  1995, \mnras, 276, 663

\bibitem[\protect\citeauthoryear{{Birnboim} \& {Dekel}}{{Birnboim} \&
  {Dekel}}{2003}]{birnboim&dekel03}
{Birnboim} Y.,  {Dekel} A.,  2003, \mnras, 345, 349

\bibitem[\protect\citeauthoryear{{B{\^i}rzan}, {Rafferty}, {McNamara}, {Wise}
  \& {Nulsen}}{{B{\^i}rzan} et~al.}{2004}]{birzanetal04}
{B{\^i}rzan} L.,  {Rafferty} D.~A.,  {McNamara} B.~R.,  {Wise} M.~W.,
  {Nulsen} P.~E.~J.,  2004, \apj, 607, 800

\bibitem[\protect\citeauthoryear{{Blandford} \& {Begelman}}{{Blandford} \&
  {Begelman}}{1999}]{blandford&begelman99}
{Blandford} R.~D.,  {Begelman} M.~C.,  1999, \mnras, 303, L1

\bibitem[\protect\citeauthoryear{{Blumenthal}, {Faber}, {Flores} \&
  {Primack}}{{Blumenthal} et~al.}{1986}]{blumenthaletal86}
{Blumenthal} G.~R.,  {Faber} S.~M.,  {Flores} R.,    {Primack} J.~R.,  1986,
  \apj, 301, 27

\bibitem[\protect\citeauthoryear{{Boehringer}, {Voges}, {Fabian}, {Edge} \&
  {Neumann}}{{Boehringer} et~al.}{1993}]{boehringeretal93}
{Boehringer} H.,  {Voges} W.,  {Fabian} A.~C.,  {Edge} A.~C.,    {Neumann}
  D.~M.,  1993, \mnras, 264, L25

\bibitem[\protect\citeauthoryear{{Bogdanovi{\'c}}, {Reynolds}, {Balbus} \&
  {Parrish}}{{Bogdanovi{\'c}} et~al.}{2009}]{bogdanovic09}
{Bogdanovi{\'c}} T.,  {Reynolds} C.~S.,  {Balbus} S.~A.,    {Parrish} I.~J.,
  2009, \apj, 704, 211

\bibitem[\protect\citeauthoryear{{Bondi}}{{Bondi}}{1952}]{bondi52}
{Bondi} H.,  1952, \mnras, 112, 195

\bibitem[\protect\citeauthoryear{{Booth} \& {Schaye}}{{Booth} \&
  {Schaye}}{2009}]{booth&schaye09}
{Booth} C.~M.,  {Schaye} J.,  2009, \mnras, 398, 53

\bibitem[\protect\citeauthoryear{{Bower}, {Benson}, {Malbon}, {Helly}, {Frenk},
  {Baugh}, {Cole} \& {Lacey}}{{Bower} et~al.}{2006}]{boweretal06}
{Bower} R.~G.,  {Benson} A.~J.,  {Malbon} R.,  {Helly} J.~C.,  {Frenk} C.~S.,
  {Baugh} C.~M.,  {Cole} S.,    {Lacey} C.~G.,  2006, \mnras, 370, 645

\bibitem[\protect\citeauthoryear{{Bridle}, {Hough}, {Lonsdale}, {Burns} \&
  {Laing}}{{Bridle} et~al.}{1994}]{bridleetal94}
{Bridle} A.~H.,  {Hough} D.~H.,  {Lonsdale} C.~J.,  {Burns} J.~O.,    {Laing}
  R.~A.,  1994, \aj, 108, 766

\bibitem[\protect\citeauthoryear{{Carilli}, {Perley} \& {Harris}}{{Carilli}
  et~al.}{1994}]{carillietal94}
{Carilli} C.~L.,  {Perley} R.~A.,    {Harris} D.~E.,  1994, \mnras, 270, 173

\bibitem[\protect\citeauthoryear{{Cattaneo}, {Dekel}, {Devriendt}, {Guiderdoni}
  \& {Blaizot}}{{Cattaneo} et~al.}{2006}]{cattaneoetal06}
{Cattaneo} A.,  {Dekel} A.,  {Devriendt} J.,  {Guiderdoni} B.,    {Blaizot} J.,
   2006, \mnras, 370, 1651

\bibitem[\protect\citeauthoryear{{Cattaneo} \& {Teyssier}}{{Cattaneo} \&
  {Teyssier}}{2007}]{cattaneo&teyssier07}
{Cattaneo} A.,  {Teyssier} R.,  2007, \mnras, 376, 1547

\bibitem[\protect\citeauthoryear{{Churazov}, {Br{\"u}ggen}, {Kaiser},
  {B{\"o}hringer} \& {Forman}}{{Churazov} et~al.}{2001}]{churazovetal01}
{Churazov} E.,  {Br{\"u}ggen} M.,  {Kaiser} C.~R.,  {B{\"o}hringer} H.,
  {Forman} W.,  2001, \apj, 554, 261

\bibitem[\protect\citeauthoryear{{De Villiers}, {Hawley}, {Krolik} \&
  {Hirose}}{{De Villiers} et~al.}{2005}]{devilliersetal05}
{De Villiers} J.,  {Hawley} J.~F.,  {Krolik} J.~H.,    {Hirose} S.,  2005,
  \apj, 620, 878

\bibitem[\protect\citeauthoryear{{Debuhr}, {Quataert}, {Ma} \&
  {Hopkins}}{{Debuhr} et~al.}{2010}]{debuhretal10}
{Debuhr} J.,  {Quataert} E.,  {Ma} C.,    {Hopkins} P.,  2010, \mnras, 406, L55

\bibitem[\protect\citeauthoryear{{Dekel} \& {Birnboim}}{{Dekel} \&
  {Birnboim}}{2006}]{dekel&birnboim06}
{Dekel} A.,  {Birnboim} Y.,  2006, \mnras, 368, 2

\bibitem[\protect\citeauthoryear{{Di Matteo}, {Colberg}, {Springel},
  {Hernquist} \& {Sijacki}}{{Di Matteo} et~al.}{2008}]{dimatteoetal08}
{Di Matteo} T.,  {Colberg} J.,  {Springel} V.,  {Hernquist} L.,    {Sijacki}
  D.,  2008, \apj, 676, 33

\bibitem[\protect\citeauthoryear{{Di Matteo}, {Springel} \& {Hernquist}}{{Di
  Matteo} et~al.}{2005}]{dimatteoetal05}
{Di Matteo} T.,  {Springel} V.,    {Hernquist} L.,  2005, \nat, 433, 604

\bibitem[\protect\citeauthoryear{{Dib}, {Bell} \& {Burkert}}{{Dib}
  et~al.}{2006}]{dibetal06}
{Dib} S.,  {Bell} E.,    {Burkert} A.,  2006, \apj, 638, 797

\bibitem[\protect\citeauthoryear{{Dolag}, {Grasso}, {Springel} \&
  {Tkachev}}{{Dolag} et~al.}{2005}]{dolagetal05}
{Dolag} K.,  {Grasso} D.,  {Springel} V.,    {Tkachev} I.,  2005, Journal of
  Cosmology and Astro-Particle Physics, 1, 9

\bibitem[\protect\citeauthoryear{{Dong}, {Rasmussen} \& {Mulchaey}}{{Dong}
  et~al.}{2010}]{dongetal10}
{Dong} R.,  {Rasmussen} J.,    {Mulchaey} J.~S.,  2010, \apj, 712, 883

\bibitem[\protect\citeauthoryear{{Dubois}, {Devriendt}, {Slyz} \&
  {Silk}}{{Dubois} et~al.}{2009}]{duboisetal09}
{Dubois} Y.,  {Devriendt} J.,  {Slyz} A.,    {Silk} J.,  2009, \mnras, 399, L49

\bibitem[\protect\citeauthoryear{{Dubois} \& {Teyssier}}{{Dubois} \&
  {Teyssier}}{2008a}]{dubois&teyssier08cluster}
{Dubois} Y.,  {Teyssier} R.,  2008a, \aap, 482, L13

\bibitem[\protect\citeauthoryear{{Dubois} \& {Teyssier}}{{Dubois} \&
  {Teyssier}}{2008b}]{dubois&teyssier08winds}
{Dubois} Y.,  {Teyssier} R.,  2008b, \aap, 477, 79

\bibitem[\protect\citeauthoryear{{Duffy}, {Schaye}, {Kay}, {Dalla Vecchia},
  {Battye} \& {Booth}}{{Duffy} et~al.}{2010}]{duffyetal10}
{Duffy} A.~R.,  {Schaye} J.,  {Kay} S.~T.,  {Dalla Vecchia} C.,  {Battye}
  R.~A.,    {Booth} C.~M.,  2010, \mnras, 405, 2161

\bibitem[\protect\citeauthoryear{{Dunkley}, {Komatsu}, {Nolta}, {Spergel},
  {Larson}, {Hinshaw}, {Page}, {Bennett}, {Gold}, {Jarosik}, {Weiland},
  {Halpern}, {Hill}, {Kogut}, {Limon}, {Meyer}, {Tucker}, {Wollack} \&
  {Wright}}{{Dunkley} et~al.}{2009}]{dunkleyetal09}
{Dunkley} J.,  {Komatsu} E.,  {Nolta} M.~R.,  {Spergel} D.~N.,  {Larson} D.,
  {Hinshaw} G.,  {Page} L.,  {Bennett} C.~L.,  {Gold} B.,  {Jarosik} N.,
  {Weiland} J.~L.,  {Halpern} M.,  {Hill} R.~S.,  {Kogut} A.,  {Limon} M.,
  {Meyer} S.~S.,  {Tucker} G.~S.,  {Wollack} E.,    {Wright} E.~L.,  2009,
  \apjs, 180, 306

\bibitem[\protect\citeauthoryear{{Dunn}, {Allen}, {Taylor}, {Shurkin},
  {Gentile}, {Fabian} \& {Reynolds}}{{Dunn} et~al.}{2010}]{dunnetal10}
{Dunn} R.~J.~H.,  {Allen} S.~W.,  {Taylor} G.~B.,  {Shurkin} K.~F.,  {Gentile}
  G.,  {Fabian} A.~C.,    {Reynolds} C.~S.,  2010, \mnras, 404, 180

\bibitem[\protect\citeauthoryear{{Fabian}, {Celotti}, {Blundell}, {Kassim} \&
  {Perley}}{{Fabian} et~al.}{2002}]{fabianetal02}
{Fabian} A.~C.,  {Celotti} A.,  {Blundell} K.~M.,  {Kassim} N.~E.,    {Perley}
  R.~A.,  2002, \mnras, 331, 369

\bibitem[\protect\citeauthoryear{{Fabian}, {Sanders}, {Allen}, {Crawford},
  {Iwasawa}, {Johnstone}, {Schmidt} \& {Taylor}}{{Fabian}
  et~al.}{2003}]{fabianetal03}
{Fabian} A.~C.,  {Sanders} J.~S.,  {Allen} S.~W.,  {Crawford} C.~S.,  {Iwasawa}
  K.,  {Johnstone} R.~M.,  {Schmidt} R.~W.,    {Taylor} G.~B.,  2003, \mnras,
  344, L43

\bibitem[\protect\citeauthoryear{{Fabian}, {Sanders}, {Taylor}, {Allen},
  {Crawford}, {Johnstone} \& {Iwasawa}}{{Fabian} et~al.}{2006}]{fabianetal06}
{Fabian} A.~C.,  {Sanders} J.~S.,  {Taylor} G.~B.,  {Allen} S.~W.,  {Crawford}
  C.~S.,  {Johnstone} R.~M.,    {Iwasawa} K.,  2006, \mnras, 366, 417

\bibitem[\protect\citeauthoryear{{Forman}, {Jones}, {Churazov}, {Markevitch},
  {Nulsen}, {Vikhlinin}, {Begelman}, {B{\"o}hringer}, {Eilek}, {Heinz},
  {Kraft}, {Owen} \& {Pahre}}{{Forman} et~al.}{2007}]{formanetal07}
{Forman} W.,  {Jones} C.,  {Churazov} E.,  {Markevitch} M.,  {Nulsen} P.,
  {Vikhlinin} A.,  {Begelman} M.,  {B{\"o}hringer} H.,  {Eilek} J.,  {Heinz}
  S.,  {Kraft} R.,  {Owen} F.,    {Pahre} M.,  2007, \apj, 665, 1057

\bibitem[\protect\citeauthoryear{{Gaibler}, {Krause} \& {Camenzind}}{{Gaibler}
  et~al.}{2009}]{gaibleretal09}
{Gaibler} V.,  {Krause} M.,    {Camenzind} M.,  2009, \mnras, 400, 1785

\bibitem[\protect\citeauthoryear{{Gastaldello}, {Buote}, {Humphrey},
  {Zappacosta}, {Bullock}, {Brighenti} \& {Mathews}}{{Gastaldello}
  et~al.}{2007}]{gastaldelloetal06}
{Gastaldello} F.,  {Buote} D.~A.,  {Humphrey} P.~J.,  {Zappacosta} L.,
  {Bullock} J.~S.,  {Brighenti} F.,    {Mathews} W.~G.,  2007, \apj, 669, 158

\bibitem[\protect\citeauthoryear{{Giodini} S.~{et
  al}}{{Giodini}}{2009}]{giodinietal09}
{Giodini} S.~{et al} .,  2009, \apj, 703, 982

\bibitem[\protect\citeauthoryear{{Gonzalez}, {Zaritsky} \&
  {Zabludoff}}{{Gonzalez} et~al.}{2007}]{gonzalezetal07}
{Gonzalez} A.~H.,  {Zaritsky} D.,    {Zabludoff} A.~I.,  2007, \apj, 666, 147

\bibitem[\protect\citeauthoryear{{Haardt} \& {Madau}}{{Haardt} \&
  {Madau}}{1996}]{haardt&madau96}
{Haardt} F.,  {Madau} P.,  1996, \apj, 461, 20

\bibitem[\protect\citeauthoryear{{H{\"a}ring} \& {Rix}}{{H{\"a}ring} \&
  {Rix}}{2004}]{haring&rix04}
{H{\"a}ring} N.,  {Rix} H.-W.,  2004, \apjl, 604, L89

\bibitem[\protect\citeauthoryear{{Hawley} \& {Krolik}}{{Hawley} \&
  {Krolik}}{2006}]{hawley&krolik06}
{Hawley} J.~F.,  {Krolik} J.~H.,  2006, \apj, 641, 103

\bibitem[\protect\citeauthoryear{{Heinz}, {Br{\"u}ggen}, {Young} \&
  {Levesque}}{{Heinz} et~al.}{2006}]{heinzetal06}
{Heinz} S.,  {Br{\"u}ggen} M.,  {Young} A.,    {Levesque} E.,  2006, \mnras,
  373, L65

\bibitem[\protect\citeauthoryear{{Hopkins}, {Lidz}, {Hernquist}, {Coil},
  {Myers}, {Cox} \& {Spergel}}{{Hopkins} et~al.}{2007}]{hopkinsetal07}
{Hopkins} P.~F.,  {Lidz} A.,  {Hernquist} L.,  {Coil} A.~L.,  {Myers} A.~D.,
  {Cox} T.~J.,    {Spergel} D.~N.,  2007, \apj, 662, 110

\bibitem[\protect\citeauthoryear{{Kennicutt}
  Jr.}{{Kennicutt}}{1998}]{kennicutt98}
{Kennicutt} Jr. R.~C.,  1998, \apj, 498, 541

\bibitem[\protect\citeauthoryear{{Kere{\v s}}, {Katz}, {Weinberg} \&
  {Dav{\'e}}}{{Kere{\v s}} et~al.}{2005}]{keresetal05}
{Kere{\v s}} D.,  {Katz} N.,  {Weinberg} D.~H.,    {Dav{\'e}} R.,  2005,
  \mnras, 363, 2

\bibitem[\protect\citeauthoryear{{Khalatyan}, {Cattaneo}, {Schramm},
  {Gottl{\"o}ber}, {Steinmetz} \& {Wisotzki}}{{Khalatyan}
  et~al.}{2008}]{khalatyanetal08}
{Khalatyan} A.,  {Cattaneo} A.,  {Schramm} M.,  {Gottl{\"o}ber} S.,
  {Steinmetz} M.,    {Wisotzki} L.,  2008, \mnras, 387, 13

\bibitem[\protect\citeauthoryear{{King} \& {Pounds}}{{King} \&
  {Pounds}}{2003}]{king&pounds03}
{King} A.~R.,  {Pounds} K.~A.,  2003, \mnras, 345, 657

\bibitem[\protect\citeauthoryear{{Krumholz}, {McKee} \& {Klein}}{{Krumholz}
  et~al.}{2004}]{krumholzetal04}
{Krumholz} M.~R.,  {McKee} C.~F.,    {Klein} R.~I.,  2004, \apj, 611, 399

\bibitem[\protect\citeauthoryear{{Lin}, {Mohr} \& {Stanford}}{{Lin}
  et~al.}{2003}]{linetal03}
{Lin} Y.,  {Mohr} J.~J.,    {Stanford} S.~A.,  2003, \apj, 591, 749

\bibitem[\protect\citeauthoryear{{Magorrian}, {Tremaine}, {Richstone},
  {Bender}, {Bower}, {Dressler}, {Faber}, {Gebhardt}, {Green}, {Grillmair},
  {Kormendy} \& {Lauer}}{{Magorrian} et~al.}{1998}]{magorrianetal98}
{Magorrian} J.,  {Tremaine} S.,  {Richstone} D.,  {Bender} R.,  {Bower} G.,
  {Dressler} A.,  {Faber} S.~M.,  {Gebhardt} K.,  {Green} R.,  {Grillmair} C.,
  {Kormendy} J.,    {Lauer} T.,  1998, \aj, 115, 2285

\bibitem[\protect\citeauthoryear{{McNamara}, {Nulsen}, {Wise}, {Rafferty},
  {Carilli}, {Sarazin} \& {Blanton}}{{McNamara} et~al.}{2005}]{mcnamaraetal05}
{McNamara} B.~R.,  {Nulsen} P.~E.~J.,  {Wise} M.~W.,  {Rafferty} D.~A.,
  {Carilli} C.,  {Sarazin} C.~L.,    {Blanton} E.~L.,  2005, \nat, 433, 45

\bibitem[\protect\citeauthoryear{{McNamara}, {Wise}, {Nulsen}, {David},
  {Carilli}, {Sarazin}, {O'Dea}, {Houck}, {Donahue}, {Baum}, {Voit},
  {O'Connell} \& {Koekemoer}}{{McNamara} et~al.}{2001}]{mcnamaraetal01}
{McNamara} B.~R.,  {Wise} M.~W.,  {Nulsen} P.~E.~J.,  {David} L.~P.,  {Carilli}
  C.~L.,  {Sarazin} C.~L.,  {O'Dea} C.~P.,  {Houck} J.,  {Donahue} M.,  {Baum}
  S.,  {Voit} M.,  {O'Connell} R.~W.,    {Koekemoer} A.,  2001, \apjl, 562,
  L149

\bibitem[\protect\citeauthoryear{{Meneghetti}, {Rasia}, {Merten}, {Bellagamba},
  {Ettori}, {Mazzotta}, {Dolag} \& {Marri}}{{Meneghetti}
  et~al.}{2010}]{meneghettietal10}
{Meneghetti} M.,  {Rasia} E.,  {Merten} J.,  {Bellagamba} F.,  {Ettori} S.,
  {Mazzotta} P.,  {Dolag} K.,    {Marri} S.,  2010, \aap, 514, A93+

\bibitem[\protect\citeauthoryear{{Mihos} \& {Hernquist}}{{Mihos} \&
  {Hernquist}}{1996}]{mihos&hernquist96}
{Mihos} J.~C.,  {Hernquist} L.,  1996, \apj, 464, 641

\bibitem[\protect\citeauthoryear{{Miller}, {Percival}, {Croom} \&
  {Babi{\'c}}}{{Miller} et~al.}{2006}]{milleretal06}
{Miller} L.,  {Percival} W.~J.,  {Croom} S.~M.,    {Babi{\'c}} A.,  2006, \aap,
  459, 43

\bibitem[\protect\citeauthoryear{{Mitchell}, {McCarthy}, {Bower}, {Theuns} \&
  {Crain}}{{Mitchell} et~al.}{2009}]{mitchelletal09}
{Mitchell} N.~L.,  {McCarthy} I.~G.,  {Bower} R.~G.,  {Theuns} T.,    {Crain}
  R.~A.,  2009, \mnras, 395, 180

\bibitem[\protect\citeauthoryear{{Morgan}, {Kochanek}, {Morgan} \&
  {Falco}}{{Morgan} et~al.}{2010}]{morganetal10}
{Morgan} C.~W.,  {Kochanek} C.~S.,  {Morgan} N.~D.,    {Falco} E.~E.,  2010,
  \apj, 712, 1129

\bibitem[\protect\citeauthoryear{{Morsony}, {Heinz}, {Br{\"u}ggen} \&
  {Ruszkowski}}{{Morsony} et~al.}{2010}]{morsonyetal10}
{Morsony} B.~J.,  {Heinz} S.,  {Br{\"u}ggen} M.,    {Ruszkowski} M.,  2010,
  \mnras, 407, 1277

\bibitem[\protect\citeauthoryear{{Navarro} \& {White}}{{Navarro} \&
  {White}}{1993}]{navarro&white93}
{Navarro} J.~F.,  {White} S.~D.~M.,  1993, \mnras, 265, 271

\bibitem[\protect\citeauthoryear{{Nayakshin} \& {Power}}{{Nayakshin} \&
  {Power}}{2010}]{nayakshin&power10}
{Nayakshin} S.,  {Power} C.,  2010, \mnras, 402, 789

\bibitem[\protect\citeauthoryear{{Ocvirk}, {Pichon} \& {Teyssier}}{{Ocvirk}
  et~al.}{2008}]{ocvirketal08}
{Ocvirk} P.,  {Pichon} C.,    {Teyssier} R.,  2008, \mnras, 390, 1326

\bibitem[\protect\citeauthoryear{{Omma}, {Binney}, {Bryan} \& {Slyz}}{{Omma}
  et~al.}{2004}]{ommaetal04}
{Omma} H.,  {Binney} J.,  {Bryan} G.,    {Slyz} A.,  2004, \mnras, 348, 1105

\bibitem[\protect\citeauthoryear{{O'Neill} \& {Jones}}{{O'Neill} \&
  {Jones}}{2010}]{oneill&jones10}
{O'Neill} S.~M.,  {Jones} T.~W.,  2010, \apj, 710, 180

\bibitem[\protect\citeauthoryear{{Owen}, {Eilek} \& {Kassim}}{{Owen}
  et~al.}{2000}]{owenetal00}
{Owen} F.~N.,  {Eilek} J.~A.,    {Kassim} N.~E.,  2000, \apj, 543, 611

\bibitem[\protect\citeauthoryear{{Parrish} \& {Quataert}}{{Parrish} \&
  {Quataert}}{2008}]{parrish&quataert08}
{Parrish} I.~J.,  {Quataert} E.,  2008, \apjl, 677, L9

\bibitem[\protect\citeauthoryear{{Parrish}, {Quataert} \& {Sharma}}{{Parrish}
  et~al.}{2009}]{parrish09}
{Parrish} I.~J.,  {Quataert} E.,    {Sharma} P.,  2009, \apj, 703, 96

\bibitem[\protect\citeauthoryear{{Parrish}, {Quataert} \& {Sharma}}{{Parrish}
  et~al.}{2010}]{parrishetal10}
{Parrish} I.~J.,  {Quataert} E.,    {Sharma} P.,  2010, \apjl, 712, L194

\bibitem[\protect\citeauthoryear{{Peirani}, {Alard}, {Pichon}, {Gavazzi} \&
  {Aubert}}{{Peirani} et~al.}{2008}]{peiranietal08}
{Peirani} S.,  {Alard} C.,  {Pichon} C.,  {Gavazzi} R.,    {Aubert} D.,  2008,
  \mnras, 390, 945

\bibitem[\protect\citeauthoryear{{Pfrommer} \& {Jonathan Dursi}}{{Pfrommer} \&
  {Jonathan Dursi}}{2010}]{pfrommer&dursi09}
{Pfrommer} C.,  {Jonathan Dursi} L.,  2010, Nature Physics, 6, 520

\bibitem[\protect\citeauthoryear{{Power}, {Nayakshin} \& {King}}{{Power}
  et~al.}{2010}]{poweretal10}
{Power} C.,  {Nayakshin} S.,    {King} A.,  2010, ArXiv e-prints

\bibitem[\protect\citeauthoryear{{Quataert}}{{Quataert}}{2008}]{quataert08}
{Quataert} E.,  2008, \apj, 673, 758

\bibitem[\protect\citeauthoryear{{Quilis}, {Bower} \& {Balogh}}{{Quilis}
  et~al.}{2001}]{quilisetal01}
{Quilis} V.,  {Bower} R.~G.,    {Balogh} M.~L.,  2001, \mnras, 328, 1091

\bibitem[\protect\citeauthoryear{{Rasera} \& {Teyssier}}{{Rasera} \&
  {Teyssier}}{2006}]{rasera&teyssier06}
{Rasera} Y.,  {Teyssier} R.,  2006, \aap, 445, 1

\bibitem[\protect\citeauthoryear{{Rephaeli} \& {Silk}}{{Rephaeli} \&
  {Silk}}{1995}]{rephaeli&silk95}
{Rephaeli} Y.,  {Silk} J.,  1995, \apj, 442, 91

\bibitem[\protect\citeauthoryear{{Reynolds}, {Heinz} \& {Begelman}}{{Reynolds}
  et~al.}{2001}]{reynoldsetal01}
{Reynolds} C.~S.,  {Heinz} S.,    {Begelman} M.~C.,  2001, \apjl, 549, L179

\bibitem[\protect\citeauthoryear{{Romeo}, {Agertz}, {Moore} \&
  {Stadel}}{{Romeo} et~al.}{2008}]{romeoetal08}
{Romeo} A.~B.,  {Agertz} O.,  {Moore} B.,    {Stadel} J.,  2008, \apj, 686, 1

\bibitem[\protect\citeauthoryear{{Ruszkowski}, {Br{\"u}ggen} \&
  {Begelman}}{{Ruszkowski} et~al.}{2004a}]{ruszkowskietal04b}
{Ruszkowski} M.,  {Br{\"u}ggen} M.,    {Begelman} M.~C.,  2004a, \apj, 611, 158

\bibitem[\protect\citeauthoryear{{Ruszkowski}, {Br{\"u}ggen} \&
  {Begelman}}{{Ruszkowski} et~al.}{2004b}]{ruszkowskietal04}
{Ruszkowski} M.,  {Br{\"u}ggen} M.,    {Begelman} M.~C.,  2004b, \apj, 615, 675

\bibitem[\protect\citeauthoryear{{Ruszkowski} \& {Oh}}{{Ruszkowski} \&
  {Oh}}{2010}]{ruszkowski&oh10}
{Ruszkowski} M.,  {Oh} S.~P.,  2010, \apj, 713, 1332

\bibitem[\protect\citeauthoryear{{Salpeter}}{{Salpeter}}{1955}]{salpeter55}
{Salpeter} E.~E.,  1955, \apj, 121, 161

\bibitem[\protect\citeauthoryear{{Shakura} \& {Sunyaev}}{{Shakura} \&
  {Sunyaev}}{1973}]{shakura&sunyaev73}
{Shakura} N.~I.,  {Sunyaev} R.~A.,  1973, \aap, 24, 337

\bibitem[\protect\citeauthoryear{{Shapiro}, {Genzel}, {F{\"o}rster Schreiber},
  {Tacconi} \& {et al,}}{{Shapiro} et~al.}{2008}]{shapiroetal08}
{Shapiro} K.~L.,  {Genzel} R.,  {F{\"o}rster Schreiber} N.~M.,  {Tacconi}
  L.~J.,    {et al,} 2008, \apj, 682, 231

\bibitem[\protect\citeauthoryear{{Sijacki}, {Pfrommer}, {Springel} \&
  {En{\ss}lin}}{{Sijacki} et~al.}{2008}]{sijackietal08}
{Sijacki} D.,  {Pfrommer} C.,  {Springel} V.,    {En{\ss}lin} T.~A.,  2008,
  \mnras, 387, 1403

\bibitem[\protect\citeauthoryear{{Sijacki}, {Springel}, {di Matteo} \&
  {Hernquist}}{{Sijacki} et~al.}{2007}]{sijackietal07}
{Sijacki} D.,  {Springel} V.,  {di Matteo} T.,    {Hernquist} L.,  2007,
  \mnras, 380, 877

\bibitem[\protect\citeauthoryear{{Silk}}{{Silk}}{1977}]{silk77}
{Silk} J.,  1977, \apj, 211, 638

\bibitem[\protect\citeauthoryear{{Simionescu}, {Roediger}, {Nulsen},
  {Br{\"u}ggen}, {Forman}, {B{\"o}hringer}, {Werner} \&
  {Finoguenov}}{{Simionescu} et~al.}{2009}]{simionescuetal09}
{Simionescu} A.,  {Roediger} E.,  {Nulsen} P.~E.~J.,  {Br{\"u}ggen} M.,
  {Forman} W.~R.,  {B{\"o}hringer} H.,  {Werner} N.,    {Finoguenov} A.,  2009,
  \aap, 495, 721

\bibitem[\protect\citeauthoryear{{Spergel}, {Verde}, {Peiris}, {Komatsu},
  {Nolta}, {Bennett}, {Halpern}, {Hinshaw}, {Jarosik}, {Kogut}, {Limon},
  {Meyer}, {Page}, {Tucker}, {Weiland}, {Wollack} \& {Wright}}{{Spergel}
  et~al.}{2003}]{spergeletal03}
{Spergel} D.~N.,  {Verde} L.,  {Peiris} H.~V.,  {Komatsu} E.,  {Nolta} M.~R.,
  {Bennett} C.~L.,  {Halpern} M.,  {Hinshaw} G.,  {Jarosik} N.,  {Kogut} A.,
  {Limon} M.,  {Meyer} S.~S.,  {Page} L.,  {Tucker} G.~S.,  {Weiland} J.~L.,
  {Wollack} E.,    {Wright} E.~L.,  2003, \apjs, 148, 175

\bibitem[\protect\citeauthoryear{{Springel}, {Di Matteo} \&
  {Hernquist}}{{Springel} et~al.}{2005}]{springeletal05}
{Springel} V.,  {Di Matteo} T.,    {Hernquist} L.,  2005, \mnras, 361, 776

\bibitem[\protect\citeauthoryear{{Springel} \& {Hernquist}}{{Springel} \&
  {Hernquist}}{2003}]{springel&hernquist03}
{Springel} V.,  {Hernquist} L.,  2003, \mnras, 339, 289

\bibitem[\protect\citeauthoryear{{Sun}, {Voit}, {Donahue}, {Jones}, {Forman} \&
  {Vikhlinin}}{{Sun} et~al.}{2009}]{sunetal09}
{Sun} M.,  {Voit} G.~M.,  {Donahue} M.,  {Jones} C.,  {Forman} W.,
  {Vikhlinin} A.,  2009, \apj, 693, 1142

\bibitem[\protect\citeauthoryear{{Sutherland} \& {Dopita}}{{Sutherland} \&
  {Dopita}}{1993}]{sutherland&dopita93}
{Sutherland} R.~S.,  {Dopita} M.~A.,  1993, \apjs, 88, 253

\bibitem[\protect\citeauthoryear{{Taylor}, {Sanders}, {Fabian} \&
  {Allen}}{{Taylor} et~al.}{2006}]{tayloretal06}
{Taylor} G.~B.,  {Sanders} J.~S.,  {Fabian} A.~C.,    {Allen} S.~W.,  2006,
  \mnras, 365, 705

\bibitem[\protect\citeauthoryear{{Teyssier}}{{Teyssier}}{2002}]{teyssier02}
{Teyssier} R.,  2002, \aap, 385, 337

\bibitem[\protect\citeauthoryear{{Teyssier}, {Moore}, {Martizzi}, {Dubois} \&
  {Mayer}}{{Teyssier} et~al.}{2010}]{teyssieretal10}
{Teyssier} R.,  {Moore} B.,  {Martizzi} D.,  {Dubois} Y.,    {Mayer} L.,  2010,
  ArXiv e-prints

\bibitem[\protect\citeauthoryear{{Tozzi} \& {Norman}}{{Tozzi} \&
  {Norman}}{2001}]{tozzi&norman01}
{Tozzi} P.,  {Norman} C.,  2001, \apj, 546, 63

\bibitem[\protect\citeauthoryear{{Tremaine}, {Gebhardt}, {Bender}, {Bower},
  {Dressler}, {Faber}, {Filippenko}, {Green}, {Grillmair}, {Ho}, {Kormendy},
  {Lauer}, {Magorrian}, {Pinkney} \& {Richstone}}{{Tremaine}
  et~al.}{2002}]{tremaineetal02}
{Tremaine} S.,  {Gebhardt} K.,  {Bender} R.,  {Bower} G.,  {Dressler} A.,
  {Faber} S.~M.,  {Filippenko} A.~V.,  {Green} R.,  {Grillmair} C.,  {Ho}
  L.~C.,  {Kormendy} J.,  {Lauer} T.~R.,  {Magorrian} J.,  {Pinkney} J.,
  {Richstone} D.,  2002, \apj, 574, 740

\bibitem[\protect\citeauthoryear{{Truelove}, {Klein}, {McKee}, {Holliman} II,
  {Howell} \& {Greenough}}{{Truelove} et~al.}{1997}]{trueloveetal97}
{Truelove} J.~K.,  {Klein} R.~I.,  {McKee} C.~F.,  {Holliman} II J.~H.,
  {Howell} L.~H.,    {Greenough} J.~A.,  1997, \apjl, 489, L179+

\bibitem[\protect\citeauthoryear{{Tweed}, {Devriendt}, {Blaizot}, {Colombi} \&
  {Slyz}}{{Tweed} et~al.}{2009}]{tweedetal09}
{Tweed} D.,  {Devriendt} J.,  {Blaizot} J.,  {Colombi} S.,    {Slyz} A.,  2009,
  \aap, 506, 647

\bibitem[\protect\citeauthoryear{{Vernaleo} \& {Reynolds}}{{Vernaleo} \&
  {Reynolds}}{2006}]{vernaleo&reynolds06}
{Vernaleo} J.~C.,  {Reynolds} C.~S.,  2006, \apj, 645, 83

\bibitem[\protect\citeauthoryear{{Vikhlinin}, {Kravtsov}, {Forman}, {Jones},
  {Markevitch}, {Murray} \& {Van Speybroeck}}{{Vikhlinin}
  et~al.}{2006}]{vikhlininetal06}
{Vikhlinin} A.,  {Kravtsov} A.,  {Forman} W.,  {Jones} C.,  {Markevitch} M.,
  {Murray} S.~S.,    {Van Speybroeck} L.,  2006, \apj, 640, 691

\bibitem[\protect\citeauthoryear{{Voigt} \& {Fabian}}{{Voigt} \&
  {Fabian}}{2004}]{voigt&fabian04}
{Voigt} L.~M.,  {Fabian} A.~C.,  2004, \mnras, 347, 1130

\end{thebibliography}

\label{lastpage}

\end{document}